\documentclass[aoas]{imsart}

\RequirePackage{amsthm,amsmath,amsfonts,amssymb}
\RequirePackage[authoryear]{natbib}

\startlocaldefs
\usepackage{graphicx}                 
\usepackage{tcolorbox}
\usepackage{bbm,color}
\usepackage{xcolor,colortbl}
\usepackage{url}
\definecolor{dred}{rgb}{1,0,0}
\definecolor{dgreen}{rgb}{0.01,.65,0.25}
\definecolor{dblue}{rgb}{0,0,1}

\def \N{\mbox{N}}
\def \v{\mbox{var}}
\def \E{\mbox{E}}

\newcommand{\bE}{\mbox{\boldmath $E$}}

\newcommand{\bY}{\mbox{\boldmath $Y$}}

\newcommand{\ba}{\mbox{\boldmath $a$}}

\newcommand{\bp}{\mbox{\boldmath $p$}}

\newcommand{\bx}{\mbox{\boldmath $x$}}
\newcommand{\by}{\mbox{\boldmath $y$}}
\newcommand{\bz}{\mbox{\boldmath $z$}}

\newcommand{\blambda}{\mbox{\boldmath $\lambda$}}

\newcommand{\bdelta}{\mbox{\boldmath $\delta$}}

\newcommand{\btheta}{\mbox{\boldmath $\theta$}}

\newcommand{\bphi}{\mbox{\boldmath $\phi$}}

\newcommand{\bomega}{\mbox{\boldmath $\bomega$}}

\newcommand{\bc}{\begin{center}}
\newcommand{\ec}{\end{center}}

\newcommand{\bi}{\begin{itemize}}
\newcommand{\ei}{\end{itemize}}

\newcommand{\be}{\begin{enumerate}}
\newcommand{\ee}{\end{enumerate}}

\newcommand{\bs}{\begin{slide*}}
\newcommand{\es}{\end{slide*}}

\newcommand{\bq}{ \begin{equation} }
\newcommand{\eq}{\end{equation}}

\endlocaldefs

\begin{document}

\begin{frontmatter}
\title{Estimating Global and Country-Specific Excess Mortality During the COVID-19 Pandemic}
\runtitle{Estimating Excess Mortality}
\begin{aug}
\author[A]{\fnms{Victoria} \snm{Knutson}\ead[label=e1]{vknuts@uw.edu}},
\author[A]{\fnms{Serge} \snm{Aleshin-Guendel}\ead[label=e2]{aleshing@uw.edu}},
\author[C]{\fnms{Ariel}\snm{ Karlinsky}\ead[label=e3]{ariel.karlinsky@mail.huji.ac.il}},
\author[D]{\fnms{William }\snm{Msemburi}\ead[label=e4]{msemburiw@who.int}},
\author[A,E]{\fnms{Jon} \snm{Wakefield}\ead[label=e5]{jonno@uw.edu}}
% Debbie Bradshaw$^3$, 
%William Msemburi$^3$, Jon Wakefield$^{1,4}$}
\address[A]{Department of Biostatistics, University of Washington, Seattle, USA}
%\address[B]{Department of Biostatistics, University of Washington, Seattle, USA}
\address[C]{Hebrew University, Jerusalem, Israel}
\address[D]{World Health Organization, Geneva, Switzerland}
\address[E]{Department of Statistics, University of Washington, Seattle, USA}
%
%\affil{$^2$Hebrew University, Jerusalem, Israel}
%
%%\affil{$^3$Burden of Disease Research Unit, Medical Research Council of South Africa, Tygerberg, South Africa}
%\affil{$^3$World Health Organisation, Geneva, Switzerland}
%\affil{$^4$Department of Statistics, University of Washington, Seattle, USA}
\end{aug}
%\begin{document}

%\maketitle
 \begin{abstract}
{\bf Abstract:} Estimating the true mortality burden of COVID-19 for every country in the world is a difficult, but crucial, public health endeavor. Attributing deaths, direct or indirect, to COVID-19 is problematic.  A more attainable target is the ``excess deaths'', the number of deaths in a particular period, relative to that expected during ``normal times'', and we estimate this for all countries on a monthly time scale for 2020 and 2021. The excess mortality requires two numbers, the total deaths and the expected deaths, but the former is unavailable for many countries, and so modeling is required for these countries. The expected deaths are based on historic data and we develop a model for producing expected estimates for all countries and we allow for uncertainty in the modeled expected numbers when calculating the excess. We describe the methods that were developed to produce the World Health Organization (WHO) excess death estimates. To achieve both interpretability and transparency we developed a relatively simple overdispersed Poisson count framework, within which the various data types can be modeled. We use data from countries with national monthly data to build a predictive log-linear regression model with time-varying coefficients for countries without data. For a number of countries, subnational data only are available, and we construct a multinomial model for such data, based on the assumption that the fractions of deaths in sub-regions remain approximately constant over time. 
Our inferential approach is Bayesian, with the covariate predictive model  being implemented in the fast and accurate {\tt INLA} software. The subnational modeling was carried out using MCMC in {\tt Stan} or in some non-standard data situations, using our own MCMC code. Based on our modeling, the point estimate for global excess mortality, over 2020--2021, is 14.9 million, with a 95\% credible interval of (13.3, 16.6) million. This leads to a point estimate of the ratio of excess deaths to reported COVID-19 deaths of 2.75, which is a huge discrepancy.
\end{abstract}

\begin{keyword}
\kwd{Bayesian inference}
\kwd{Global health}
\kwd{Poisson framework}
%\kwd{Mortality}
\kwd{Subnational modeling}
\end{keyword}

\end{frontmatter}

\section{Introduction}\label{sec:introduction}

The World Health Organization (WHO) has been tracking the impact of COVID-19 as the pandemic has evolved over time. Aggregate case and COVID-19 death numbers are reported to the WHO by countries, and the data have been made publicly available at \url{https://covid19.who.int/}. For a number of reasons, these reported data neither provide a complete picture of the health burden attributable to COVID-19, nor of how many lives have been lost, both directly and indirectly, due to the pandemic. Some deaths that are attributable to COVID-19 have not been certified as such because tests had not been conducted prior to death.  Deaths may also have been mistakenly certified as COVID-19, though this is less likely. The latter does not affect our estimates of excess mortality based on all-cause mortality (ACM) data, however, but does cause  the ratio of excess mortality to reported COVID-19 deaths to be lower than if such mistaken certification did not occur.
%There have also been variations in the death certification rules countries have applied in the presence of comorbidities and COVID-19. 
There have also been variations in the death certification rules countries have applied in regards to COVID-19 \citep{garcia2021differences,riffe2021data}.
The impact of the pandemic is far reaching. Beyond the deaths directly attributable to it are those that can be linked to the conditions that have prevailed since the pandemic began and have led to some health systems being overwhelmed or some patients avoiding healthcare.  In countries where COVID-19 spread was limited, due to lockdown measures or otherwise, some potential causes of death  have decreased, such as those attributable to air pollution, or  traffic accidents, or from other communicable diseases such as influenza like illness, resulting in negative excess or deficit deaths \citep{kung2020reduced, WMD}. 
In light of the challenges posed by using reported COVID-19 data, excess mortality is considered a more objective and comparable (across countries) measure of the mortality impact of COVID-19 \citep{leon2020covid}. The WHO defines excess mortality as, ``the mortality above what would be expected based on the non-crisis mortality rate in the population of interest'' (\url{https://www.who.int/hac/about/definitions/en/}). Knowledge of the excess deaths not only paints a clearer picture of the pandemic, but can also aid in implementing public health initiatives. The world also has a moral obligation to count the number of deaths attributable to the pandemic.

The ACM counts in country  $c$ and in month $t$ (for months in 2020 and 2021) are denoted by $Y_{c,t}$. These counts, in addition to the contribution from expected deaths, are assumed to be a result of the direct effects of COVID-19 (i.e.,~deaths attributable to it) and the indirect knock-on effects on health systems and society, along with deaths that were averted. 
 The choice of a monthly time scale gives sufficient temporal resolution for most public health purposes. The hypothetical or ``counterfactual'' no-COVID-19 scenario uses the expected death numbers $E_{c,t}$, which have been forecasted to month $t$, using historic (prior to the pandemic) deaths data, usually over 2015--2019. Excess deaths are defined as:
\begin{equation}\label{eq:excess}
\delta_{c,t} = Y_{c,t} - E_{c,t}
\end{equation}
for country $c$ where $c=1,\dots,194$, and in month $t$ where $t=1,\dots,24$, represent months in  2020 and 2021.

The exercise of determining excess deaths for all countries is non-trivial, because the required ACM counts $Y_{c,t}$ are currently unavailable for many country/month combinations. Routine mortality data is often received by the WHO a year or more after the year of death. In addition, differential reporting capacity and variable data quality across countries has resulted in many nations lacking the systems to provide good quality routine data even historically \citep{mikkelsen_global_2015, adair_2018, gbd_2019, un_demographic_yearbook, icdr_2021}.
% has resulted in many nations lacking the systems to provide good quality routine data even in the past.
Correspondingly, these countries lack the capacity  required to monitor ACM during the unprecedented COVID-19 pandemic. Hence, a number of countries are unable to contribute to the centralized systematic mortality surveillance that would be needed to measure global, regional and country level excess mortality by the WHO.

In this paper we describe our ongoing methods development to produce the WHO excess mortality estimates. In Section \ref{sec:data} we discuss data sources, before describing models for estimation of  the expected numbers in Section \ref{sec:expected}. Section \ref{sec:observed:national} describes our national covariate model and in Section \ref{sec:observed:subnational} we outline the models we used for countries with subnational monthly data, national annual data, or a combination. Section  \ref{sec:results} provides the main results, with more extensive summaries appearing in the Supplementary Materials. 
Two other sets of global estimates of excess deaths have been produced by The Economist and the Institute for Health Metrics and Evaluation (IHME) with the latter being described in \cite{wang:2022:covid}. We fully describe and critique these methods in  Section \ref{sec:other}.
%; for the pandemic period both approaches are centered on directly modeling pandemic excess mortality rates, whereas we specify a distribution for the raw counts. 
%We have different models  for different data scenarios but in each case the starting point is the Poisson distribution. 
The paper concludes with a discussion in Section \ref{sec:discussion}.%Some potential causes of death, such as traffic accidents and communicable diseases, may have been avoided ti give a negative excess, in countries in which the pandemic was not felt.
\clearpage
\section{Data Sources}\label{sec:data}

\subsection{Mortality Data}\label{sec:mortality}

Excess mortality cannot be directly measured for all countries due to many not having the required ACM data. The WHO usually receives routine mortality data on an annual basis in  the year after the year of death or perhaps after an even greater lag. Civil registration and vital statistics (CRVS) systems differ greatly across countries with varying timelines and quality control measures for compiling unit record cause-of-death numbers into aggregates identified by cause, age, sex, place, and period of death. In addition, differential reporting coverage, the absence of electronic surveillance systems in some locations and limited investments in CRVS systems has resulted in many nations lacking the structures necessary to provide good quality routine data, even before the COVID-19 pandemic. This lack of capacity and the data required to monitor ACM has been exacerbated during the unprecedented pandemic. Therefore, many countries are unable to contribute to a centralized systematic mortality surveillance that would be needed to measure global, regional and country level excess mortality by the WHO.
%This results in the greater number of countries left unable to contribute to a centralized systematic mortality surveillance that would be needed to measure global, regional and country level excess mortality by the WHO.

All countries report their official COVID-19 death count, but for many countries we would not expect this to be accurate, and for many countries we would expect serious underestimation, for the reasons already outlined and for political reasons. However, the official count does provide an interesting summary for comparison with the estimated excess, and the COVID-19 death rate is used as a covariate in our ACM estimation model.

For this study, our main sources of data are reports of ACM  as collected and reported by countries' relevant institutions -- from national statistics offices, ministries of health, population registries, etc. These have been collected in several repositories such as the data routinely shared with WHO as part of its standing agreement with member states, Eurostat, The Human Mortality Database (HMD) as part of the Short-Term Mortality Fluctuations (STMF) project \citep{nemeth2021open} and the World Mortality Dataset (WMD), as described in \cite{WMD}. 
Monthly data are included after accounting for delayed registration either by adjusting for registration delay (Australia, Brazil, United States) or by not-including highly incomplete months.
%Data are only included after late registration has been considered.
%The analyzed time frame is from 2015 up to the latest available data in 2021, with this time frame varying by country. 

\begin{table}[htp]
\begin{center}
\begin{tabular}{|l|c|c|c|c|c|c|}
Region & Full & Partial & Subnational   & No & Total&Proportion of\\ 
& National &National& and/or Annual & Data&&Population \\\hline
AFRO &4&2&0&41&47&0.13\\
AMRO &12&11&4&8&35 &0.92\\
EMRO &4&5&0&12&21 &0.32\\
EURO &46&5&1&1&53&0.98 \\
SEARO &1&1&3&6&11&0.67 \\
WPRO &6&3&2&16 &27&0.95\\
 \hline
Global &73 &27&10& 84 &194&0.70
\end{tabular}
\end{center}
\caption{Country data availability summary for 2020 and 2021. Full national countries have data over all 24 months and partial  national have data for less than 24 months; for example, 83 countries have data for at least the first 18 months, and 96 countries have data for at least the first 12 months. ``Subnational/Annual Data'' refers to countries with subnational monthly data for some period (4 countries), national annual data  (5 countries) or a combination (China). WHO regions: African Region (AFRO), Region of the Americas (AMRO), Eastern Mediterranean Region (EMRO), European Region (EURO), South-East Asian Region (SEARO), Western Pacific Region (WPRO). The ``Proportion Population'' column denotes the proportion of the population that is contained in the available database, and is calculated at the country-month level. This proportion includes the contribution from subnational sources, where we estimate the proportion of deaths that occur in a month in the subnational regions, and multiply this by the country population. It also includes the countries for which we have annual data. The Supplementary Materials include a table that lists the type of the data available for each country.}\label{tab:datamortsummary}
\end{table}%

In this paper we report the current state of data at our disposal. This project is ongoing and data is added as soon as available.
%The work we report on here is a snapshot of the current state of data availability and over time the situation will improve.
%collection and in the future data availability will improve. 
Table \ref{tab:datamortsummary}  shows the breakdown of data availability by WHO region.
Just over a half (100) of the 194 countries provide monthly national data from at least some of the pandemic period, while 10 other countries provide subnational monthly data, national annual data, or a combination of the two (this includes Argentina which has partial national and subnational data, so could be placed in the partial or subnational/annual data boxes). It is immediately clear that there is a huge regional imbalance in data availability, with the EURO region being very well represented (with 52 out of 53 countries providing data), the AMRO region having data from 77\% of the countries, and other regions being more poorly represented. For example, in the AFRO region we only have data from 6 out of 47 countries. The WPRO region is dominated, population-wise, by China for which we have annual data. For those countries with data in month $t$, we assume that the ACM part of the excess $\delta_{c,t}$, as defined in (\ref{eq:excess}), is known exactly. Hence, we do not account for inaccuracies in the reported deaths, beyond accounting for delayed registration and under-reporting. With respect to the latter, when data are reported to the WHO, certain checks are carried out to determine whether the data are complete, via comparison with census data, for example \citep{WHO2020}. Based on these checks a scaling of the raw counts may be performed. We provide further discussion on this issue in Section \ref{sec:discussion}.

%For all countries we do, however, account for uncertainty in the expected numbers, as discussed in Section \ref{sec:expected}.
%Countries without data are referred to as Tier 2 countries.

%\begin{table}[htp]
%\begin{center}
%\begin{tabular}{|l|c|c|c|}
%Region & Total Countries & With Data 2020& With Data 2021\\ \hline
%EMRO & 21 & 8&6\\
%AFRO & 47 & 3&3\\
%EURO & 53 & 50&49\\
%AMRO & 35 &22&13\\
%WPRO & 27 & 10&9 \\
%SEARO & 11 & 3 &3\\ \hline
%Global & 194 & 96 & 83
%\end{tabular}
%\end{center}
%\caption{Country data availability summary for 2020, and up to June 2021.}\label{tab:datamortsummary}
%\end{table}%

\subsection{Covariate Data}

For countries with no data, we predict the ACM count using a log-linear covariate model.
A range of covariates were considered, including a high income country binary indicator, the COVID-19 test positivity rate, the COVID-19 death rate, temperature, 
population density, a socio-demographic index (SDI), the human development index (HDI), stringency (index for lockdown restrictions and closures, overall government response), economic measures (including measures such as income support and debt relief), containment, and the historic (from 2019): non-communicable disease rate,  cardiovascular disease rate, HIV rate, diabetes prevalence, life expectancy, proportion of the population under-15, proportion of the population over-65. The containment measure combines ``lockdown'' restrictions and closures with measures such as testing policy and contact tracing, short term investment in healthcare, as well investments in vaccines -- it is calculated using all ordinal containment and closure policy indicators and health system policy indicators, for further details see \cite{hale2020variation}. Some of the covariates are time-varying (COVID-19 test positivity rate, COVID-19 death rate, temperature, stringency, overall government response, containment), while the remainder are constant over time. A number of the covariates  were not available by month for all countries and so their values were imputed. Specifically, (WHO) regional medians were used for countries with missing data. The historic country-level covariates are taken from \cite{gbd2020global} and so are modeled. Some of the covariates are modeled also. For more details on the covariates, see the ``Data detail'' tab at \url{https://msemburi.shinyapps.io/excessvis/}.

\section{Expected Mortality Modeling}\label{sec:expected}

%Overdispersion is the norm and to address this add either gamma or normal random effects twe use either negative binomial or add normal random effects to address this.

A key component of  the excess mortality calculation is the ACM count that would be expected in non-pandemic times, for each country and month. We describe models for two types of countries: those that have historic {\it monthly} ACM data, and those that have historic {\it annual} ACM data only -- there are 100  and 94 countries in these categories, respectively. In terms of the period upon which we base the expected numbers, it is usually 2015--2019 for countries with monthly historical data, and is usually 2000--2019 for countries with annual historical data.

\subsection{Countries with Monthly Data}

We consider first those countries with monthly ACM data over multiple years. For country $c$,  $Y_{c,t}$ represents the ACM count for country $c$ and month $t$, for $t=1,\dots,M_c$, where $M_c$ is the number of historic months for which we have data.  We assume the sampling model for $Y_{c,t}$ is,
$$
Y_{c,t} | \mu_{c,t} \sim \mbox{NegBin}(\mu^\text{E}_{c,t}, \phi^\text{E}_c),
$$
parametrized in terms of the
mean, $\mu^\text{E}_{c,t}$, and the overdispersion parameter, $\phi^\text{E}_c$, such that  $\v(Y_{c,t} | \mu^\text{E}_{c,t},\phi^\text{E}_c) =\mu^\text{E}_{c,t}(1+\mu^\text{E}_{c,t}/\phi^\text{E}_c)$, with the Poisson model being recovered as $\phi^\text{E}_c \to \infty$.
We let $v[t]$ index the year in which month $t$ occurred (for example, labeled $1,\dots,5$ when data are available for 2015--2019) and $m[t]$ be the month (labeled $1,\dots,12$), so that given $v,m$ we can
%The scale parameter is assigned a prior, and estimated. Hence, we can
find $t$ as $t = 12(v-1)  + m$.
The mean is modeled as,
\begin{equation}\label{eq:eta}
%\log (\E[Y_{c,t}] ) =
\eta_{c,t}= \log (\mu_{c,t}) = f^\text{y}_{c}( v[t] ) +f^\text{m}_{c}(m[t])
\end{equation}
where  $f^\text{y}_{c}(\cdot)$ models the {\it annual trend}, and $f^\text{m}_{c}(\cdot)$ is a smooth function of time $t$ which accounts for {\it within-year} seasonal variation.  The yearly trend is modeled with a thin-plate spline  and  within-year variation with a cyclic cubic spline \citep{rivera2020excess}. In both cases we use the {\tt gam} function in the {\tt mgcv} package with REML  used to select smoothing parameters (and with the default settings). The spline model is fitted separately for each country. Algeria, Iraq and Sri Lanka have less than three years of historical data, and so a linear term is used for modeling yearly variation.
This model is used to obtain predictions of the expected deaths ${\mu}^\text{E}_{c,t}$ for all $t$ in  2020 and 2021, with both a point estimate and a standard error being produced, and these can be viewed as summaries of the posterior distribution, see Section 6.10 of Wood (2017)\nocite{wood2017generalized} for details.

\subsection{Countries with Annual Data}\label{sec:expected:annual}

For countries with only annual historic data, the goal is to predict expected numbers by month $t$ for $t =1, \dots, 24$. We summarize our strategy for producing expected numbers for countries with annual data only, before giving details:
\begin{enumerate}
\item Fit a negative binomial spline model to the countries with annual counts only. Use the spline to predict the total annual ACM for 2020 and 2021, for these countries.%, using {\tt mgcv}.
\item In a separate exercise, fit the multinomial model to all of the countries with monthly data, with deaths being attributed via the log-linear temperature model.% (\ref{eq:multloglin}). This produces an estimate  $\widehat\beta$.
\item Combine the spline model with the multinomial model using monthly temperature apportionment to obtain expected numbers for the countries without monthly data.
\end{enumerate}

\begin{figure}[h]
\centering
\includegraphics[scale=.5]{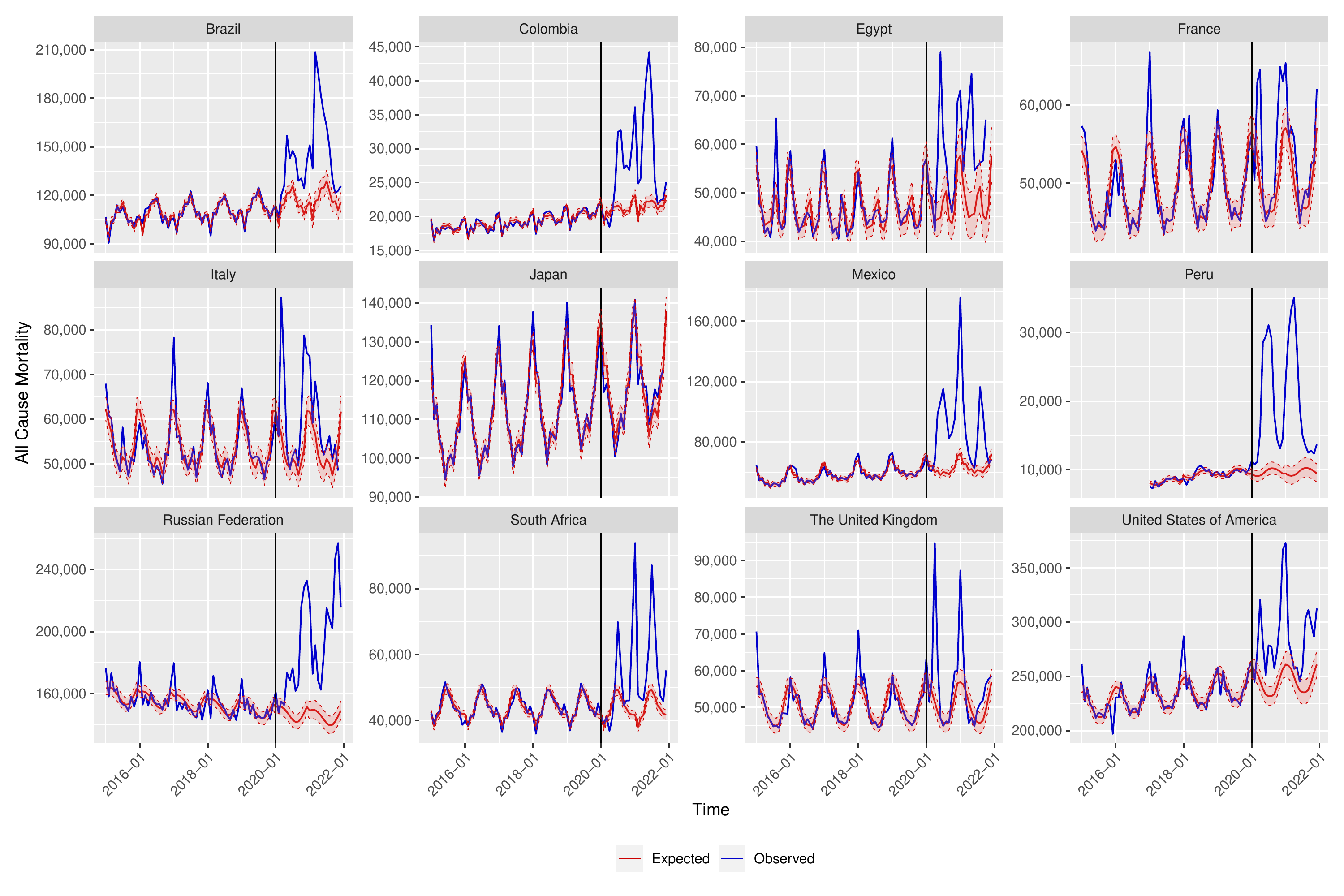}
\caption{Monthly time series of all cause mortality: expected counts in red and observed counts in blue,  for selected countries. The black vertical line is drawn at the start of 2020. The dashed red bands denote 95\% credible intervals for the mean expected numbers. For these countries, ACM counts are available for all months apart from Egypt, for which the last month is missing. We emphasize that the credible bands are for the mean function, and so we would not expect 95\% of the observed points to fall within these intervals.}
\label{fig:acm_monthly_tier1_plot}
\end{figure}

The annual trend can be estimated for each country using the method we described in the previous section minus the monthly term, i.e.,~by using a spline for year. To apportion the yearly totals to the months, we use the fact that a collection of Poisson random variables conditioned on their sum produce a multinomial distribution with within-year variation modeled using temperature, which is acting as a surrogate for seasonality. It is well-known that mortality is associated with temperature (see for example \cite{parks2018national}), and we wanted a relatively simple model, using a well-measured variable.
This relationship is learned from countries with historic monthly data. We use a smooth series of monthly temperatures since 2015. 
Let $\bY_{c,v}=\{ Y_{c,v,m}, m=1,\dots,12 \}$ be the vector  that contains the ACM counts by month in year $v$, $v=1,\dots,5$. Suppose each of the 12 constituent counts are Poisson with mean $\zeta_{c,v,m}$, for $m=1,\dots,12$.  Then, within year $v$, conditional on the total ACM, $Y^+_{c,v}$,
$$\bY_{c,v} | Y^+_{c,v} , \bp_{c,v} \sim\mbox{Multinomial} ( Y^+_{c,v} , \bp_{c,v}) ,$$
where $\bp_{c,v}=\{ p_{c,v,m}, m=1,\dots,12 \}$ with
$$p_{c,v,m}  =\frac{\zeta_{c, v,m}} {\sum_{m'=1}^{12}  \zeta_{c,v,m'}},$$
We assume,
\begin{equation}\label{eq:multloglin}
\log (\zeta_{c,v,m} )= 
z_{c,v,m} \beta
\end{equation}
%where $r[c]$ represents the WHO region for country $c$, and 
where $z_{c,v,m}$ is the temperature and $\beta$ is the associated log-linear coefficient; no intercept is needed in the log-linear model, since when we take the ratio, to form the multinomial probabilities, if included, it would cancel.
  The multinomial model can be fitted in {\tt INLA} using the Poisson trick \citep{baker:94} which involves fitting the Poisson model for the data in country $c$, month $m$:
$$ Y_{c,v,m} | \lambda_{c,v} \sim \mbox{Poisson}( ~\lambda_{c,v}\mbox{e}^{z_{c,v,m}\beta }~),$$
where  the $\lambda_{c,v}$ parameters are given (improper) priors $\pi(\lambda_{c,v}) \propto 1/\lambda_{c,v}$.  We use the default {\tt INLA} prior for $\beta$, which is a normal with a large variance.  Further details of the Poisson trick may be found in the Supplementary Materials.
The estimated mean expected counts are shown in red in Figure \ref{fig:acm_monthly_tier1_plot}, for selected countries. 

\subsection{Modeling Uncertainty in the Expected Numbers}

For all countries the expected numbers appear directly in the excess calculation, (\ref{eq:excess}). In addition, for countries with no pandemic ACM data, the Poisson model we adopt for covariate modeling includes the expected number as an offset. For all countries and months, we obtain not just an estimate of the mean expected mortality but also a measure of the uncertainty (due to uncertainty in estimating the spline model) in this estimate. We now describe how the uncertainty in the mean expected count is accounted for in our modeling.

For countries with monthly data, we  use the spline model to predict the log of the mean expected number of deaths. Asymptotically, the estimator for the log of the mean expected numbers is normally distributed. Let $\widehat{\eta}_{c,t'}$ and  $\widehat \sigma^2_{c,t'}$ represent the mean and  standard deviation of the log prediction for pandemic months, labeled as $t'=1,\dots,24$. 
We  simulate $S$ samples from the asymptotic normal sampling distribution with mean $\widehat{\eta}_{c,t'}$ and standard deviation $\widehat \sigma_{c,t'}$; denote these samples by  $\eta^{(s)}_{c,t'}$, $s=1,\dots,S$. We then transform the samples so that we have samples for the expected numbers $E^{(s)}_{c,t'}=\exp(\eta^{(s)}_{c,t'})$, for $s=1,\dots,S$. We then use the method of moments to fit a gamma distribution to these $S$ samples with shape $\tau_{c,t'}$ and rate $\tau_{c,t'}/E_{c,t'}$. In particular, letting $m_{c,t'}$ denote the sample mean, and $V_{c,t'}$ denote the sample variance, we set $\widehat E_{c,t'}= m_{c,t'}$ and $\widehat \tau_{c,t'}=m_{c,t'}^2/V_{c,t'}$.  We approximate the distribution of the expected numbers as gamma, since this is conjugate to the Poisson, and so allows efficient inference with INLA \citep{rue:etal:09} using a negative binomial, as we describe in Section \ref{sec:observed:national}. Effectively, we are approximating the sampling distribution of the mean expected count by a gamma distribution.
	
We now consider a generic country $c$ with yearly data only. In pandemic year $v'$, we use the spline model to predict the log of the expected number of deaths. Let $\widehat{\eta}_{c,v'}$ and  $\widehat \sigma^2_{c,v'}$ represent the mean and  standard deviation of the prediction, for $v'=1,2$ (the two pandemic years).
 We then simulate $S$ samples from a normal distribution with mean
$\widehat{\eta}_{c,v'}$ and standard deviation $\widehat \sigma_{c,v'}$; denote these samples by  $\eta^{(s)}_{c,v'}$, $s=1,\dots,S$. We then transform the samples so that we have samples for the expected numbers $E^{(s)}_{c,v'}=\exp(\eta^{(s)}_{c,v'})$, for $s=1,\dots,S$. 
We then apply the monthly temperature model to produce predictions of the proportion of deaths in each month in each year, i.e.,~for a given pandemic month $m'$, we have $S$ samples of the predicted proportion of deaths in month $m'$ of year $v'$, $p^{(s)}_{c,v',m'}$, for $s=1,\dots,S$. Converting to pandemic months $t'=12(v'-1)+m'$ we then produce samples of the expected number of deaths in month $t'$, as $E^{(s)}_{c,t'}=E^{(s)}_{c,v'}\times p^{(s)}_{c,v',m'}$. We then use the method of moments to fit a gamma distribution to these $S$ samples as for the countries with monthly data. To summarize, in both cases we have a distribution for $E_{c,t'}$ which is 
%Gamma$(\widehat E^2_{c,t'}/V_{c,t'},\widehat E_{c,t'}/V_{c,t'})$.
Gamma$(\widehat{\tau}_{c,t'},\widehat{\tau}_{c,t'}/\widehat{E}_{c,t'})$.
The Supplementary Materials provide comparisons of the true distribution of the mean expected counts and the approximating gamma distributions, and illustrates that the latter are accurate. 
We also experimented with including negative binomial sampling variability in the calculation of the expected numbers, but it made little additional contribution to the intervals for the excess.%, $\delta_{c,t}$.

In the next section we describe a Bayesian model for the ACM counts in the pandemic, for countries without data. As we have describe above, inference for the expected numbers is an approximation to a Bayesian analysis. We sample from the asymptotic normal distribution of the prediction estimator which will approximate a Bayesian analysis with (improper) flat priors. Hence, when we combine the two components in the excess (\ref{eq:excess}) we view the resultant inference as Bayesian.

We next describe how we model ACM -- we have different models  for different data scenarios but in each case the starting point is the Poisson distribution. 

\section{National Mortality Models for Countries with No Data}\label{sec:observed:national}

For countries with observed monthly national ACM data, $Y_{c,t}$, we use these directly in the excess calculation.
For the countries with no data we need to estimate the ACM count. We follow a Bayesian approach so that for countries without data we obtain a predictive distribution over this count and this, when combined with the gamma distribution for the expected numbers, gives a distribution for the excess $\delta_{c,t}$. 
%For clarity, we let $Y_{c,t}$, $c \in s$ and $\widetilde Y_{c,t}$, $c \in  r$ denote the ACM counts for countries with and without data availability. Hence, the total global excess is
%$$\sum_{c \in s} (Y_{c,t} - E_{c,t})  + \sum_{c \in r} (\widetilde Y_{c,t}  - E_{c,t}) .$$

\begin{figure}
\centering
\includegraphics[scale=.5]{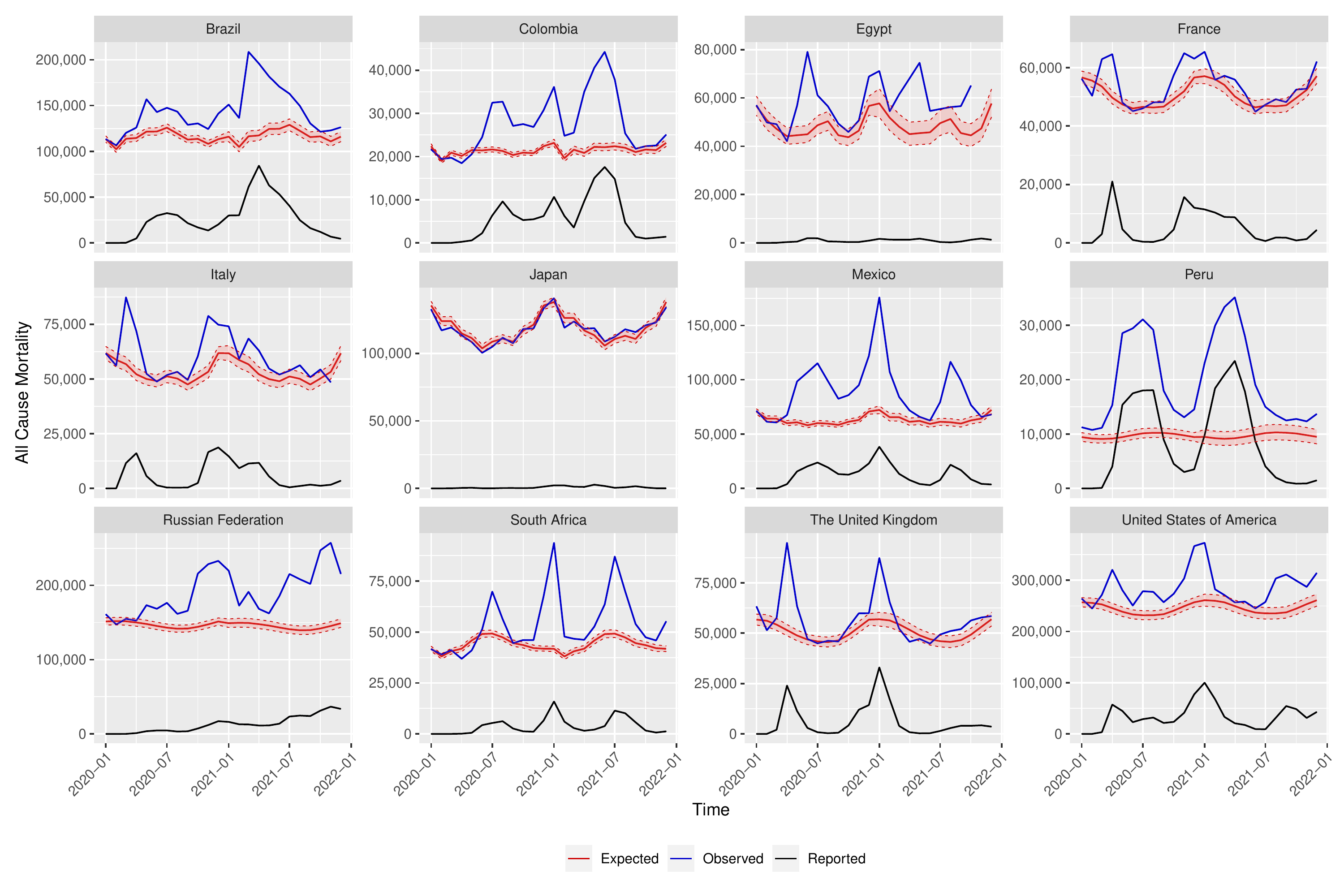}
\caption{Monthly time series of ACM counts, expected counts (with 95\% interval estimates) and reported COVID-19 mortality counts, for selected countries. ACM counts are available for all months apart from Egypt, for which the last month is missing.}
\label{fig:EstReportedExpected_by_Country}
\end{figure}

\begin{figure}
\centering
\includegraphics[scale=.5]{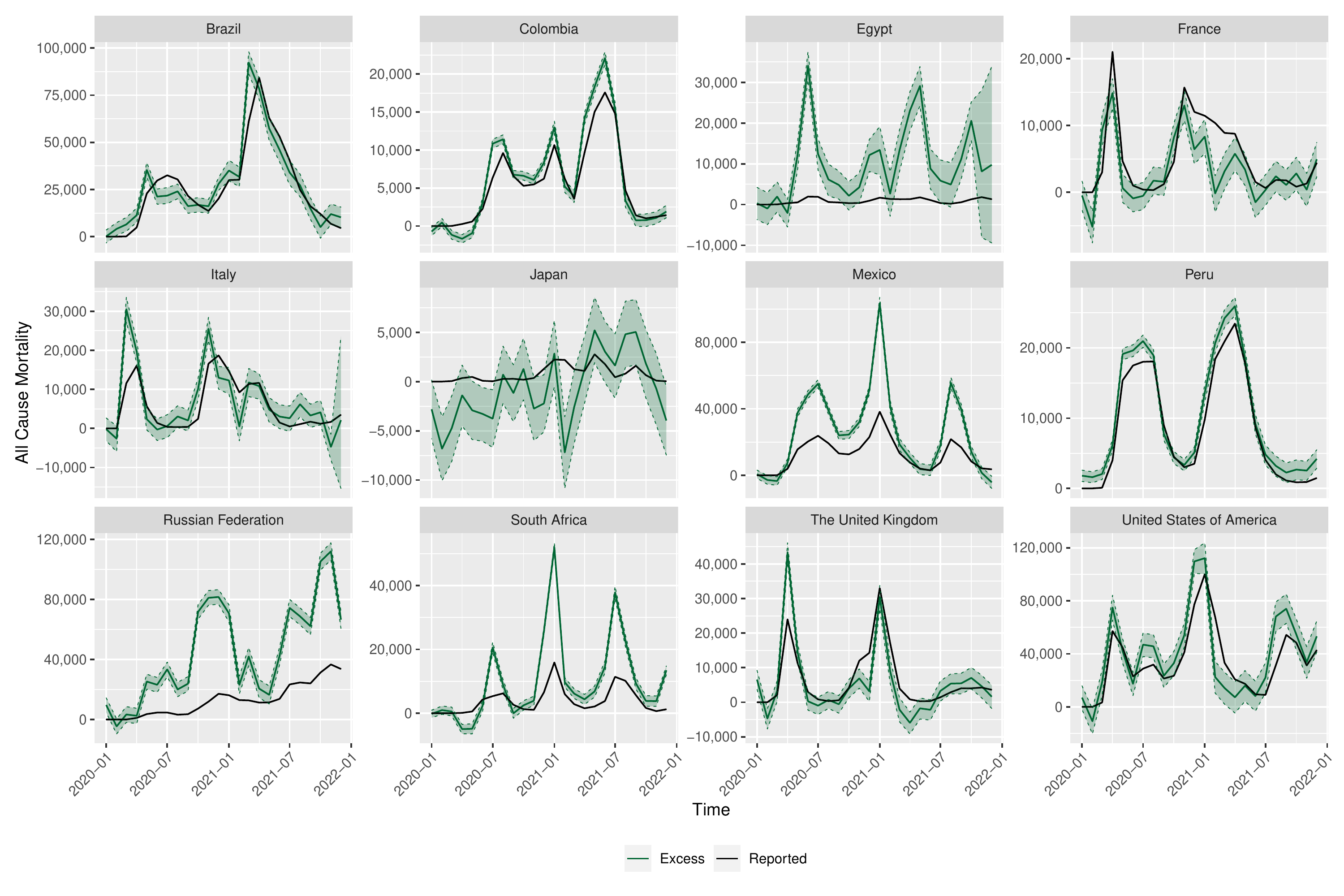}
\caption{Monthly time series of excess mortality, along with reported COVID-19 mortality counts. ACM counts are available for all months apart from Egypt, for which the last month is missing. For this month, the covariate prediction model is used for the point and interval estimates.}
\label{fig:EstExcessReported_byCountry}
\end{figure}

In Figure \ref{fig:EstReportedExpected_by_Country}  we plot the monthly counts for a range of countries with monthly  ACM data, along with the reported COVID-19 deaths and the expected numbers. We see very different scenarios in different countries. In all countries but Japan there is a clear large difference between the observed and the expected, though within each country this difference shows large fluctuations over time.
In Figure \ref{fig:EstExcessReported_byCountry}, again for countries with monthly ACM data, we plot the excess $\delta_{c,t} = Y_{c,t}-E_{c,t}$, as a function of month $t$ (including uncertainty in the expected numbers), along with the reported COVID-19 deaths. As expected, $\delta_{c,t}$ is greater than the reported overall in general, except in Japan, and for most countries displayed the difference between the excess and the reported shows a complex temporal pattern.

While complex models that attempt to pick up data nuances are desirable, given the  idiosyncrasies of the different data sources described in Section \ref{sec:data}, any modeling exercise is fraught with difficulties, and we resort to a relatively simple model in which we build an overdispersed Poisson log-linear regression model for the available monthly ACM data to predict the monthly ACM in those countries with no data. We cannot overemphasize  the regional imbalance of the missing ACM data -- in the AFRO region in particular, our estimates should be viewed with extreme caution, since they are predicted from data which overwhelmingly is from other regions.

The basic starting model is
\begin{equation}\label{eq:poissss}
Y_{c,t} | E_{c,t},\theta_{c,t}  \sim \mbox{Poisson}(E_{c,t}\theta_{c,t}),\end{equation}
so that $\theta_{c,t}>0$ is a relative rate parameter, with $\theta_{c,t}>1$ / $\theta_{c,t}<1$ corresponding to a higher/lower ACM rate than expected, based on historic data.
Recall, from Section \ref{sec:expected}, that we model the distribution of the expected counts $E_{c,t}$ as Gamma$(\widehat{\tau}_{c,t},\widehat{\tau}_{c,t}/\widehat{E}_{c,t})$. When combined with (\ref{eq:poissss}),
we obtain the sampling model,
\begin{eqnarray*}
Y_{c,t} | \theta_{c,t} \sim \mbox{NegBin}( \widehat{E}_{c,t}\theta_{c,t},\widehat{ \tau}_{c,t})
\end{eqnarray*}
with known overdispersion parameter $\widehat{ \tau}_{c,t}$ to give $\v( Y_{c,t} | \theta_{c,t}  ) = \widehat{E}_{c,t}\theta_{c,t}(1+\widehat{E}_{c,t}\theta_{c,t}/ \widehat{ \tau}_{c,t})$.  The mean is $\E[ Y_{c,t} | \theta_{c,t}] = \widehat E_{c,t}\theta_{c,t}$ and the relative rate parameter $ \theta_{c,t} $ is modeled as,
\begin{eqnarray}\label{eq:log-linear}
\log \theta_{c,t} &=& \alpha + 
\sum_{b=1}^B  \beta_{bt}  X_{bct} + \sum_{g=1}^G \gamma_g Z_{gc} + \epsilon_{c,t}.
%\log \theta_{c,t} &=& \alpha^\text{\tiny{H}}_{t} 1( c \in \mbox{ high }) + \alpha^\text{\tiny{N}}_{t} 1( c \in \mbox{ not high }) \\
%&+& \beta^\text{\tiny{H}}_{t} 1( c \in \mbox{ high }) X_{c,t} + \beta^\text{\tiny{N}}_{t} 1( c \in \mbox{ not high }) X_{c,t}
\end{eqnarray}
The model details are:
\bi
\item  The intercept  is $\alpha$.
\item The time-invariant covariates (e.g.,~historic cardiovascular and diabetes rates) have fixed association parameters $\gamma_g$.
\item We have $B$ time-varying covariates (e.g.,~sqrt(C19 death rate), test positivity rate, containment), and we allow the associations for these variables, $\beta_{bt}$, to be time-varying via a random walk of order 2 (RW2) prior \citep{rue:knorrheld:05} which has variance $\sigma^2_\beta$. These parameters include a sum-to-zero constraint, since we include a fixed effect for the overall association (across months) -- these are included in the $G$ time-invariant part of the model. 
\item There are two sources of excess-Poisson variation in our model. The negative binomial component, with known $\widehat{ \tau}_{c,t}$, arises because of the uncertainty in the expected numbers, while the $\epsilon_{c,t} \sim \N(0,\sigma_\epsilon^2)$ adjustments allow for overdispersion, given a fixed value of the expected numbers.
%\item We also experiment with including interactions between the association parameters and the binary high/non-high income variable.
%\item Each of $ \alpha^\text{\tiny{H}}_{t} , \alpha^\text{\tiny{N}}_{t},\beta^\text{\tiny{H}}_{t} , \beta^\text{\tiny{N}}_{t}$ have RW2 priors.
\item
The Bayesian model is completed by prior specifications on the regression coefficients of the log-linear model and any hyperparameters. We use default priors (normal with large variance) on the intercept and fixed association parameters, and penalized complexity (PC) priors on the RW2 standard deviations and on $\sigma_\epsilon$ \citep{simpson:etal:17}. Specifically, letting $\sigma_\beta$ denote a generic RW2 standard deviation parameter, the PC priors are such that $\Pr(\sigma_\beta > 1)=0.01$, and the PC prior on the overdispersion parameter $\sigma_\epsilon$ has $\Pr(\sigma_\epsilon > 1)=0.01$.
\ei

%$$
%\log  \widehat \theta^\star_{c,t} = \widehat \alpha + 
%\sum_{b=1}^B \widehat  \beta_{bt}  X_{bct} + \sum_{g=1}^G\widehat  \gamma_g Z_{gc}
%$$

Each country will clearly have its own specific temporally correlated baseline, as a result of unobserved covariates and model misspecification, but we did not include terms to model such a baseline  (using a RW2 or a spline, for example), since fits from this model are not being used  to estimate the excess for countries with data. Rather, we are using this model to predict the ACM for countries with no data. Hence, we 
did not use RW2 intercepts as these would dilute the covariate effects, due to confounding by time \citep{kelsall1999frequency}, and it is these covariate effects that are key to prediction for countries with no data. If we had included a RW2 baseline, then a country-specific RW2 model would give estimated contributions of zero in countries with no data and so would not provide any benefit.
 This is but one of the model assumptions that are forced upon us by the limited data we have available.
The country-level  model was fitted using the INLA method  \citep{rue:etal:09} and accompanying {\tt R} implementation.

%For countries without observed ACM we obtain the posterior for 
%$\widehat Y_{c,t} = E_{c,t}  \theta_{c,t}
%$
%using (\ref{eq:log-linear}).
%
For countries with no ACM data, we obtain a predictive distribution by averaging the negative binomial model with respect to the posterior via,
$$
\Pr( Y_{c,t} | \by ) = \int \underbrace{\Pr(  Y_{c,t} |  \theta_{c,t}   ) }_{\text{Negative Binomial}}\times \underbrace{p( \theta_{c,t}  | \by )}_{\text{Posterior}}~d \theta_{c,t} .$$
We use INLA to fit the covariate model, and then use the posterior sampling feature to  produce samples for the components of (\ref{eq:log-linear}), which in turn produces samples $\theta^{(s)}_{c,t}  \sim p(  \theta_{c,t} | \by )$ from the posterior.
%which can be approximated by
%$$
%\Pr( Y_{c,t} | \by ) \approx \frac{1}{S}  \sum_{s=1}^S \Pr(  \widetilde Y_{c,t}|  \theta_{c,t}^{(s)}  )
%$$
%where $\theta_{c^\star t} ^{(s)} \sim p( \theta_{c,t} | \by )$ are samples from the posterior. This approximates the discrete predictive distribution, 
%To obtain samples from this predictive we:
%\begin{enumerate}
%\item Simulate $\theta^{(s)}_{c,t}  \sim p(  \theta_{c,t} | \by )$ from the posterior.
We then simulate $Y^{(s)}_{c,t} | \theta^{(s)}_{c,t}  $ from the negative binomial, for $s=1,\dots,S$.

%We should report some measure of error such as:
%- average (or median - let's look at both) absolute bias - we could report this on the rate and count scales

Partial monthly data is available for 27 countries, and for these we require a switch from observed data to the covariate modeled ACM. The naive application of the covariate model will lead to the possibility of unrealistic jumps (up or down) when we switch from the observed data to the covariate model, and to alleviate this problem we benchmark the predictions to the last observed data point. We let $T^{(1)}_{c}$ represent the number of observed months of data and $T^{(2)}_{c}$ be the number of months for which there is no data, for country $c$. For a country with partial data, let $\by^{(1)}_{c}=
[y_{c,1},\dots,y_{c,T^{(1)}_{c}}]$ represent the observed partial data. We then wish to predict the ACM counts
$\by_{c}^{(2)}=
[y_{c,T^{(1)}_{c}+1},\dots,y_{c,T^{(1)}_{c}+T^{(2)}_{c}}]$ 
for the missing period. 
The model for the missing data period is,
\begin{equation}\label{eq:bench}
y^{(2)}_{c,t} | \by^{(2)}_{c} , \theta_{c,t} ,f_c \sim \mbox{NegBin}( \widehat{E}_{c,t}\theta_{c,t}f_{c} , \widehat{\tau}_{c,t}),
\end{equation}
for $t=T^{(1)}_{c}+1,\dots,T^{(1)}_{c}+T^{(2)}_{c}$, where $\theta_{c,t}$ is a function of the covariates  in the missing data period (specifically given by (\ref{eq:log-linear})), and the benchmarking factor is,
$$f_{c} =  f_c\left( \theta_{c,T^{(1)}_{c}}\right)= \frac{
y_{c,T^{(1)}_{c}}
}{
\widehat E_{c,T^{(1)}_{c}}
 \theta_{c,T^{(1)}_{c}}
 },
$$
%where $\widehat y_{c,T^{(1)}_{c}} = \widehat E_{c,T^{(1)}_{c}}  \widehat \theta_{c,T^{(1)}_{c}} $.
where $ \theta_{c,T^{(1)}_{c}}$ is given by equation (\ref{eq:log-linear}).
This factor matches the last observed death count to the covariate model projected back to the last observed count. This factor is applied subsequently to all of the missing data months. To implement the benchmark, samples from the posteriors for $\theta_{c,t}$ and $f_c$ are used in (\ref{eq:bench}), and then negative binomial counts are drawn.

\section{Observed Mortality Subnational and Annual Data Modeling}\label{sec:observed:subnational}
        
For a small number of countries for which national ACM data are not available (Argentina, India, Indonesia and Turkey) we instead have ACM data from subregions, with the number of regions with data potentially changing over time. 
For other countries we obtain national annual ACM data only.
%, while for China we have subnational monthly and national annual data. 
In this section we describe the models we use in these situations.
For the subnational scenario we construct a statistical model building on, and expanding, a method previously proposed by \cite{karlinsky_subnational} that is based on a  proportionality assumption.

\subsection{Subnational Data Model}\label{sec:subnational}
%\section{Multinomial Version}

%For a small number of countries we have no national data, but subnational data that can inform on the national totals. 
For Turkey we have subnational monthly data over the complete two years of the pandemic, while for Indonesia we have monthly subnational data for 2020 and for the first six month of 2021. Argentina has observed data for 2020 and subnational monthly data for 2021. India has data from up to 17 states and union territories (from now on, states) over the pandemic period (out of 36), but this number varies by month.

\begin{figure}[!htbp]
        \center{\includegraphics[scale=0.47]
        {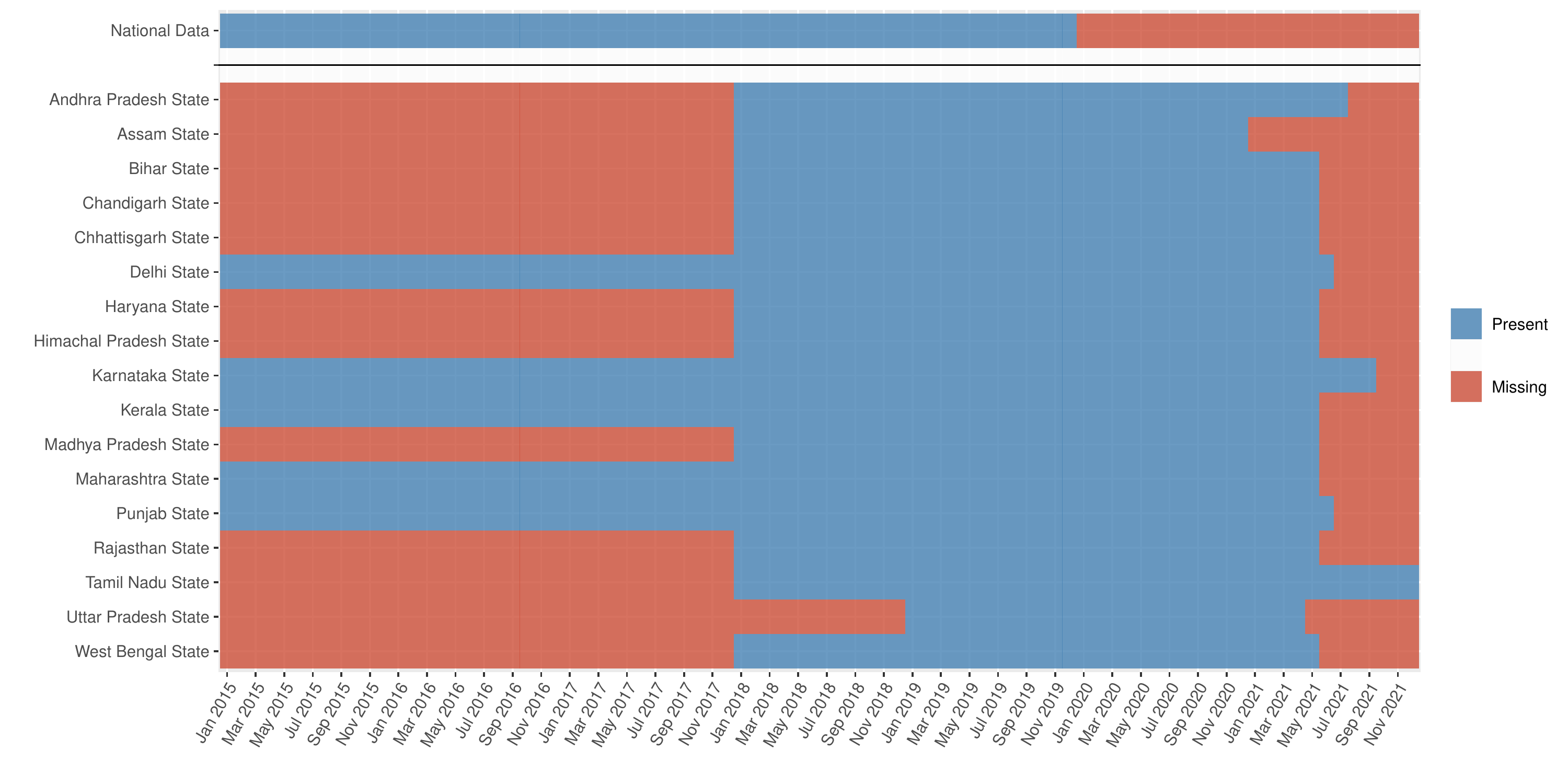}}
%        \caption{\label{fig:my-label} Plot of Missingness in Subnational Data across 2015-2021}
        \caption{Plot of missingness in subnational data for India across 2015--2021.}\label{fig:missing}
\end{figure}
%        \begin{table}[htp]
%\begin{center}
%\begin{tabular}{l|cc|ccc|c}
%Year&Region 1&Region 2 &Region 3&Region 4 &Region 5 & National\\ \hline
%Historic 1&\cellcolor{blue}&\cellcolor{red}&\cellcolor{red}&\cellcolor{red}&\cellcolor{red}&\cellcolor{blue}\\
%Historic 2&\cellcolor{blue}&\cellcolor{blue}&\cellcolor{blue}&\cellcolor{red}&\cellcolor{red}&\cellcolor{blue}\\
%Historic 3&\cellcolor{red}&\cellcolor{blue}&\cellcolor{blue}&\cellcolor{red}&\cellcolor{red}&\cellcolor{blue}\\
%Historic 4&\cellcolor{blue}&\cellcolor{blue}&\cellcolor{red}&\cellcolor{blue}&\cellcolor{blue}&\cellcolor{blue}\\
%%Historic 5&\cellcolor{blue}&\cellcolor{blue}&\cellcolor{blue}&\cellcolor{blue}&\cellcolor{blue}&\cellcolor{blue}\\
%\hline
%Pandemic 1&\cellcolor{blue}&\cellcolor{blue}&\cellcolor{red}&\cellcolor{red}&\cellcolor{red}&\cellcolor{red}\\
%Pandemic 2&\cellcolor{blue}&\cellcolor{red}&\cellcolor{red}&\cellcolor{red}&\cellcolor{red}&\cellcolor{red}\\
%\end{tabular}
%\end{center}
%\caption{Counts in {\bf \dblue{blue}} are observed and counts in {\bf \dred{red}} are unobserved.}
%\label{tab:notsosimplesub}
%\end{table}%

\begin{figure}[!htbp]
        \center{\includegraphics[scale=0.5]{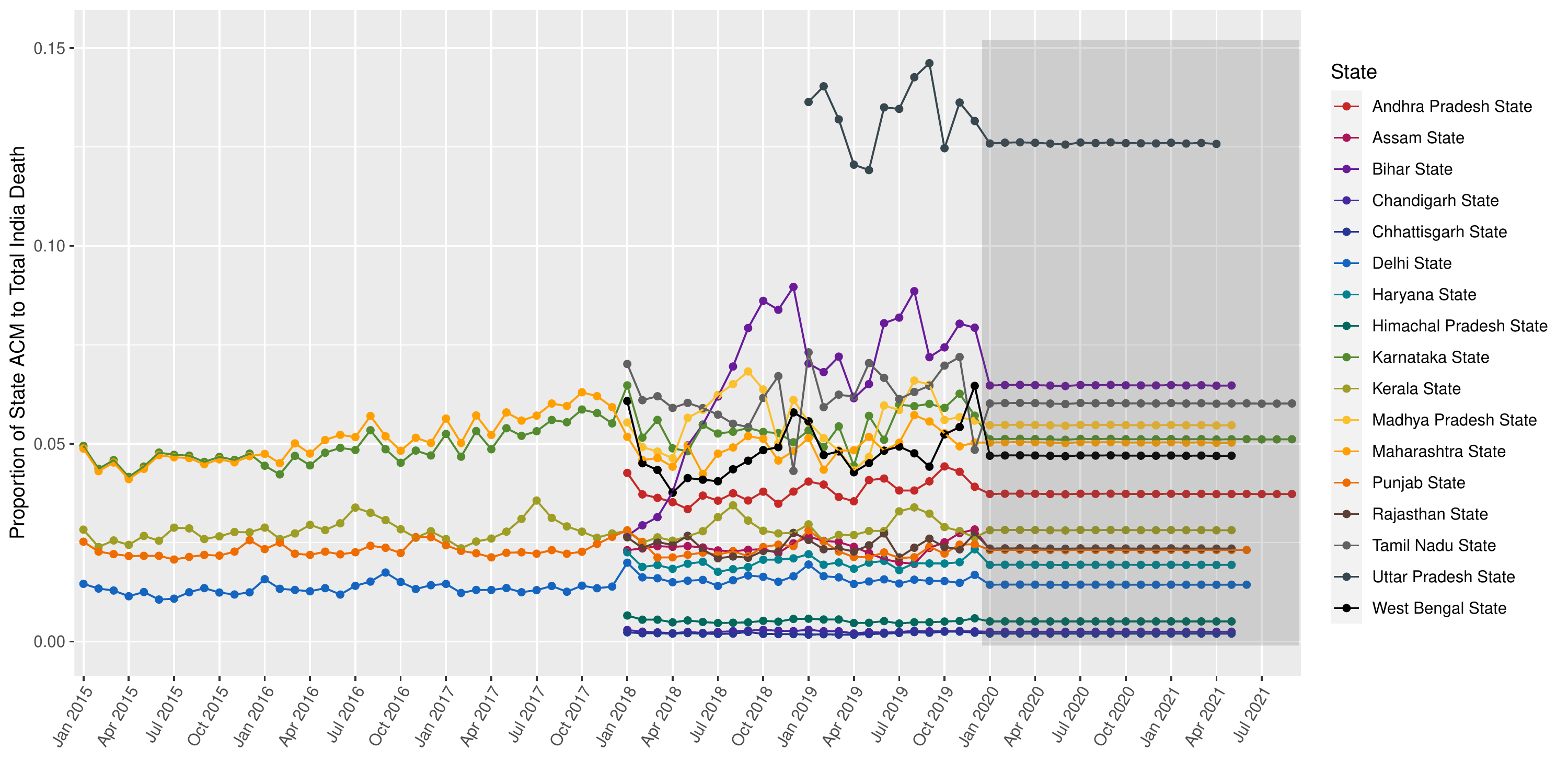}}
        \caption{Plot of estimated proportion of subnational deaths to national deaths in pre-pandemic and  pandemic periods (grey rectangle). The horizontal flat lines are the point estimates for the fraction for the respective states during the pandemic months.}\label{fig:IndiaStates_Proportions} 
\end{figure}

We consider the most complex subnational scenario in which the number of regions with monthly data varies by month, using India as an example. For India, we use a variety of sources for registered number of deaths at the state and union-territory level. The information was either reported directly by the states through official reports and automatic vital registration, or by journalists who obtained death registration information through Right To Information requests (see the Supplementary Materials for full details). The available data we have for India is summarized in Figure \ref{fig:missing}. We assume in total that there are $K$ regions that contribute data at any time. We develop the model for a generic country and hence drop the $c$ subscript.
% This situation is interesting for two reasons:
%\begin{enumerate}
%\item We may wish to model dependencies between the regions, or have distinct sampling schemes for different data sources.
%\item The number of regions for which data are available may change over time.
%\end{enumerate}
%
For the historic data in month $t$ we have total deaths counts along with counts over regions, $Y_{t,k}$, $k \in K_t$, so that in period $t$, $|K_t|$ is the number of regions that provide data with $k \in K_t$ being the indices of these areas from $1,\dots,K$. 
%In Table \ref{tab:notsosimplesub}, we have $K_1=1$, $K_2=3$, $K_3=2$, $K_4=4$, and then in pandemic years, $K_5=2,K_6=1$ with the national total unobserved. 
We let region $0$ denote all other regions, which are not observed in pandemic times, at time $t$ and $S_t = \{ 0 \} \cup K_t $. 
We assume, in month $t$:
\begin{eqnarray*}
Y_{t,k}|  \lambda_{t,k}&\sim &\mbox{Poisson} (N_{t,k} \lambda_{t,k} ),\qquad k \in S_t,
\end{eqnarray*}
where $N_{t,k}$ is the population size, and $\lambda_{t,k}$ is the rate of mortality. Hence,
\begin{eqnarray*}
Y_{t,+} |  \lambda_{t,k} , k \in S_t &\sim& \mbox{Poisson} \left( \sum_{k \in S_t} N_{t,k} \lambda_{t,k} \right).
\end{eqnarray*}
%In year $t^\star$ (e.g.,~2020):
%$
%Y_{t^\star+} \sim \mbox{Poisson} \left( N_{t^\star1} \lambda_{t1} + N_{t^\star 2} \lambda_{t2} \right)
%$
If we condition on the total deaths,
we obtain,
$$\bY_t | \bp_t \sim \mbox{Multinomial}_{|S_t|}( Y_{t,+} , \bp_t ),$$
with $\bp_t=\{ p_{t,k}, k \in S_t\}$, with
$$p_{t,k} = \Pr( \mbox{ death in region $k$ } | \mbox{ month $t$, death }) = \frac{N_{t,k}\lambda_{t,k}}{N_{t,
+}\lambda_{t,+}},$$
Our method hinges on this ratio being approximately constant over time.
If, over all regions, there are significant changes in the proportions of deaths in the regions as compared to the national total, or changes in the populations within the regions over time, then the approach will be imprecise for that region.
However, with multiple regions, we gain some robustness since it is the cumulative departure from the constant fractions that is relevant.
For India, the fractions of the total ACM by state are shown in Figure \ref{fig:IndiaStates_Proportions}. There are certainly deviations from constancy for some states, but in general the assumption appears tenable, at least in pre-pandemic periods. Of course, the great unknown is whether the assumption remains reasonable over the pandemic. To  address this, we carry out extensive sensitivity and cross-validation analyses (reported in the Supplementary Materials).

%which is equivalent to $_t \approx p$. 
%The key assumption upon which the method depends is,
%\begin{equation}\label{eq:keyassumption}\frac{\lambda_{t1}}{\lambda_{t2}} = \mbox{const},\end{equation}
%i.e.,~that the ratio of the death rates in the two regions is constant over time.

%We allow $p_t$ to ``wobble" as a function of month $t$ via:
%$$\log\left( \frac{p_{t}}{1-p_t} \right) = \alpha + \epsilon_{t},
%$$
%with $\epsilon_{t} \sim \N(0,\sigma^2_\epsilon)$, and we examine the size and temporal structure of the error terms $\epsilon_t$, to assess the proportionality assumption, at least over the available data. 

We model the monthly probabilities as,
%reparameterize and  assume the model for month $t$ is,
%$$\bY_t | \bp_t \sim \mbox{Multinomial}_{K_t+1}( Y_{t+} , \bp_t ),$$
%with $\bp_t=\{ p_{t,k}, k \in S_t\}$, with
%$$p_{t,k} = \Pr( \mbox{ death in region $k$ } | \mbox{ period $t$, death }),$$
%and
\begin{equation}\label{eq:multinomial}
\mbox{log}\left(\frac{p_{t,k} }{p_{t,|K_t|+1}}\right)= \alpha_k 
%+ \sum_{b=1}^B  \delta_{bt}  X_{bct} 
+ e_{t},\qquad k \in S_t,
\end{equation}
where the $\alpha_k $ parameters are unrestricted and $e_{t} \sim \N(0,\sigma^2_\epsilon)$, and we can examine the size and temporal structure of the error terms $e_t$, to assess the proportionality assumption, at least over the available pre-pandemic period.  We emphasize that we do not use any covariates in the subnational model, but infer the national ACM from the subnational contributions.

To specify the model, we take a multinomial with a total number of categories that corresponds to all regions that appear in the data, $K$, along with a final category for the unknown remainder. We specify the likelihood over all months by exploiting the property that a multinomial collapsed over cells is also multinomial. Hence, in  year $t$ we have a multinomial with $|K_t|+1$ categories with constituent probabilities constructed from the full set of $K+1$ probabilities.

To derive the predictive distribution, we abuse notation and let $Y_{t,1}$ denote the total number of observed subnational deaths at time $t$,  and $Y_{t,2}$ the total number of unobserved subnational deaths at time $t$, with $Y_{t,+} =Y_{t,1} +Y_{t,2}$ being the total (national) number of deaths at time $t$. Hence, at time $t$,  $Y_{t,1}|p_t, Y_{t,+}\sim\text{Binomial}(Y_{t,+}, p_{t})$, where $p_{t}=\sum_{k\in K_t}p_{t,k}$.
 In order to fit the multinomial model in a Bayesian framework and predict the total number of deaths in 2020--2021, we need to specify a prior for $Y_{t,2}$ or, equivalently, for $Y_{t,+} $, where $t$ indexes months in this period.
 We will use the prior $p(Y_{t,+} )\propto1/Y_{t,+} $, which is a common non-informative prior for a binomial sample size \citep{link:13}, and has the desirable property that the posterior mean for $Y_{t,2}$, conditional on $p_t$, is $\E[Y_{t,2}|p_{t}]=Y_{t,1}  (1-p_{t})/p_{t}$, i.e.,~of the same form as the simple frequentist ``obvious'' estimator, which leads to the naive estimate of the ACM, $Y_{t,1}+\widehat{Y}_{t,2}=Y_{t,1}/p_t$.

To give more details for implementation we will use a general result.
Suppose \begin{eqnarray*}
Y_{t,1} | Y_{t,+} ,p_t& \sim&\text{Binomial}(Y_{t,+},p_t)\\
p(Y_{t,+})&\propto&1/Y_{t,+},\end{eqnarray*}
so that, in particular, the marginal distribution of $Y_{+t}$ does not depend on $p_t$. Then the posterior for the missing ACM count, conditional on $p_t$, is
$${Y_{t,+}|Y_{t,1},p_t \sim Y_{t,1} + \text{NegBin}(Y_{t,+}, 1-p_t)},$$
or, equivalently,
$$Y_{t,+}-Y_{t,1}|Y_{t,1},p_t \sim\text{NegBin}(Y_{t,1}, 1-p_t).$$ 
This  links to one of the usual motivations for a negative binomial (number of trials until we observe a certain fixed number of events) --- making inference for the number of total deaths it takes to produce $Y_{t,1}$ deaths in the sub-regions.
We implement this model in {\tt Stan}. In the Supplementary Materials we detail a simulation study that validates the method in the situation in which the missing data follow the assumed form.

For the other countries with subnational data, the number of subregions is constant over time, and so in the above formulation the multinomial is replaced by a binomial. Details for these countries are in the Supplementary Materials. For Indonesia we have subnational data from only Jakarta at the monthly level and historic national ACM at the annual level. Hence, we fit a binomial subnational model to the annual historic data, summing the monthly subnational historic data to the annual level, and then predict the monthly national ACM for 2020--2021 using the $p_t$ fit on the historic annual data.

%we have annual subnational data for 2020 and we fitted an annual binomial model to these data

%and so we use a national binomial model and then apportion the counts using the multinomial temperature model described in Section \ref{sec:expected:annual}.
  
 \subsection{Annual Data Model}
  
We have  annual national ACM counts for  Viet Nam, Grenada, Sri Lanka, Saint Kitts and Nevis, and Saint Vincent and the Grenadines. For these countries we estimate the monthly counts using a multinomial model. This model  is derived from the overdispersed Poisson model (\ref{eq:poissss})  that is used for countries with no pandemic data.
Conditioning on an annual total leads to a multinomial model for the monthly ACM within-year counts with apportionment probabilities $E_{c,t}\theta_{c,t}/\sum_{t'=1}^{12} E_{c,t'}\theta_{c,t'}$ where $\theta_{c,t}$ is given by  the log-linear covariate model (\ref{eq:log-linear}).
To obtain counts for these countries, we sample expected numbers $E_{c,t}$  and rates $\theta_{c,t}$ and then sample multinomial counts with these probabilities.

%take the predicted relative rate parameters for these countries from the Poisson covariate year and push them through the implied multinomial model, combined with the gamma distributed expecteds in which the log-linear covariate model (\ref{eq:log-linear}) is used to apportion the total count to months.

%  For China, we have annual national data and also subnational monthly data for the first 9 months of each of 2020 and 2021. In the Supplementary Materials we describe a model for combining the two types of data and an MCMC implementation.

%Surinam (one time point dropped)

\section{Results}\label{sec:results}

In this section we summarize the excess mortality results, further results are available in the Supplementary Materials, and 
a ShinyApp is available (\url{https://msemburi.shinyapps.io/excessvis/}) that allows access to the full results.  The aim is to build a covariate prediction  model for the countries with no ACM data, using (\ref{eq:log-linear}).  The covariate model choice exercise was carried out in an empirical fashion. In an ideal world, we would have had region-specific models, but the paucity of data in many of the regions (as summarized in Table \ref{tab:datamortsummary}) did not allow for this. Instead, for all of the time-varying covariates (COVID-19 test positivity rate, COVID-19 death rate, temperature, stringency, overall government response, containment) we added an interaction with the binary country-level variable, low/middle or high income. We examined plots of the covariates by availability in the ACM observed/unobserved countries, and discarded a number of covariates (historic HIV rate, and over-65 and under-15 proportions of the population) that had little overlap over countries with/without ACM data (meaning, for example, that the countries with high HIV rates tended to be those without observed ACM data, making extrapolation hazardous). On a contextual basis we then formed a covariate model with time-varying covariates: containment, square root COVID-19 death rate (the square root transforms helps in preventing the association being driven by a few countries), temperature and COVID-19 positivity rate. The constant covariates we use are: historic diabetes rate and historic cardiovascular rate. We took this model as our starting point and added and removed variables to examine the sensitivity of the predictions. We evaluated the models using cross-validation and various metrics that are described in the Supplementary Materials. We found that the predictions were quite robust to covariate models and so only report the results for the model described above.

  In the Supplementary Materials we describe our approaches to model assessment and model comparison. We assessed the frequentist coverage of our procedure using cross-validation. In particular, we performed two experiments: in one we left out all data from a country, and in the other we left out all data from one month (systematically going through all countries and all months, respectively, in the two schemes). The model was fitted to the remaining data and was used to produce predictive intervals for the left out data, which can then be compared with the left out data. The empirical coverage at levels, 50\%, 80\%, 95\%, was calculated by summarizing across all left out data. For the leave-one-country out analysis the coverages were 59.3\%, 82.7\%, 91.6\%, and for the leave-one-month out analysis 57.8\%, 83.7\%, 92.9\%. From these summaries, we would conclude that the model is reasonably well calibrated, at least for countries which ``look like'' those with observed data.  Using the same cross-validation strategies we also evaluated the relative and absolute relative bias of the ACM rate. The relative biases from the country and monthly leave out strategies were 1.98\% and 1.84\%, respectively.  The absolute relative biases from the country and monthly leave out strategies were 10.08\% and 10.18\%, respectively. The absolute relative bias tells us that the point estimates are reasonable overall, though for any one country are typically off by 10\%. It is interesting that leaving out countries or complete months give very similar results.
  The Supplementary Materials contain comparisons of fitted versus observed, both in-sample and out-of-sample, along with residual plots over time.
  %contain these summaries for other covariate models considered, which showed significantly poorer coverage estimates.
  
\begin{figure}[htbp]
\centering
\includegraphics[scale=.46]{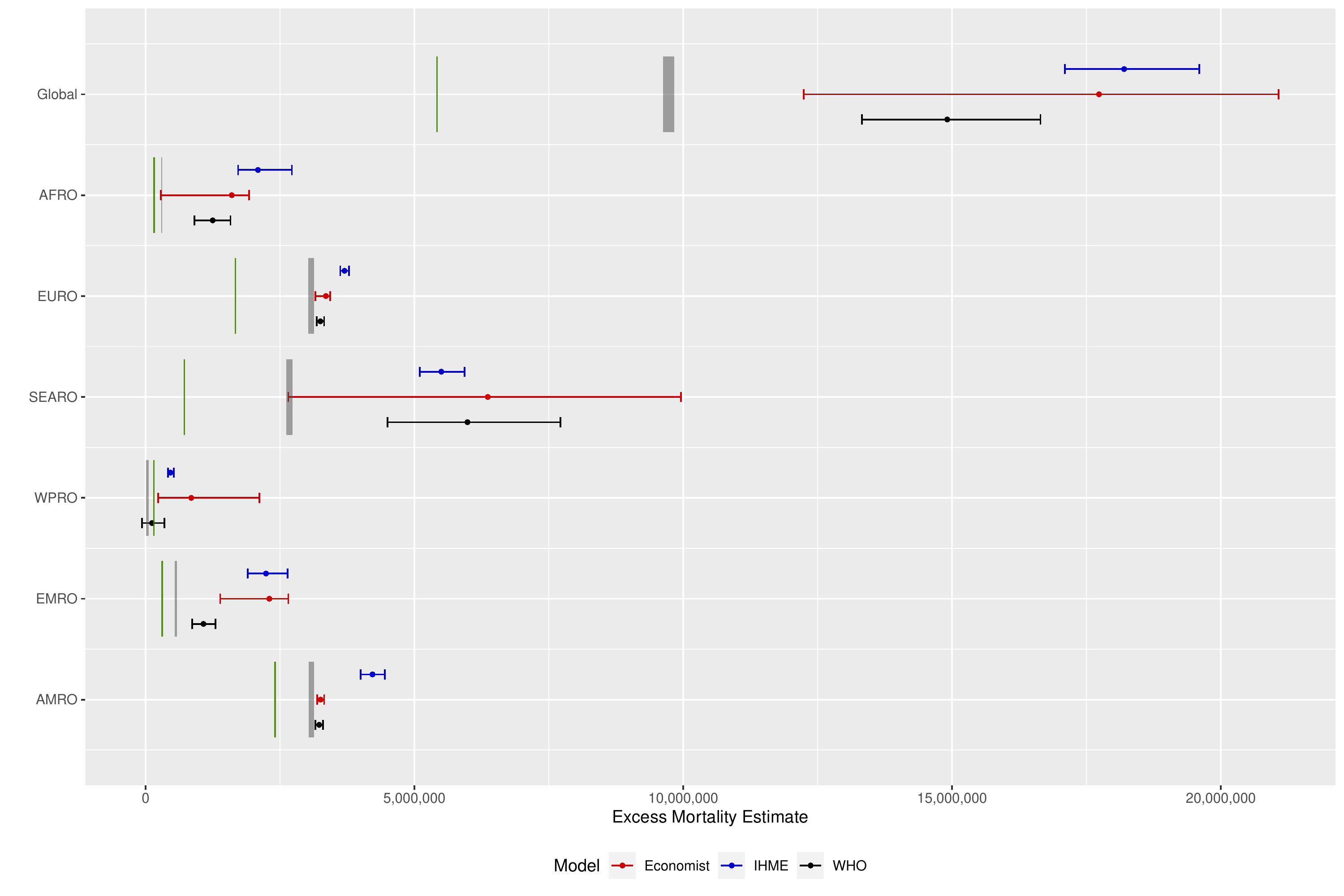}
\caption{Global and regional point excess mortality estimates and 95\% intervals from WHO, The Economist and IHME. The grey vertical thin rectangles correspond to the excess from those countries with observed ACM death, so the only uncertainty comes from the expected numbers (the width of these rectangles reflects this uncertainty). These grey rectangles include subnational and annual only contributions. The green vertical lines show the reported COVID-19 deaths. 
%Regional estimates of excess mortality with 95\% credible intervals. Covariate Model 1: Positivity Rate, Sqrt Covid Rate, Containment.  Covariate Model 2: Positivity Rate, Sqrt Covid Rate, Containment, High Income. Covariate Model 3: Positivity Rate, Sqrt Covid Rate, Containment, High Income, High Income interactions with Positivity Rate and Covid Rate. Covariate Model 4: Positivity Rate, Sqrt Covid Rate, Containment, High Income, High Income interactions with Positivity Rate, Covid Rate and Containment. IF WE HAD AN ACCOMPANYING TABLE WITH COVARIATES AND LCPO AND ERROR SCORES WE COULD REFERENCE THAT.
}\label{fig:RegionalEstimates}
\end{figure}

 Our point estimate for the excess mortality over 2020--2021 is 14.9 million with a 95\% credible interval of (13.3, 16.6) million.  In Figure \ref{fig:RegionalEstimates} we plot global and regional estimates. The excess estimates based purely on countries with observed data are also plotted, with uncertainty, which is due to the expected numbers, as grey rectangles. Note that we  include the data from countries with subnational and annual data in the rectangles.  Globally, and with respect to our estimate, around two-thirds of the contribution to the excess is from observed data, and a third from modeling (this is for the cumulative annual estimates, those countries with only annual data lead to more uncertainty in the monthly counts). Subnational data in India makes a substantial contribution to the total -- we estimate that we catch approximately  63\% of the deaths over the pandemic. In order to estimate the proportion of the excess we capture for the subnational data, we multiply the total national expected estimate by our estimate of the fraction of deaths we capture. This further emphasizes which region's estimates are based primarily on observed data (EURO and AMRO) and those that are not. It is interesting that the IHME estimates for EURO and AMRO are relatively higher than the rectangles, even though the excess is observed for the majority of country-month combinations. Our global estimate is the lowest of the three. In general, The Economist confidence intervals are widest and those of IHME are the narrowest. As we discuss in Section \ref{sec:discussion}, in terms of the procedures used, the IHME intervals are not based on any statistical principles, and The Economist intervals are based on a bootstrap procedure whose validity has not been shown for the gradient boosting approach used.
The narrowness of the IHME SEARO interval is particularly striking, given the uncertainty over India's excess mortality.

For our estimates, the width of the intervals depend on the available information (as indicated by the grey rectangles) and on the mean-variance relationship that is implied by our overdispersed Poisson framework (narrower intervals if the mean is lower). Neither IHME or The Economist assume such a mean-variance relationship since they model the log excess rate and excess rate, respectively, and do not weight observations in a way that is consistent with an overdispersed Poisson model (see Section \ref{sec:discussion} for further details).
 %For regions with excess mortality, we would expect these to be a lower bound.
% , so it is surprising that the IHME interval estimates go below these values in the AMRO and EURO regions. This is less surprising given the non-standard  manner in which the interval estimates are calculated (see Section \ref{sec:discussion}).

\begin{figure}
\centering
\includegraphics[scale=.5]{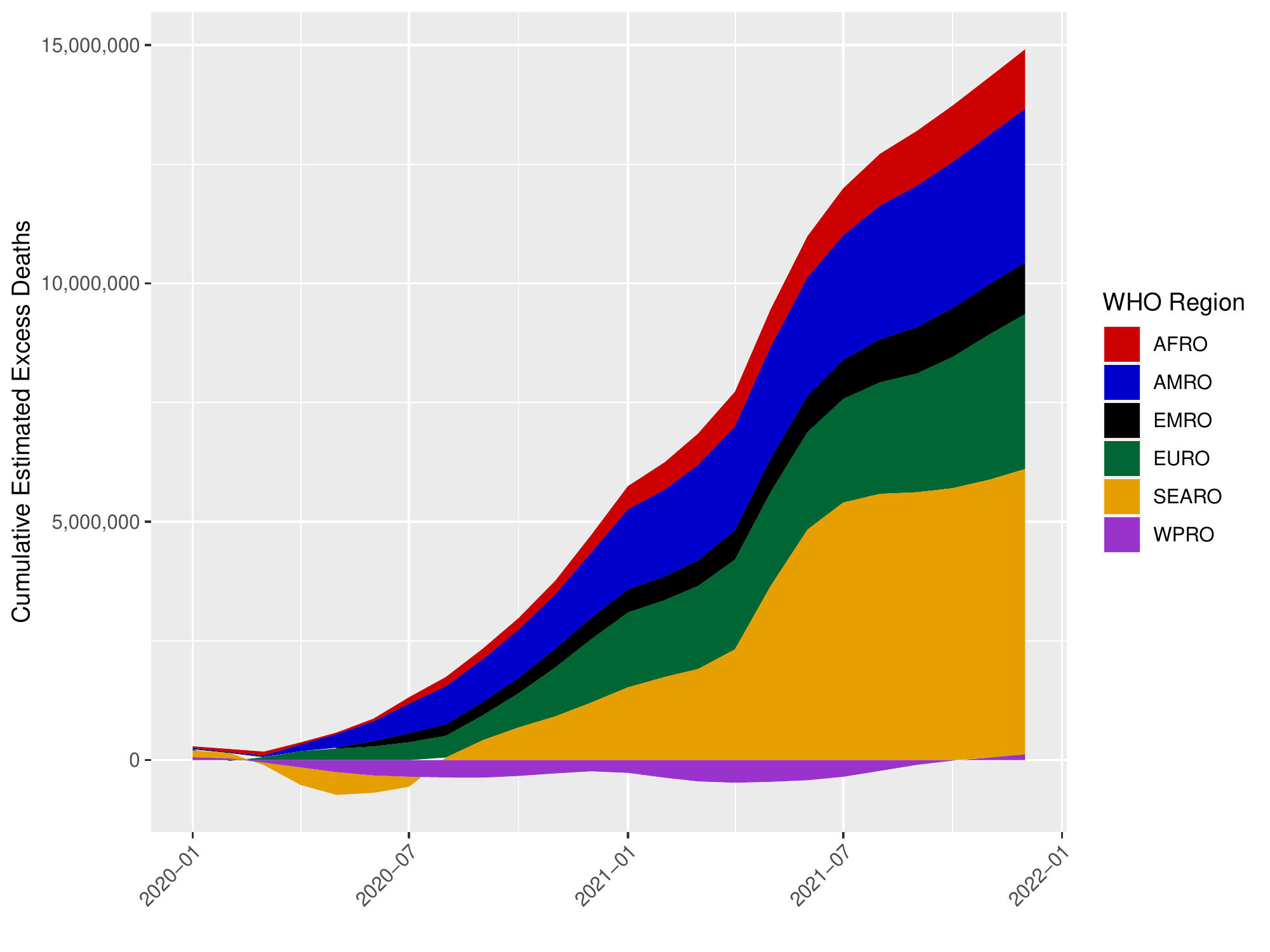}
\caption{Cumulative excess deaths over 2020--2021 for all countries, by region.}
\label{fig:CumulativeDeaths1}
\end{figure}

Figure \ref{fig:CumulativeDeaths1} gives the cumulative estimated by month and by region. The impact of the surge of deaths in India (which is in the SEARO region) in May 2021 is apparent. The WPRO region has a number of countries with negative excess (because of strong lockdown policies leading to the avoidance of certain types of death), and in this region, the mortality impact of the pandemic was smallest according to our analysis up to the end of 2021.

\begin{figure}
\centering
\includegraphics[scale=.25]{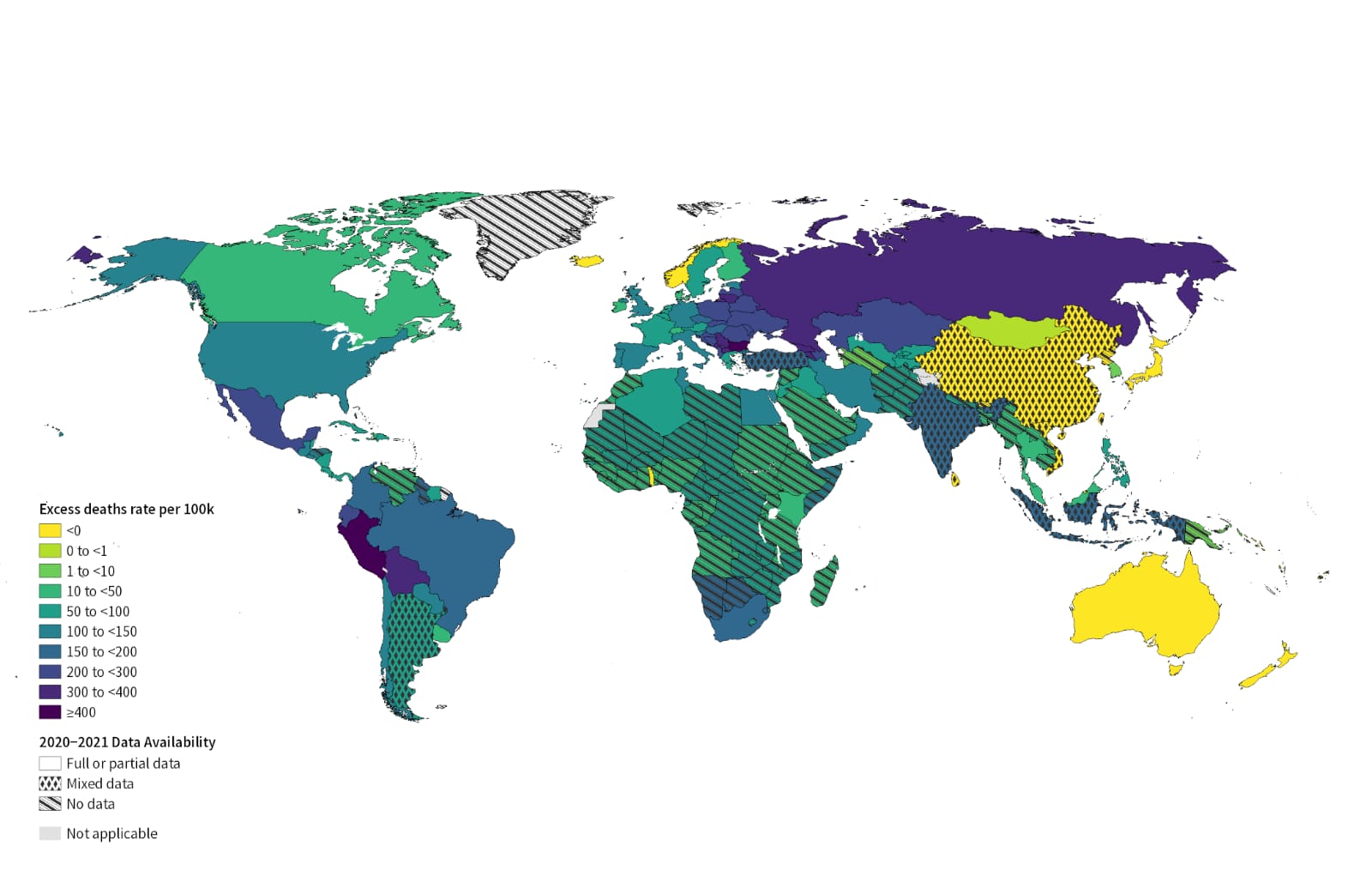}
\caption{Excess death rate, 
%$(Y_{c,t}-E_{c,t})/N_{c,t}$ 
per 100,000 by country. Countries with no hatching have monthly observed data, and the two  types of symbols indicate other data types.}
\label{fig:rel_excess_map}
\end{figure}

\begin{figure}
\centering
\includegraphics[scale=.25]{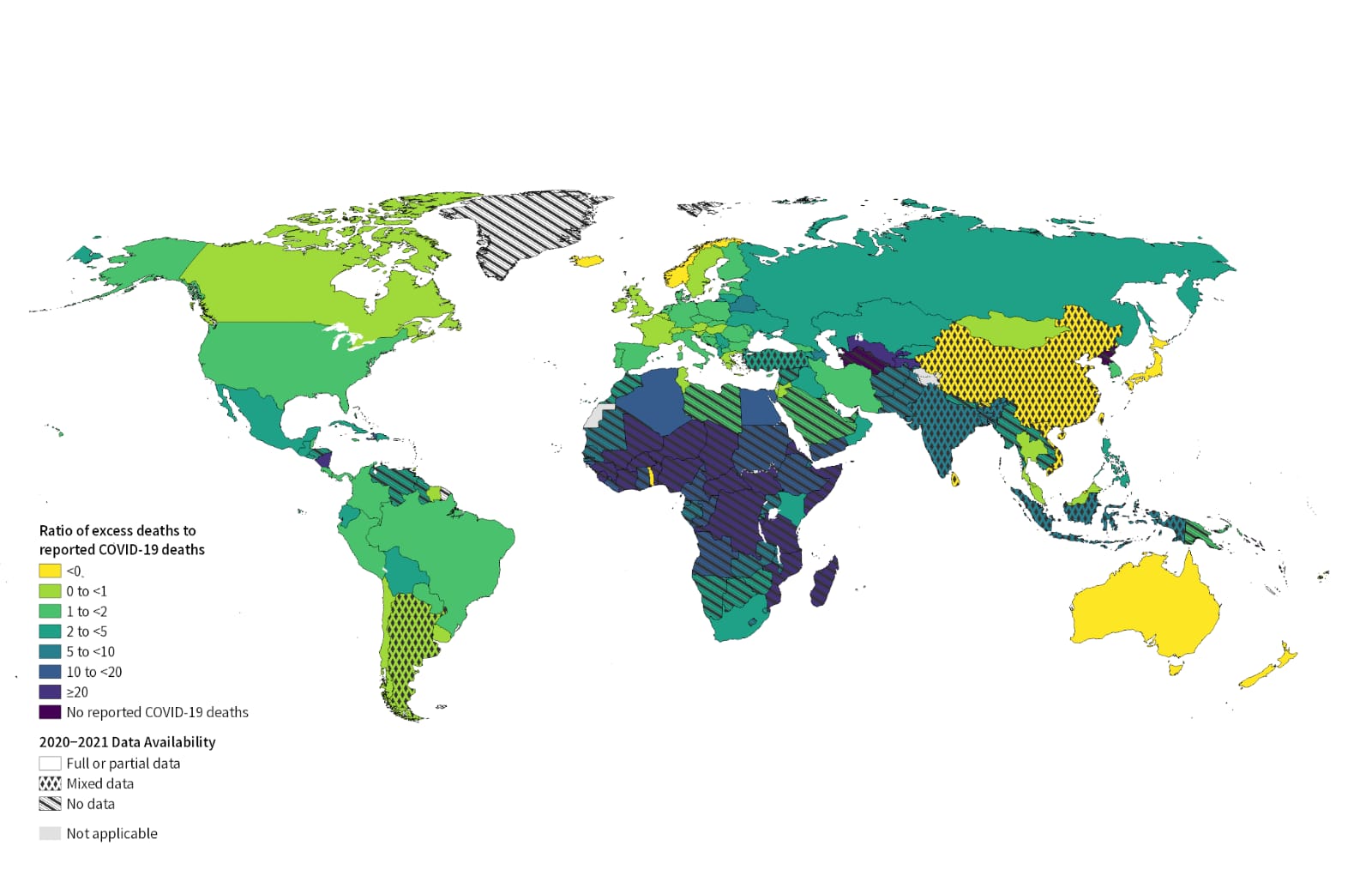}
\caption{Ratio of excess death rate to reported COVID-19 death rate, per 100,000 by country. Countries with no hatching have monthly observed data, and the two types of symbols indicate other data types.}
\label{fig:excess_covid_r_map}
\end{figure}

Figure \ref{fig:rel_excess_map} shows the global excess death rates, where countries with no data are highlighted with hatching and countries with subnational and annual data with diamond symbols. The paucity of full or partial data in AFRO and SEARO in particular is apparent. The countries with the  highest estimated excess yearly death rates (per 100,000 population, and 95\% credible intervals) are: Peru with 437 (431, 442), 
Bulgaria with 415 (399, 432)
and Bolivia with 375 (370, 379). These rankings should be viewed cautiously -- 
rankings of countries in terms of the excess death rate are examined more fully in the Supplementary Materials; in particular, the uncertainty in a country's placement in any list is highlighted.
%The countries with the  highest estimated ratios of excess deaths to reported COVID-19 deaths were: Peru (868), Bulgaria (834), Bolivia (744)
Countries with negative excess estimates include Australia, China, Japan, South Korea, Vietnam and New Zealand.  Figure \ref{fig:excess_covid_r_map} maps the ratio of excess deaths to reported COVID-19 deaths. There is a huge range of this excess, with many countries in the AFRO region having high ratios, and countries in Western Europe having ratios closer to 1 (with some, such as France, having values below 1). Globally, over January 2020--December 2021, there were 542,0534 reported COVID-19 deaths, and according to our estimates, the ratio of excess to reported COVID-19 deaths is 2.75, with a 95\% interval estimate of (2.46, 3.07), which is a huge discrepancy.

%Since we do not have national data to check this assumption we instead examine the sensitivity to dropping different states data out, along with other cross-validation exercises -- the Supplementary Materials contain full details.

\begin{figure}[htbp]
\centering
\includegraphics[scale=.4]{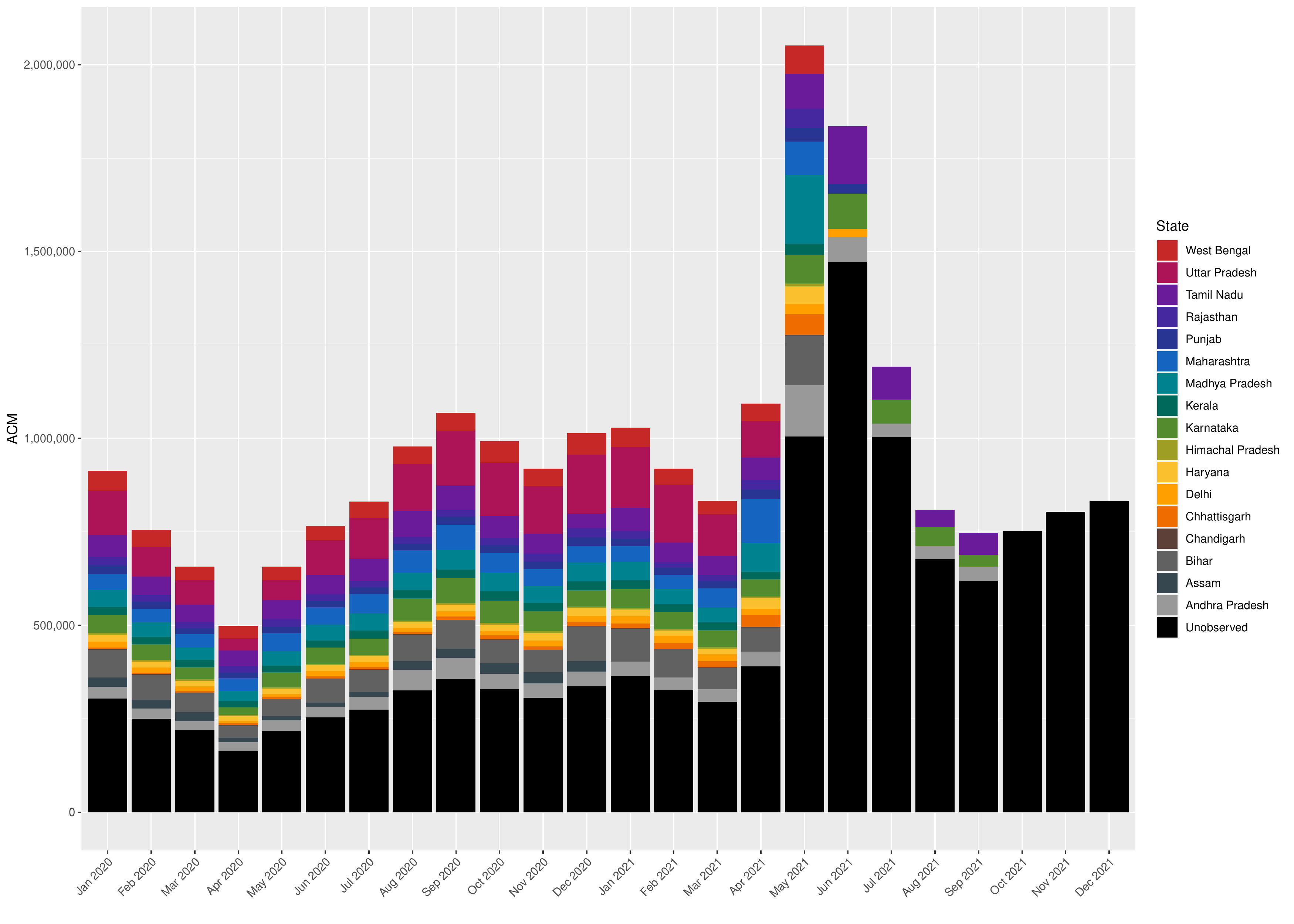}
%\hspace{-.8in}
%\includegraphics[height = 6.3cm]{ClusterLocs_Admin2.pdf}
\caption{All-cause mortality by month in the pandemic. Black rectangles are estimated while colored rectangles are observed.}
\label{fig:IndiaPlot2}
\end{figure}

We estimate that India has the highest cumulative excess of 4.7 million deaths, with a 95\% credible interval of (3.31, 6.48) million. 
Figure \ref{fig:IndiaPlot2} shows the ACM counts by states, with the black rectangles showing the estimated excess over the states that we have no data from, based on the fraction of deaths in each state, as estimated from the pre-pandemic period (the Supplementary Materials contain a similar plot for the pre-pandemic period, where the national total is also known).  For the final 3 months of 2021 there is data from a single state (Tamil Nadu) only  available, and for these 3 months the counts appear high, and so we do not use these data and instead use a simple predictive model. 
Specifically, we model $\log(Y_t/E_t)$ (using the estimated $Y_t$ for the first 21 months and weighting by the variance of the estimate) using an autoregressive order 1 (AR1) model, in INLA and then predict the final 3 months. More details on the AR1 model are contained in the Supplementary Materials.
Recall that these estimates are based on subnational data, and hinge on the assumption that at any month, the sum of the available states proportions are close to those observed historically. We cannot check this assumption and so we interpret our results with caution. The choice is between using the global covariate model, or the subnational the Supplementary Materials contain a sensitivity analysis in which we remove data from different states and examine the excess mortality estimates from the subsets only. We also provide a comparison between our estimates and those of different groups, which shows our estimates are consistent with previous studies.

%\begin{figure}
%\centering
%\includegraphics[scale=.5]{VictoriaFigures/India_ACMPlot2.pdf}
%\caption{Estimates with 95\% uncertainty for India. The final 3 months are based on an AR1 model.}
%\label{fig:India_ACMPlot}
%\end{figure}

\section{Comparison to Alternative Methods}\label{sec:other}

The Economist and IHME also produce country and global excess mortality estimates and The Economist update their estimates daily (which is not our objective). The Economist method is the more transparent and defensible of the two.
The Economist estimates excess deaths for all countries \citep{economist:21} using  methods described at \cite{economist:method:21}.  The Economist is not a peer-reviewed publication. From the start of the work, The Economist's methods and code have been freely available. The response is taken as excess deaths per 100k population, per day and the regression approach is gradient boosting \citep{friedman:01}, with regression trees applied to a very large collection of variables (144 in total) at the 7-day average level, when available. Since the excess is modeled, negative excess is possible; as we describe shortly, the IHME approach models log excess, so that negative values are not possible. A weighting of log population is taken in The Economist approach, though this choice is arbitrary, beside having the desirable property of having weights that increase with increasing population size. The weights are reduced by 50\% for subnational data sources, which is also arbitrary. The loss function is taken as mean squared error. An alternative would be to take the negative log likelihood of a Poisson as the loss function as described, for example, in Section 7.2 of \cite{buhlmann2007boosting}. 
%\url{https://www.economist.com/graphic-detail/2021/05/13/how-we-estimated-the-true-death-toll-of-the-pandemic}. 
The trees are grown based on the unpublished work of \cite{lunde2020information}.
 Model assessment is based on 10-fold cross-validation and a non-parametric bootstrap is used to assess (frequentist) uncertainty, based on 200 datasets sampled with replacement from the full data (with random sampling of countries first, and then observations within the sampled country). The expected numbers are modeled using the method described in \cite{WMD}. Specifically, the number of deaths is modeled as linear in year, with weekly (or monthly or quarterly, if weekly data not available) intercepts, using data from 2015--2019. These expected numbers are used directly in the calculation of the excess, when ACM data are observed. For countries without ACM data the excess (i.e.,~$\delta_{c,t}$) is predicted directly,  though the 2019 WHO ACM counts are used as one of the covariates. Uncertainty in the expected numbers in the overall uncertainty for the excess mortality is not accounted for.  The models  update daily, with two new models trained on the latest data every morning, replacing old models and then used for improved central estimates and estimates of uncertainty. In Figure \ref{fig:RegionalEstimates} we saw that The Economist confidence intervals are relatively wide when compared to those from our model. A benefit of the parametric approach that we describe is that  inference is efficient (to give narrower interval estimates) if the model assumptions are appropriate. The boosting algorithm approach provides a more flexible mean function, but the flexibility can lead to wide interval estimates. A more fundamental point is that the boosting estimator is a potentially sparse estimator, even when using trees, and the limiting distribution is a complicated object which may not be continuous, due to the selection of the covariates (since all of the covariates may not be always selected). As far as we know, no theoretical justification for the use of the bootstrap for gradient boosting exists. For further discussion see \cite{gine1990bootstrapping} and \cite{dezeure2015high}.

 IHME also  produce estimates of excess mortality with methods described in the Appendix of  \cite{wang:2022:covid}. Expected mortality is estimated using an ensemble approach in which six different models are used to model the expected numbers.  The expected ACM is only calculated for time periods not affected by late registration, which if not accounted for, would lead to underestimation of excess mortality rate. An out of sample prediction is then carried out for each of the models, and then the final predicted expected number is a weighted combination of the six models, with weights proportional to the mean squared error of prediction, as estimated from a leave-out exercise. While superficially this approach has elements in common with the super learner prediction algorithm \citep{vanderlaan:07}, it differs in key elements and does not share the optimality properties of super learner -- in summary, the weighting is ad hoc. %To select the covariates to appear in the final model the lasso \citep{tibshirani:96} is used. 

 An unweighted analysis is used, with response the log excess cumulative mortality rate:
 $$Z_{c}=\log [(Y_{c} - E_{c})/N_{c}]$$
 where $Y_{c}$, $E_{c}$ and $N_{c}$ are the observed cumulative counts, expected cumulative counts and population size respectively, for data in country $c$, all over the relevant observed period. The modeling of this difference does not seem as natural as modeling %$\log (Y_{c,t}/E_{c,t})$, 
 the log of observed over expected mortality which would be an approximation to the response we have used (though we model over time also).
The uncertainty in the true rate of excess is highly dependent on the population size, but this information is not used, since the model implicitly assumes each data point has the same uncertainty attached to COVID-19. If we assume that $\E[Y_{c}]=N_{c}\phi_{c}$ and $\v(Y_{c})=\kappa \E[Y_{c}]$ then, the delta method gives $\v(Z_{c}) \approx \kappa N_{c}\phi_{c}/(Y_{c} - E_{c})^2,$ which would give weights approximately proportional to $N_{c}$ (variance proportional to $1/N_c$), illustrating the inadequacy of the constant variance assumption.
 The covariates are also included based on the expected direction of the association, but this expected direction is presumably with respect to univariate models, and in a predictive model with multiple covariates it seems overly restrictive. Covariates are selected in an initial step using the log cumulative excess, as defined above, and the lasso \citep{tibshirani:96}. Since cumulative rates are used, a weighted average (e.g.,~using population) of the covariates is taken.
 
The uncertainty in this initial covariate selection phase is not accounted for, so that we would expect, all else being equal, the final predictive intervals to be too narrow. With the selected covariates (16 are listed in Section 4.2.2 of the Appendix of \cite{wang:2022:covid}), the log of the excess rate is modeled (using the expected ACM rate from the ensemble step and the observed ACM rate). We might also expect the modeling of the log excess to in some cases push estimates of the excess rate that are close to zero upwards.
%The uncertainty in the excess mortality rate is highly dependent on the population size, but this information is not used, since the model explicitly assumes each data point has the same uncertainty attached to it, which is a serious deficiency (in the Supplementary Materials, a delta method calculation shows that with quasi-Poisson sampling the variance of the log excess would be proportional to the population size). 
At this stage, Global Burden of Disease (GBD) defined regional and super regional residuals \citep{gbd_2019} are generated, and their mean is added to the prediction -- it is not clear why fixed (or random) effects are not added to the log excess rate model directly. This would make the calculation of uncertainty measures more straightforward. 

%CUMUALTIVE 

We describe the estimation of the excess rate for four different data scenarios:
\begin{itemize}
\item For countries with observed ACM data over the whole 2-year period, the only uncertainty arises from the modeling of the expected numbers -- the uncertainty in this step comes from parameter uncertainty, and not Poisson variation. For each of the six constituent models 100 draws are taken from the asymptotic normal distribution of the estimators, and then a weighted combination of the resultant predicted expected numbers is taken.  
%CHECK with SASAHA the REGMOD uncertainty.
% This is similar to the approach we take, albeit with a different model and a different calculation of uncertainty for the expected numbers.
\item For countries with no ACM data, similar to The Economist method, the expected numbers are not calculated, but instead the model directly predicts the excess rate using the estimated regression coefficients of the model. The uncertainty here comes from the random covariates and from the expected numbers modeling, not from any parameter uncertainty for any one fit. However, 100 fits are carried out with 100 different expected numbers. There is also no sampling uncertainty, analogous to our negative binomial uncertainty for ACM. This, combined with the lasso pre-selection of covariates, would indicate that the interval estimates would be too narrow, perhaps substantially so.

\item For countries with partial data, the cumulative excess rate over the missing (customized to each country) period is obtained from the regression model, adjusted by the residuals (as described above), and then taking random covariates for the missing period.
\item  It is not possible to obtain negative estimates from the log excess rate model, and so the only way for negative excess to arise is from countries with observed ACM data (for example, Iceland,
Australia, Singapore, New Zealand). The rationale is that there are few locations with a cumulative negative excess rate, and so they wish to avoid making predictions of negative excess.
\end{itemize}
 %A resampling approach is used to determine point and interval estimates (although some elements of the ensemble model are Bayesian, the overall uncertainty measures are presumably frequentist). The procedure involves sampling excess mortality (using the ensemble model) and covariates (some of the covariates are modeled) and then combining these inputs in the log excess mortality rate regression model, to obtain a prediction.
 The overall approach (which has not been peer-reviewed in the statistical literature) is more algorithmic than statistical in nature, and it would be impossible to determine its operating characteristics. In particular, the uncertainty estimates are unlikely to be well-calibrated -- we saw they were relatively narrow in Figure \ref{fig:RegionalEstimates}.

\section{Discussion}\label{sec:discussion}

The estimation of excess mortality during the COVID-19 pandemic is hamstrung by the lack of national ACM data for almost half the countries of the world, with EURO and AMRO being well-represented in the databases, but other regions more poorly. We have presented a relatively simple Poisson modeling framework, as we wanted to strive for transparency and leverage a well-understood Bayesian hierarchical structure. We stress that, within the Poisson framework, though we have different models for countries with different data types, the estimates for each country are comparable, and so side-by-side comparisons can be made (with the caveat that the range of uncertainty in the estimates for different countries varies considerably).
We deliberately avoid breaking down excess mortality into that directly attributable to COVID-19 and that not, since we believe the information required to do this accurately is unavailable.

%We have not considered late registration and pay more attention to under-registration.
%We have based our monthly estimates for China on the reported national ACM data, but we have not accounted for under-registration, which may be present due to the enormous pressures of the pandemic and a zero COVID 19 policy.

We did not adjust  the observed ACM on the basis of heatwaves, as done by \cite{WMD} and \cite{wang:2022:covid}, and neither did we adjust for conflicts (The Economist adjusts for conflict by excluding ACM data from places which entered large conflicts in the period). Another inadequacy of our modeling is that we are missing covariates in some countries, and regional values are used instead, we do not account for this uncertainty in our modeling.
We also do not currently account for the modeling of some of the covariates, and would like to address this aspect also. This is considered, albeit in an ad hoc, unvalidated procedure, by \cite{wang:2022:covid}.

Estimating excess mortality by month over the pandemic is a dynamic process and the results we have shown are a snapshot, given the current version of the model and the currently available data. As new data become available we will continue to both update our estimates, and  refine our model. Another aspect we will explore is the use of  spatial modeling, though we approach this with hesitancy.

A crucial component of the excess calculation is the estimation of the expected number of deaths. There are two elements to the calculation, the mortality data upon which it is based and the model that is adopted. First, with respect to the data, as mentioned in Section \ref{sec:mortality}, the WHO adjust the raw mortality counts, if there is perceived to be a completeness issue (and the scaling value may be carried forward to the pandemic period). We note that as part of the process to produce excess estimates, country consultation is carried out, in which  the adjusted country numbers are shared with government, who are asked to ``sign off" on the adjusted counts. Second, for the expected counts modeling, we used splines both for the annual trend and for the within-year seasonal variation, see equation (\ref{eq:eta}). A country for which the completeness adjustment and spline modeling provided a less than satisfactory excess estimates was Germany. Under the default data process/spline modeling the excess estimate was 195K with 95\% credible interval (161K, 2290K). However, on closer examination this excess estimate was too high due to a combination of data/model issues. For Germany, ACM in 2016--2018 were scaled up   due to the completeness assessment, which lead to a dip in the ACM sequence in 2019. The long-term spline fit to these adjusted data produced expected numbers that were too low (and therefore an excess that was too high). Hence, we reanalyzed the Germany data with unadjusted data and a linear term $f_c^\text{y}(\cdot)$ in equation (\ref{eq:eta}), rather than a spline. This produced a more realistic excess estimate of 122K with a 95\% interval of (101K, 143K). More details for the Germany analysis are contained in the Supplementary Materials.

For Sweden, we were concerned there were similar issues due to an unnecessary completeness adjustment of the raw mortality figure reported to the WHO in 2019 (the mortality count was
lower than recent counts).
On closer scrutiny, we decided that this adjustment was not necessary and we redid the analysis for Sweden, which also included using a linear term for the annual trend instead of a spline 
 (for the same reasons as described for Germany). The details are in the Supplementary Materials, but it resulted in an estimate for Sweden that was virtually unchanged, giving a point and interval of 13.4K (11.7K, 15.2K). The changes in the excess estimates for Germany and Sweden do not change the global or EURO figures. As a side note, for both these countries when using the unadjusted data, both the model with a linear term for the annual trend and the model with a spline for the annual trend produced similar excess estimates. However, using a spline for the annual trend can lead to sensitivity to the last year of pre-pandemic data, and 
a priority going forward is to systematically compare and evaluate different models for producing the expected numbers, building on recent work \citep{scholey2021robustness}. For the next round of estimates we will also revisit the under-reporting adjustment procedure.

To reiterate: the biggest limitation to our study is the lack of any observed monthly national mortality data in just under half of the countries of the world, which requires us to predict these counts based on a model built with data from countries which are not representative of the missing countries, or using subnational data. In Section \ref{sec:results}, we reported coverage estimates, calculated via cross-validation, that were reasonably close to the nominal. However, given the aforementioned regional imbalance in  countries for which we have data, we would not expect the coverage to be as accurate for the missing countries in, for example, the AFRO region.
%
% but if the frequentist coverage could be evaluated (hypothetically) for the countries that we have no data for, we would not expect the predictive intervals to be as well calibrated.
Improvements in death registration systems is vital to understand and react to pandemics in a timely manner, and obviate the need to carry out such modeling.

\section*{Acknowledgments}

The authors would like to thank the WHO Technical Advisory Group on COVID-19 Mortality Assessment, for helpful feedback during the model development, and also  Sondre Ulvund Solstad and Haidong Wang for generously sharing details on The Economist and IHME approaches, respectively. We would also like to thank the editor and three referees for their comments. The authors alone are responsible for the views expressed in this article and they do not necessarily represent the views, decisions or policies of the institutions with which they are affiliated.

\clearpage
\bibliographystyle{natbib} 
\bibliography{/Users/jonno/Dropbox/BibFiles/spatepi}

\end{document}

\begin{frontmatter}
\title{Supplementary Materials for ``Estimating Global and Country-Specific Excess Mortality During the Covid-19 Pandemic''}
\runtitle{Estimating Excess Mortality}
\begin{aug}
\author[A]{\fnms{Victoria} \snm{Knutson}\ead[label=e1]{vknuts@uw.edu}},
\author[A]{\fnms{Serge} \snm{Aleshin-Guendel}\ead[label=e2]{aleshing@uw.edu}},
\author[C]{\fnms{Ariel}\snm{ Karlinsky}\ead[label=e3]{karlinsky@gmail.com}},
\author[D]{\fnms{William }\snm{Msemburi}\ead[label=e4]{msemburiw@who.int}},
\author[A,E]{\fnms{Jon} \snm{Wakefield}\ead[label=e5]{jonno@uw.edu}},
% Debbie Bradshaw$^3$, 
%William Msemburi$^3$, Jon Wakefield$^{1,4}$}
\address[A]{Department of Biostatistics, University of Washington, Seattle, USA}
%\address[B]{Department of Biostatistics, University of Washington, Seattle, USA}
\address[C]{Hebrew University, Jerusalem, Israel}
\address[D]{World Health Organisation, Geneva, Switzerland}
\address[E]{Department of Statistics, University of Washington, Seattle, USA}
%
%\affil{$^2$Hebrew University, Jerusalem, Israel}
%
%%\affil{$^3$Burden of Disease Research Unit, Medical Research Council of South Africa, Tygerberg, South Africa}
%\affil{$^3$World Health Organisation, Geneva, Switzerland}
%\affil{$^4$Department of Statistics, University of Washington, Seattle, USA}
\end{aug}

\end{frontmatter}

\section{Modeling Framework}

In this section we discuss the overview of the modeling framework. The excess is defined as the observed all-cause mortality (ACM) minus the expected ACM. Modeling is required to obtain the latter. If ACM data were available for all countries during historic and pandemic then we would only need to model the historic data to produce expected numbers (assuming that there were no issues of undercount or late registration).  
%As an aside, this was the situation considered by \cite{konstantinoudis2022regional} in which for five European countries, historic ACM data were modeled using a Bayesian hierarchical Poisson model (with log population as offset) with spatial smoothing, a weekly random walk of order 1 (RW1) model (to account for seasonality), yearly intercepts, a flexible model for the association of mortality with temperature, and an indicator for public holidays. Each of the countries had observed data during the pandemic so the model was only fitted to 2015--2019 data and then predicted forward, to produce the expected numbers.

One approach, for countries have full ACM data, would be to model the historic and pandemic data simultaneously. For example, one could include in a model seasonable terms and a long-term trend, and then have an indicator for pandemic times. However, the aim of our modeling of pandemic data (for those countries that have it), is to construct a predictive model for those countries who have not provided ACM data in the pandemic, using country-specific covariates. Many of these covariates vary by month, and we wish the ``signal'' during the pandemic to be attributed to the covariates, rather than country-specific seasonality terms. Hence, we first model the historic period data (to create expected numbers), and then, conditional on the expected numbers, model the pandemic data. The expected numbers modeling is done for all countries, regardless of the data they have available in the pandemic. Recall, the excess
$$\delta_{c,t} = y_{c,t} - E_{c,t},$$
and the first term on the RHS is known for countries with full data, in which case the posterior for $\delta_{c,t}$ is based on the posterior for $E_{c,t}$ only.

\begin{table}[htp]
\begin{center}
\begin{tabular}{l|c|c|c|c||c}
& Full Data & Subnational Data& Annual Data & No Data&All Data\\ \hline
Historic & $\by^\text{hf}$ & $\by^\text{hs}$ & $\by^\text{ha} $& $\by^\text{hn}$& $\by^\text{h}$\\
Pandemic &$ \by^\text{pf}$ &$ \by^\text{ps}$ &$ \by^\text{pa}$ & $\by^\text{pn}$& $\by^\text{p}$\\ \hline
Expected &$\bE^\text{f} $&$\bE^\text{s} $&$\bE^\text{a} $&$\bE^\text{n} $&$\bE$\\ \hline
Covariates & $\bx^\text{f}$ & Not used & $\bx^\text{a}$&$\bx^\text{n}$& $\bx$
\end{tabular}
\end{center}
\caption{Notation by data types in historic and pandemic times; $\by^\text{pn}$ is unobserved.}
\label{framedata}
\end{table}%

Table \ref{framedata} summarizes the notation for the data we have available.
%Let $\by^\text{h}=\{ y^\text{h}_{c,t} \}$ and $\by^\text{p}=\{ y^\text{p}_{c,t} \}$ denote the vector of all historic and pandemic data, respectively. 
We let $\bphi$ represent the parameters of the models that are used to calculate the expected numbers and $\blambda$ the parameters that are used to model the pandemic data. 
The posterior for the unknown parameters is:
\begin{eqnarray*}
\pi(\bphi, \blambda | \by^\text{h},\by^\text{p},\bE,\bx) &\propto& p( \by^\text{h},\by^\text{p} | \bphi, \blambda,\bE,\bx) \times \pi (\bphi, \blambda)\\&=& 
p( \by^\text{h}| \bphi ) \pi(\bphi) \times p(\by^\text{p} |\blambda , \bE,\bx)  \pi ( \blambda)
\end{eqnarray*} 
where $\bE = \bE(\bphi,\by^\text{h})$. We next discuss the modeling of each of the historic and pandemic data.

The posterior for the parameters $\bphi$ in the expected numbers model is:
$$p (\bphi | \by^\text{h} ) \propto  \underbrace{p( \by^\text{h}  | \bphi )}_{\text{{Expecteds Model}}} \times \pi (  \bphi  ).$$
This posterior is approximated using the {\tt mgcv} library {\tt gam} function \citep[Section 6.10]{wood2017generalized}.

%Let  the vectors $\by^\text{hf}=\{ y^\text{hf}_{c,t} \}$ and $\by^\text{pf}=\{ y^\text{fp}_{c,t} \}$ denote the observed {historic} and  {pandemic} mortality data, respectively fwhere we are collecting together the data over months. We let $\bE_c = \bE_c(\by_c^\text{fh})$ represent the expected numbers.
%For countries with observed ACM data for all months of the pandemic and for all countries. 
%
%
%We now extend to derive the posterior for the different data types. We let $\by^\text{pf},\by^\text{hf}$ denote the pandemic and historic data for countries with full data, and where we have collected together the  data across countries. Similarly,  let $\by^\text{sp},\by^\text{sh}$ denote the pandemic and historic data for countries with subnational data, and where we have again collected together the  data across countries. Finally, let  $\by^\text{nh}$ denote the historic data for countries with no pandemic mortality data. 
We let $\blambda=(\btheta,\bp)$ where $\btheta$ are the parameters of the covariate model and $\bp$ the parameters of the subnational multinomial model. For the annual data we use a mulitnomial model with the probabilities depending on the covariate model.
The posterior is,
\begin{eqnarray*}
\pi (  \btheta ,\bp| \by^\text{pf},\by^\text{ps},\by^\text{pa},\bE^\text{f},\bE^\text{a},\bx^\text{f},\bx^\text{a}) &\propto& 
p( \by^\text{pf},\by^\text{ps},\by^\text{pa}|  \btheta, \bp, \bE^\text{f},\bE^\text{a},\bx^\text{f},\bx^\text{a}) \times \pi (  \btheta,\bp)\\
&=&  \underbrace{p( \by^\text{pf} | \btheta,\bE^\text{f},\bx^\text{f} ) \times  \pi (\btheta ) }_{\text{Covariate Model}}\\
&\times & \underbrace{p ( \by^\text{ps} | \bp )\times  \pi(\bp) }_{\text{Subnational Model}}  \\
\\&\times & 
\underbrace{p ( \by^\text{pa} | \btheta,\bE^\text{a},\bx^\text{a} ) }_{\text{Annual Model}}.
% \\&\times & \underbrace{p ( \by^\text{fh},\by^\text{sh}, \by^\text{nh} | \bphi ) \times \pi( \bphi )  }_{\text{Spline Model}} 
\end{eqnarray*}
We can factor the above into two pieces: the $\btheta$ parameters from the covariate model and annual models, and the $\bp$ parameters from the subnational model.
Note that there are very few annual data countries and for practical reasons we estimate the $\btheta$ parameters from the covariate model, and then used samples from the posterior to estimate the monthly contributions to the annual data.
Predictive distributions are constructed  for the total ACM  counts for the no data countries, using 
$$p(\by^\text{pn} | \btheta,\bx^\text{n}),$$
which is a negative binomial distribution.

To obtain the posterior for the excess $\bdelta^\text{f},\bdelta^\text{s},\bdelta^\text{a},\bdelta^\text{n}$ for countries with full, subnational, annual and no data, respectively, note that $\bdelta^\text{f} = \bdelta(\bphi)$, $\bdelta^\text{s} = \bdelta(\bphi,\bp), \bdelta^\text{a}=\bdelta(\bphi,\btheta)$ and $\bdelta^\text{n} = \bdelta(\bphi,\btheta)$. With sampling-based methods (including INLA, which allows samples to be taken from the posterior), it is straightforward to obtain samples for $\bdelta^\text{f},\bdelta^\text{s},\bdelta^\text{a},\bdelta^\text{n}$ using samples from $\bphi,\btheta$, respectively.

Notes:
\begin{itemize}
\item For the multinomial within-year expected mortality model we use temperature only to account for seasonality of deaths. This is a simple model and so the month to month variation in the excess for the countries that use this model will not be as accurate as the cumulative yearly totals that use the observed mortality counts.
\item For the subnational model, we emphasize that do not use any covariates.
\item The above is slightly simplified, because in reality many countries switch between data types, such as having full data for part of the pandemic and then annual data only or no data. We could write down a more detailed posterior distribution but it would be complicated and no insight would be gained. As described in Section 4 of the main paper, we ``benchmark" the join between full and no data, so that there are no sudden jumps.
\item In a similar vein, China has subnational and annual national data, and we could write down the posterior, but it would not be useful.
\end{itemize}

%We 
%model as: 
%\begin{eqnarray*}
%p( \by^\text{p} , \by^\text{h} |   \btheta,\bphi) &=&p( \by^\text{p} , \bE |  \btheta,\bphi)\\& =& \underbrace{p( \by^\text{p} | \bE , \btheta)}_{\text{{Pandemic Mortality Model}}}\times \underbrace{p(\bE | \bphi)}_{\text{{Spline Model}}}.
%\end{eqnarray*}
%
%When pandemic data are observed, we use them directly, and then use predictive distributions for those countries with unobserved ACM.
%To simplify a little, let $\by_o^F,\by_o^S$ denote countries with full and subnational {observed} data and $\by_m^S,\by_m^N$ denote the {missing} data from countries with subnational and no data.
%
%We have {predictive distribution:}
%$$p(\by_m^S,\by_m^N | \by_o^F,\by_o^S,\bE) = \underbrace{p(\by_m^S | \by_o^S)}_\text{{Subnational Model}} \times
%\underbrace{p(\by_m^N | \by_o^F,\bE)}_\text{{Covariate Model}}.
%$$

\begin{figure}[htbp]
\centering
\includegraphics[scale=.63]{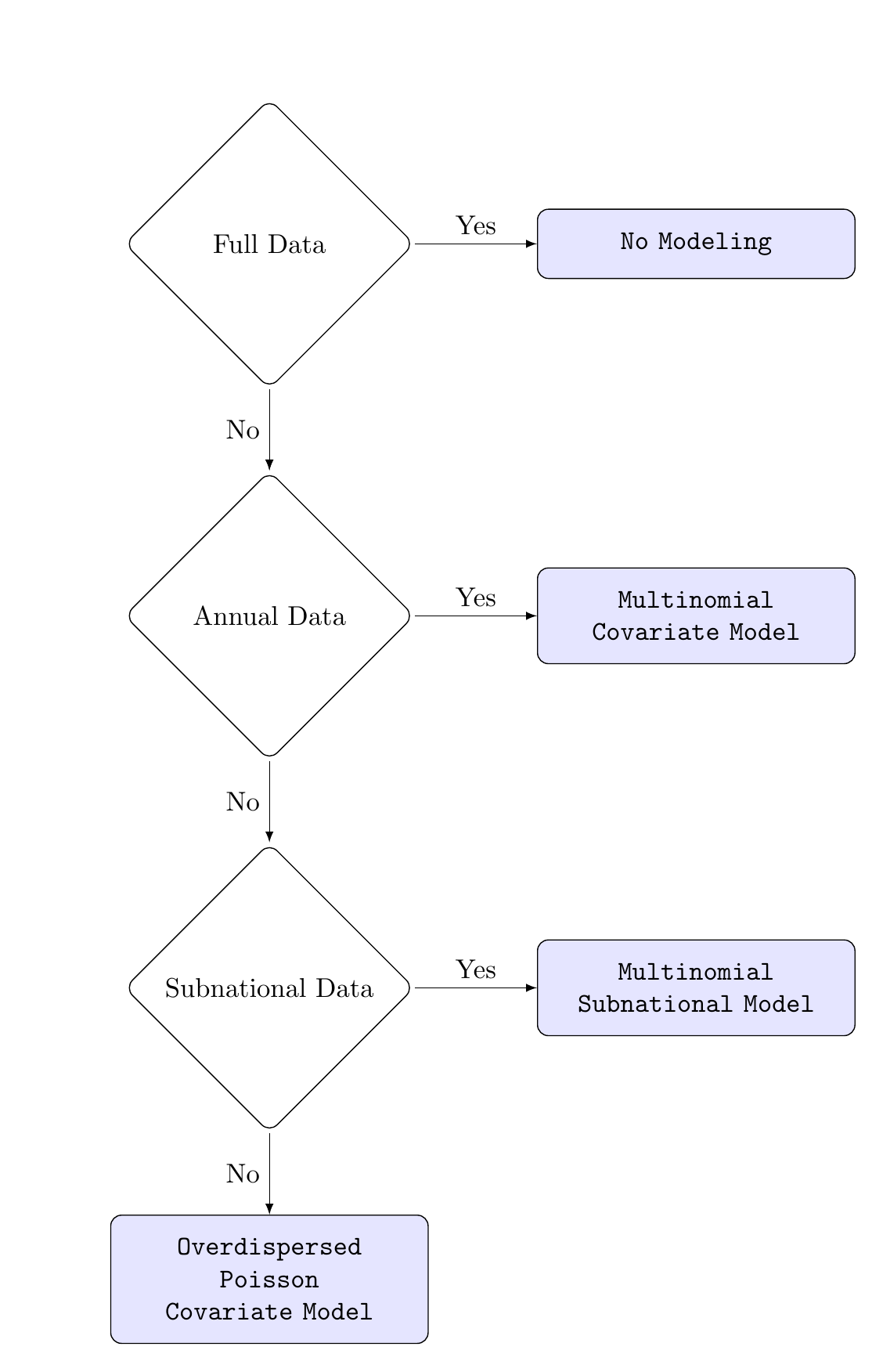}
\caption{Flowchart of ACM model choices based on data availability. The bottom box corresponds to the ``No Data situation''.}
\label{fig:flowchart}
\end{figure}

\begin{figure}[htbp]
\centering
\includegraphics[scale=.63]{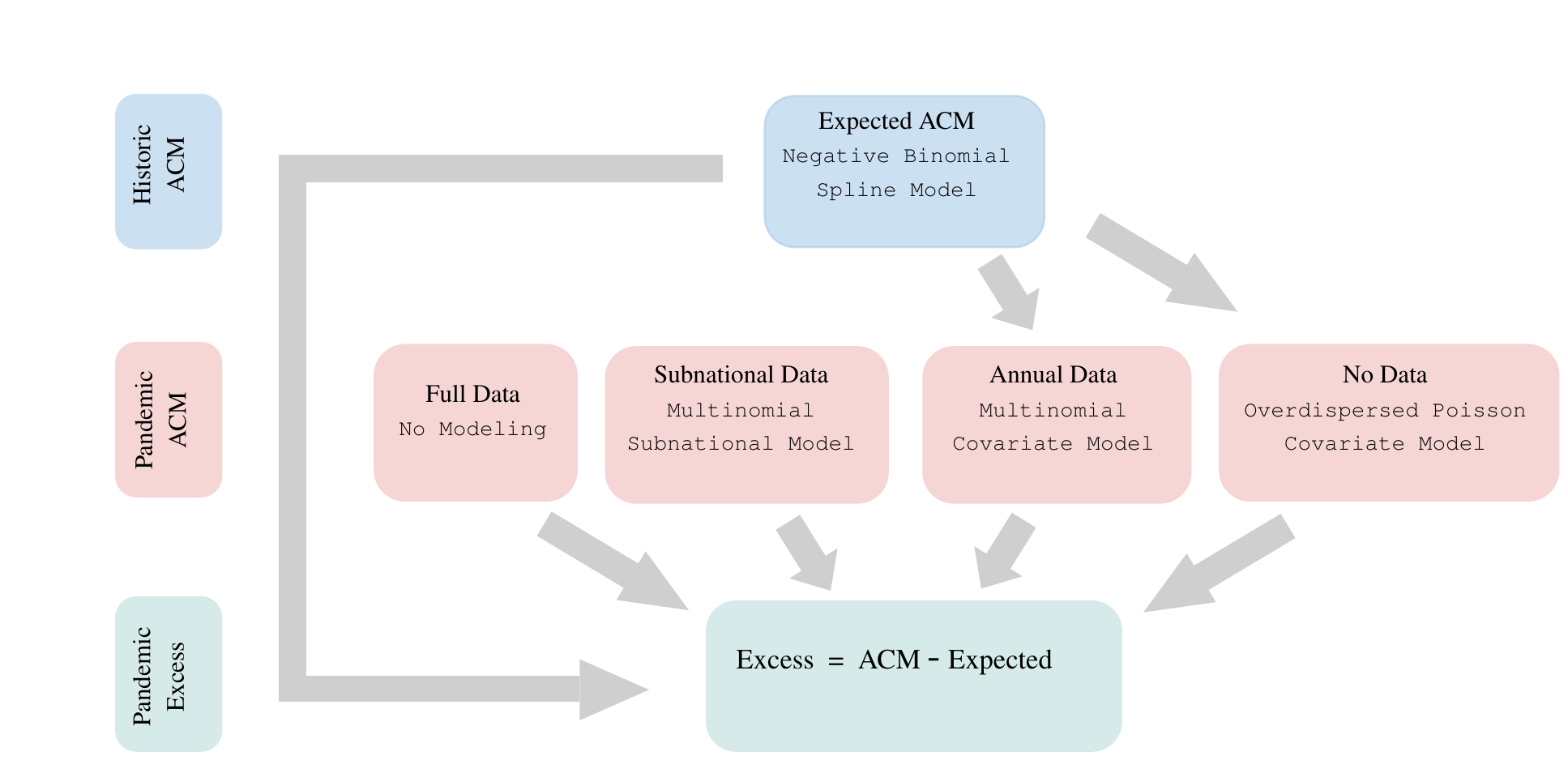}
\caption{Overview of modeling strategy.}
\label{fig:COVIDModelFlowChart}
\end{figure}

Figure \ref{fig:flowchart} summarizes, via a decision tree, how the models are chosen for the different types of data, and Figure \ref{fig:COVIDModelFlowChart} shows the relationship between the different models, and how they feed into the excess calculation.

\clearpage
\section{Supplementary Materials: Evidence from Countries without All-Cause Mortality Data}

%Details on imputation.

%All Cause Mortality data necessary for estimating excess mortality is lacking in many countries. 
%I list here and cite some additional evidence from researchers, journalists and other sources which suggest significant excess mortality and undercounting of COVID-19 in some of these countries. Currently, evidence is presented for the countries: {\CountriesAdd}
%\end{abstract}
%
%\section{InLine}

We briefly review
%Unfortunately, we were not able to obtain sufficient ACM data from many countries, and thus have to project these deaths using the countries for which we have information. However, while lacking sufficient data, there's 
preliminary evidence from researchers, journalists and various officials of significant excess mortality in many of the countries for which we have no ACM data. This includes, but is not limited to, evidence from Honduras \citep{efe_honduras_2021}, Haiti \citep{degennaro_2021}, Pakistan \citep{bbc_pakistan}, Tanzania \citep{parkinson_tanzania_2021}, Bangladesh \citep{hanifi_etal_2021, icddrb_burials, icddrb_survey}, Zambia \citep{lusaka_mourges_2021, lusaka_mourges_2}, Syria \citep{watson2021leveraging}, Yemen \citep{besson2021excess}, Sudan \citep{sudan_exces, omdurman_sudan_2021} and Papua New Guinea \citep{png_2021}. 
Also, evidence from territories which are not member countries show significant excess that tracks reported COVID-19 deaths. These territories neighbor many countries for which we lack data and are suggestive of similar patterns in their regions. This includes territories in the Pacific such as New Caledonia and French Polynesia, in the Caribbean such as Aruba, Guadeloupe, Martinique and Guadeloupe, and in Africa such as R\'{e}union and Mayotte \citep{WMD}.	

%\section{Additional Evidence}
%\begin{itemize}
%    \item Americas
%    \begin{itemize}
%        \item Honduras: The news agency EFE reported in September 2021 that the Funeral Homes Association of Honduras data suggests 19,830 people died of COVID-19, more than double the official figure at the time of 9,679 \citep{efe_honduras_2021}.
%        \item Haiti: \citet{degennaro_2021} estimate 16,204 excess deaths from March to October 2020 in Port-au-Prince alone. They surveyed 18 funeral homes and compared pre-COVID monthly number of funerals with that during COVID, and then projected it to the entirety of Port au Prince using the estimated annual death count of 23,282. This estimate of excess mortality is over 70 times higher than the reported COVID-19 deaths in all of Haiti.
%    \end{itemize}
%    \item Asia
%    \begin{itemize}
%        \item Pakistan: BBC in Urdu reported that in several cities in Pakistan such as Karachi and Lahore, the number of burials in June 2020 was 2-3 times higher than that in June 2019 \citep{bbc_pakistan}.
%    \end{itemize}
%    \item Africa
%    \begin{itemize}
%        \item Tanzania: \citet{parkinson_tanzania_2021} has interviewed doctors, officials and graveyard managers regarding all cause mortality during COVID-19. While unable to provide a concrete figure, the conclusion is that "Covid-19 has likely killed thousands'', in comparison to 750 reported deaths.    
%    \end{itemize}
%
%\end{itemize}
\clearpage
\section{Supplementary Materials: Poisson-Multinomial Trick}

Multinomial data cannot be directly fitted in INLA but can be modeled using the Poisson-Multinomial trick \citep{baker:94}.
For the multinomial model we use in Section 3.4 of the paper we can fit the Poisson model:
$$Y_{c,v,m} | \lambda_{c,v},\beta \sim \mbox{Poisson}( \lambda_{c,v} \exp(z_{c,v,m}\beta) ),$$ 
with the default prior (normal distribution with large variance) for $\beta$ and the improper prior $1/
lambda_{c,v}$ for $\lambda_{c,v}$.
We let $g_{c,v,m} = g_{c,v,m}(\beta) = \exp(z_{c,v,m}\beta)$ and $G_{c,v}=\sum_{m'} g_{c,v,m'}$. Then
\begin{eqnarray*}
\Pr( \bY | \beta) &=& \prod_c \prod_v\int  \prod_m \exp( -\lambda_{c,v} \times g_{c,v,m} ) (\lambda_{c,v}g_{c,v,m})^{y_{c,v,m}} \times \lambda_{c,v}^{-1}~d\lambda_{c,v} \times g_{c,v,m}^{y_{c,v,m}}\\
&\propto&  \prod_c \prod_v G_{c,v}^{-y_{c,v,+}} \prod_m g_{c,v,m}^{y_{c,v,m}}\\
&=& \prod_c \prod_v \prod_{m} \left(
\frac{g_{c,v,m}}{G_{c,v}}
\right)
^{y_{c,v,m}},
\end{eqnarray*}
i.e.,~a multinomial with probabilities $g_{c,v,m}/G_{c,v}$, which is the model we wish to fit.

\section{Supplementary Materials: Subnational Model Simulation Study} 

To examine the behavior of the Multinomial model that we use for subnational modeling in Section 5.1 of the paper,  we perform a simulation study.
%described on page 22, in the prediction of total country All Cause Mortality from reported subnational All Cause Mortality, we conceive of the following setup whereby 
We simulate ACM national data in month $t$ from the model
$$Y_{+,t} = 1000 + 0.1\times \left[1+\sin \left( W_t \right)\right],$$
for $t=1,\dots,24$ where $W_t = 0, \frac{\pi}{6}, \frac{\pi}{3},\dots,4\pi$.
% so we consider some sinusoidal pattern to simulate fluctuant patterns in All Cause Mortality under COVID. 
Then from this underlying true total mortality we simulate subnational counts for $K=5$ regions from our Multinomial model where,
$$\bY_t | \bp_t \sim \mbox{Multinomial}_{K+1}( Y_{t,+} , \bp_t ),$$
with $\bp_t=\{ p_{t,k}, k \in S \}$, 
$$p_{t,k} = \Pr( \mbox{ death in region $k$ } | \mbox{ period $t$, death }),$$
and
$$\mbox{log}\left(\frac{p_{t,k} }{p_{t,K+1}}\right)= \alpha_k + \epsilon_t,$$
where the error term is given by $\epsilon_t \sim \N(0,0.5^2)$ and  the $\alpha_k $ parameters are given in Table \ref{tab:simulation_alphas}.

\begin{table}[htp]
\begin{center}
\begin{tabular}{|l|c|c|c|c|c|}
 & Region 1 & Region 2 & Region 3 & Region 4 & Region 5\\ \hline
$\alpha_k$ & -0.25 & -1.3 & -1.15 & -2.5 & 1.75
\end{tabular}
\end{center}
\caption{Values of $\alpha_k$ used in the simulation study for the subnational model, $k=1,\dots,5$.}
\label{tab:simulation_alphas}
\end{table}

Next, we randomly sample  a number $J$ of region-time points  to remove as missing subnational values, to replicate a random pattern of missingness, similar to what we see in the subnational data for India. We choose $J=20$ region-time observations, from the available total of $24 \times 5 = 120$. Finally, we fit the multinomial model described above using this data with the first $t=1,\dots,18$ time points as observed subnational and total national mortality for fitting the model, and the remaining $t=19,\dots,24$ time points for prediction and validation. We then compare the predicted values from this subnational model to the true values of $Y_{+,t}$ which are known in this simulation setting. 

\begin{figure}[htbp]
\centering
\includegraphics[scale=.4]{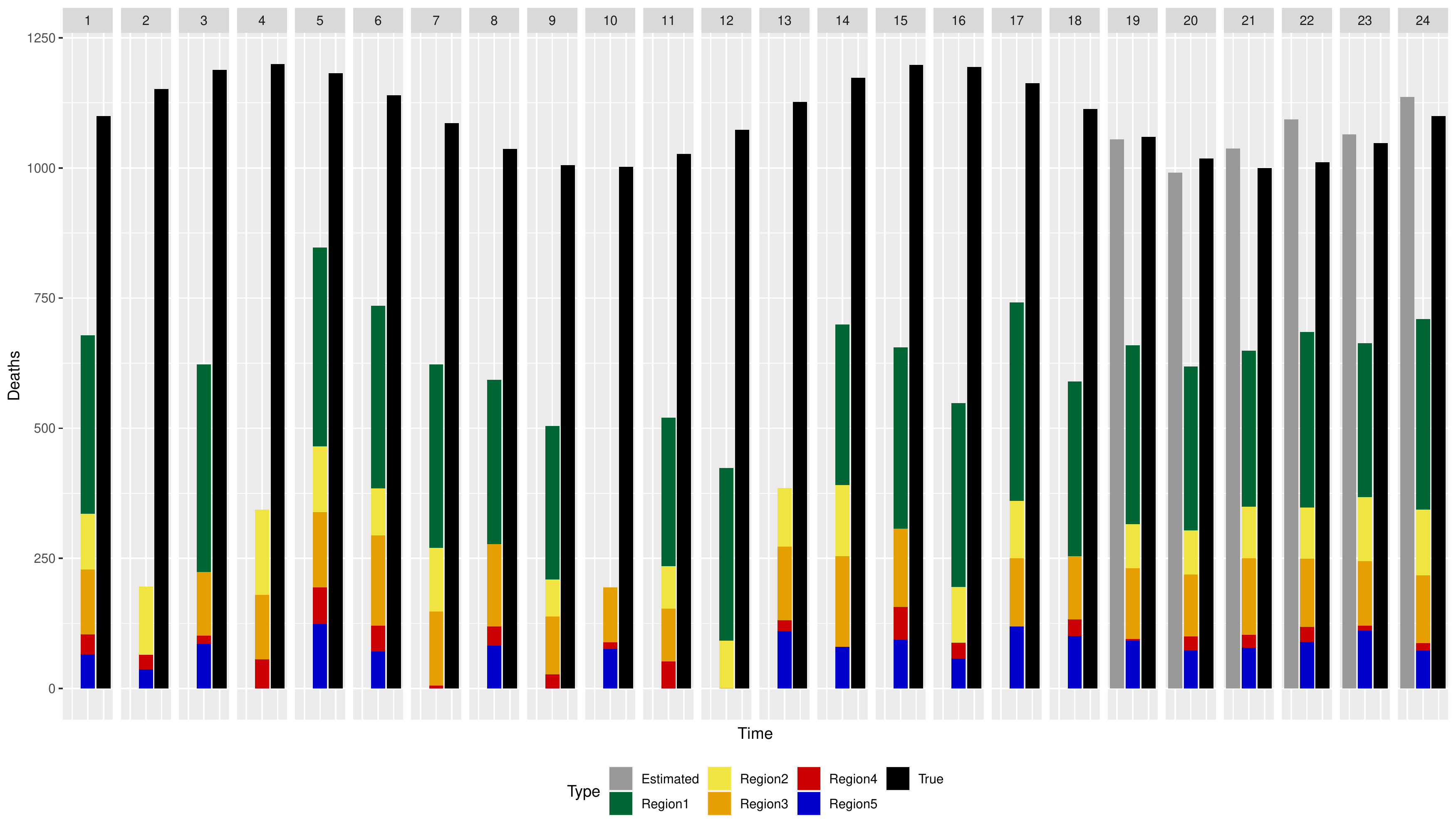}
\caption{Simulated regional and national data across all 24 time points, with the last 6 time points showing the estimated and true deaths.}\label{fig:Multinomial_SimulationResults}
\end{figure}

\begin{figure}[htbp]
\centering
\includegraphics[scale=.4]{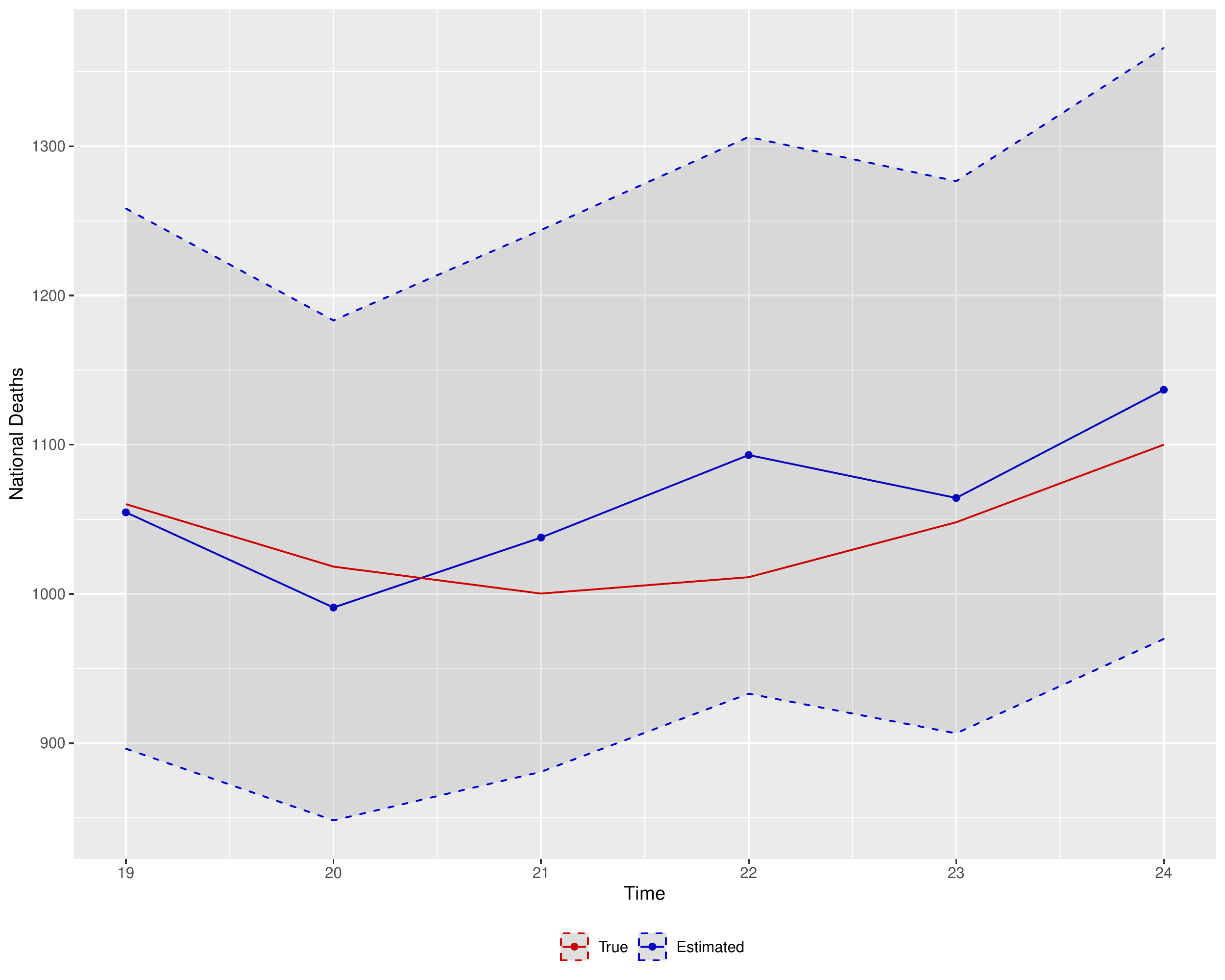}
\caption{Subnational simulation: True and predicted national deaths during the prediction window (6 time points), along with 95\% credible interval.}\label{fig:Multinom_Simulation_Ests}
\end{figure}

In Figure \ref{fig:Multinomial_SimulationResults}, we show the data that are available, along with the predictions and truth in the final 6 time periods. In Figure \ref{fig:Multinom_Simulation_Ests}, we show the observed national ACM in the last 6 months, along with the point and interval estimate from our model. The model captures the trend in the mortality well, though the interval estimates are quite wide.

\clearpage
\section{Supplementary Materials: Gamma Expected Numbers}

In Section 3 of the paper we describe how the expected numbers are modeled using a negative binomial spline model. From this model we obtain Monte Carlo samples from the predictive distribution of the expected numbers in 
pandemic months. Note that we obtain the predicted for the mean count (so that we do not include negative binomial uncertainty). We approximate the distribution of the Monte Carlo samples using gamma distributions, which can then be marginalized over in our predictive log-linear Poisson model (since they are conjugate) -- see Section 4 of the main paper for details.

\begin{figure}[htbp]
\centering
\includegraphics[scale=.4]{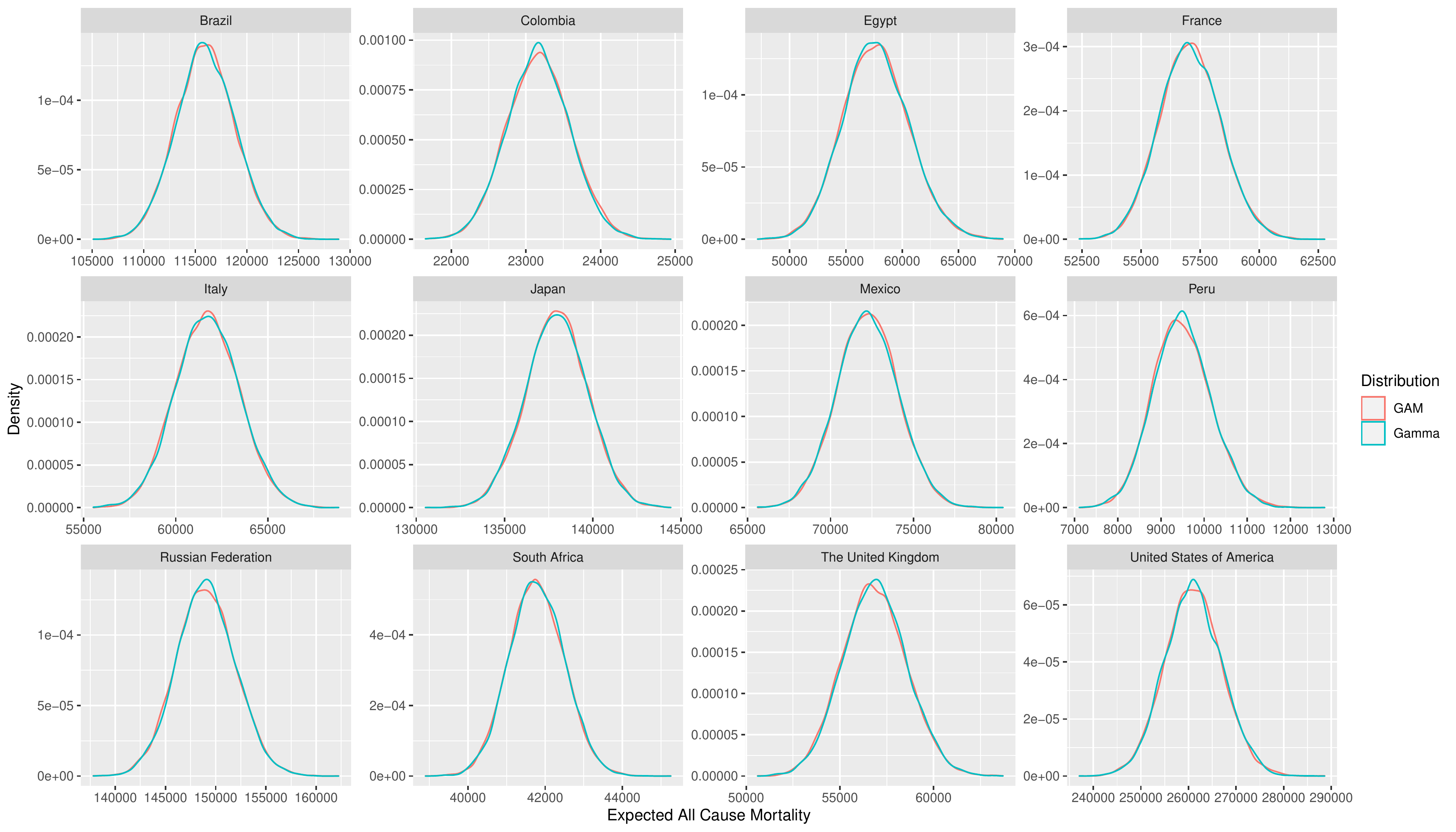}
\caption{Gamma fits to the predictive distributions of the expected numbers, for a range of countries, in December 2021.}\label{fig:expected_check_dec_2021}
\end{figure}

In Figure \ref{fig:expected_check_dec_2021} we show the gamma fits to the predicted distribution of the expected numbers, in December 2021, for a selection of countries. The differences between the predictions and the gamma approximations are small in all cases. This is a randomly chosen month, and is representative of the accuracy of the approximation.

\clearpage
\section{Supplementary Materials: Further Results}

\subsection{Global Summaries}
We first discuss the estimates for the association parameters in the log-linear model described in Section 4 of the paper:
\begin{eqnarray}\label{eq:log-linear}
\log \theta_{c,t} &=& \alpha + 
\sum_{b=1}^B  \beta_{bt}  X_{bct} + \sum_{g=1}^G \gamma_g Z_{gc} + \epsilon_{c,t}.
%\log \theta_{c,t} &=& \alpha^\text{\tiny{H}}_{t} 1( c \in \mbox{ high }) + \alpha^\text{\tiny{N}}_{t} 1( c \in \mbox{ not high }) \\
%&+& \beta^\text{\tiny{H}}_{t} 1( c \in \mbox{ high }) X_{c,t} + \beta^\text{\tiny{N}}_{t} 1( c \in \mbox{ not high }) X_{c,t}
\end{eqnarray}
Apart from the high income indicator, all covariates are standardized to have mean 0 and standard deviation 1, which aids in prior specification and when comparing estimated coefficients. This should be borne in mind as associations are examined. 

\begin{figure}[htbp]
\centering
\includegraphics[scale=.5]{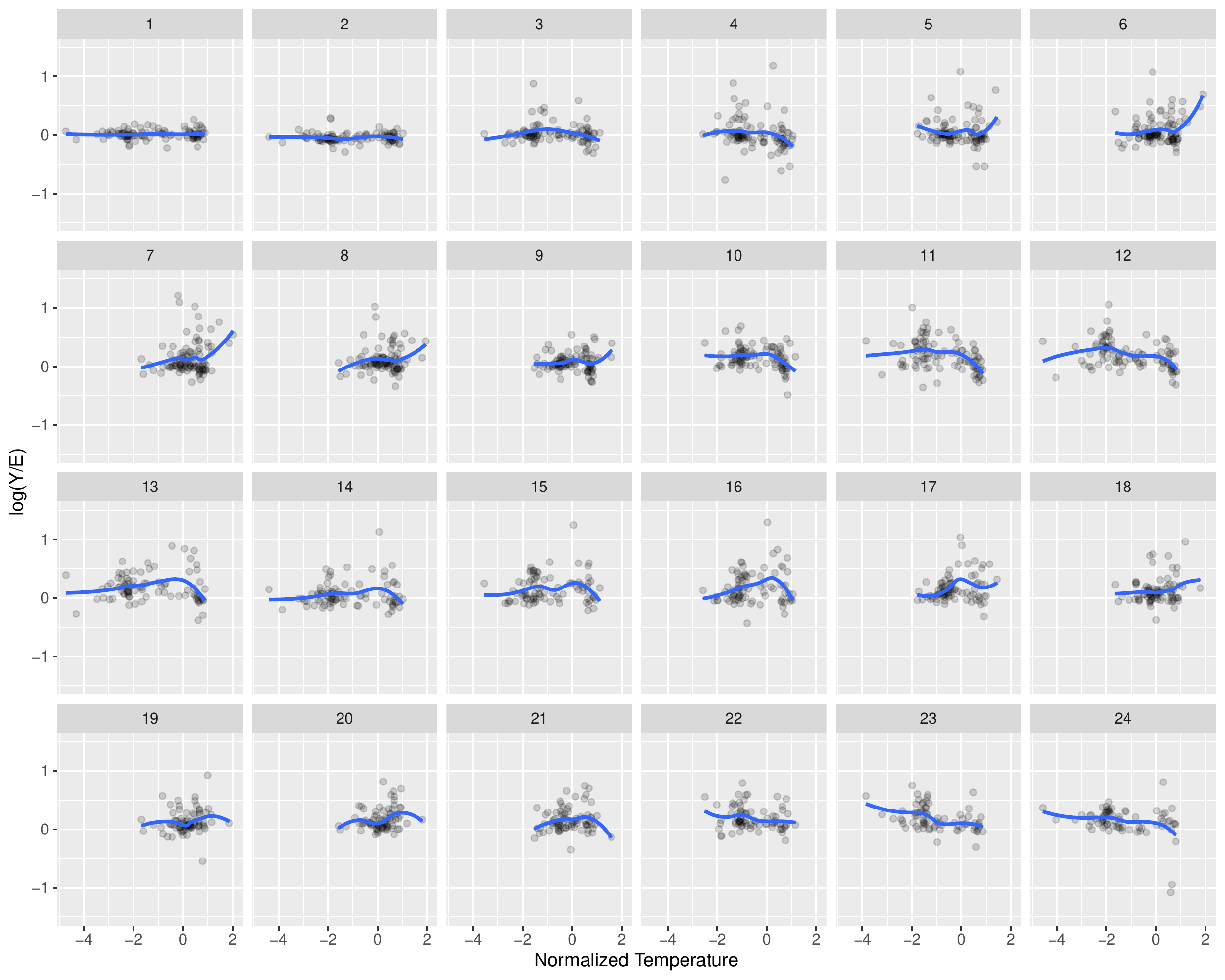}
\caption{Association between $\log(Y/E)$ by month and temperature, with smoothers.}
\label{fig:covariate_plot_temperature_1}
\end{figure}

\begin{figure}[htbp]
\centering
\includegraphics[scale=.4]{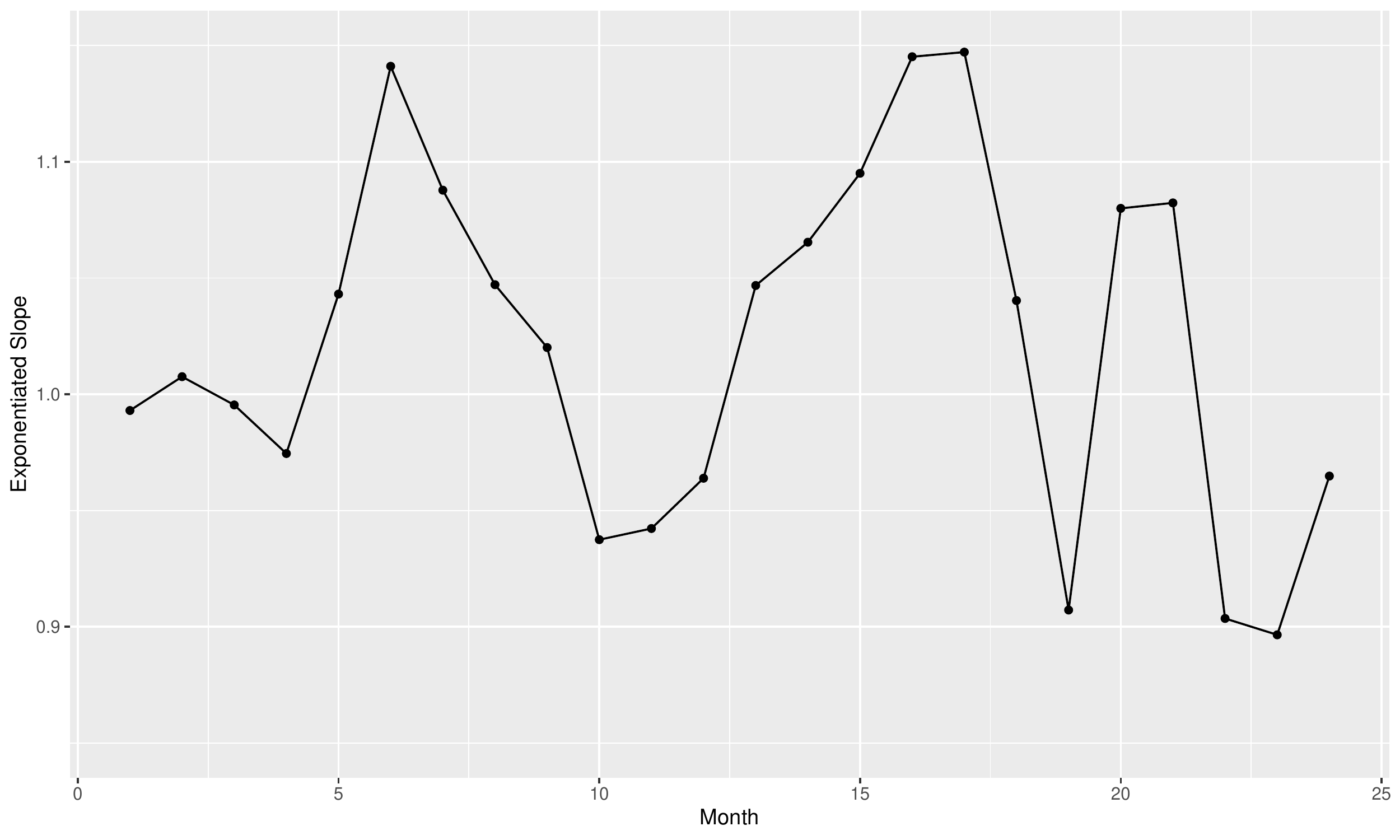}
\caption{Estimates of relative rate parameters  associated with temperature, by month.}
\label{fig:covariate_plot_temperature_3}
\end{figure}

\begin{figure}[htbp]
\centering
\includegraphics[scale=.55]{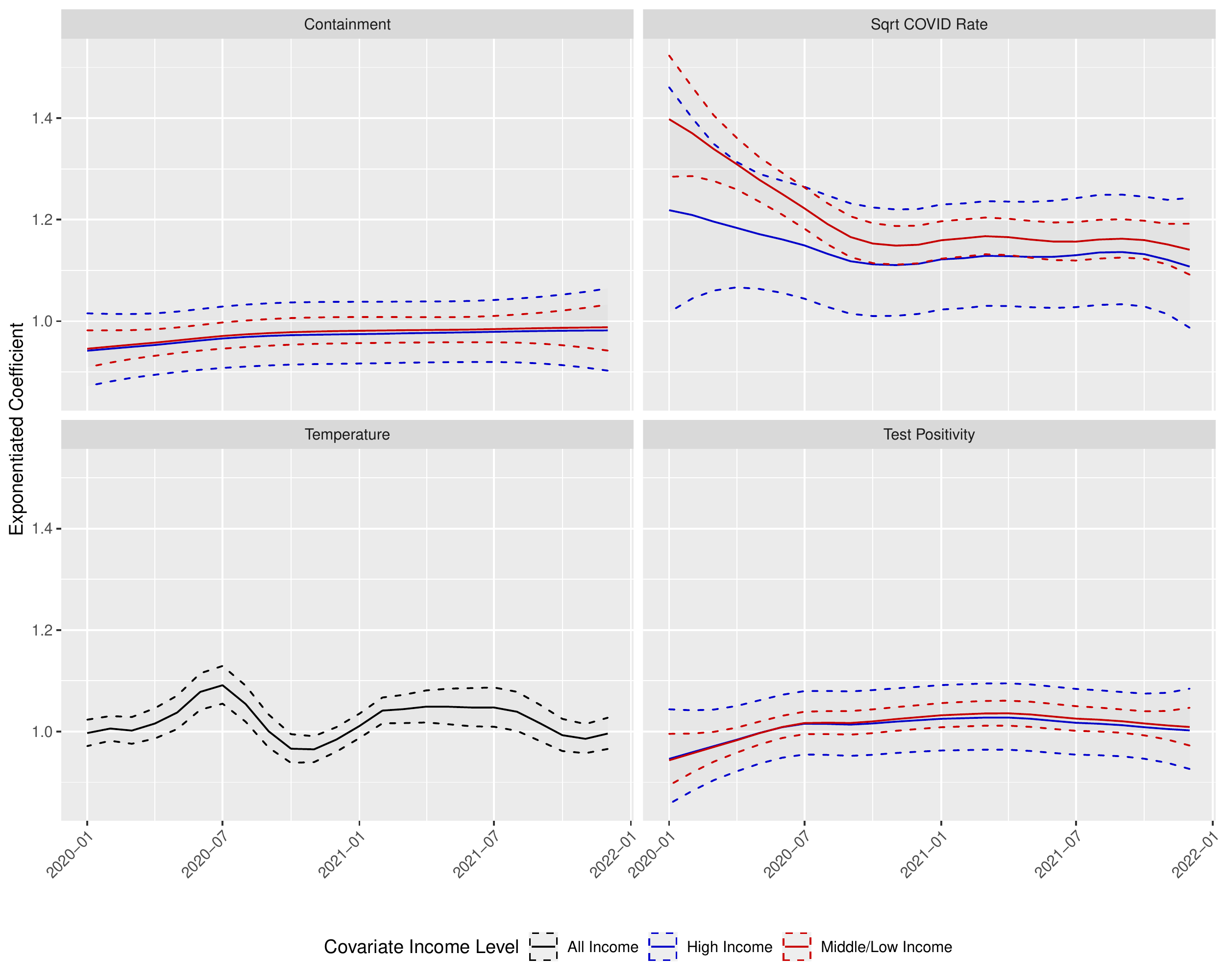}
\caption{Time-varying associations versus month, with 95\% interval estimates. We plot the multiplicative change in the relative rate of the excess associated with a 1-unit increase in the relevant covariate, against month.}
\label{fig:TimeVarying2}
\end{figure}

We first look at empirical univariate associations. We model the temperature association as time-varying and so in Figure \ref{fig:covariate_plot_temperature_1} we plot $\log(Y_{c,t}/E_{c,t})$ versus temperature $x_{c,t}$ (for countries with observed data) and for each month $t=1,\dots,24$.  Starting with the simpler model $Y_{c,t} | \theta_{c,t} \sim \mbox{Poisson}(E_{c,t} \theta_{c,t})$ we derive the variance  for $\log(Y_{c,t}/E_{c,t})$ as proportional to $Y_{c,t}^2/E_{c,t}$ and we take the reciprocal of this in a weighted least squares analysis. In Figure \ref{fig:covariate_plot_temperature_3}, we plot the exponentiated slopes from simple linear models fitted to each month, to give an indication of the associations. We see a clear time-varying association. The pattern is consistent with the bottom left panel of Figure \ref{fig:TimeVarying2}, where the exponentiated fit from the overdispersed Poisson model is shown, albeit smoothed which is induced by the RW2 prior on the log relative rates associated with temperature.

\begin{figure}[htbp]
\centering
\includegraphics[scale=.5]{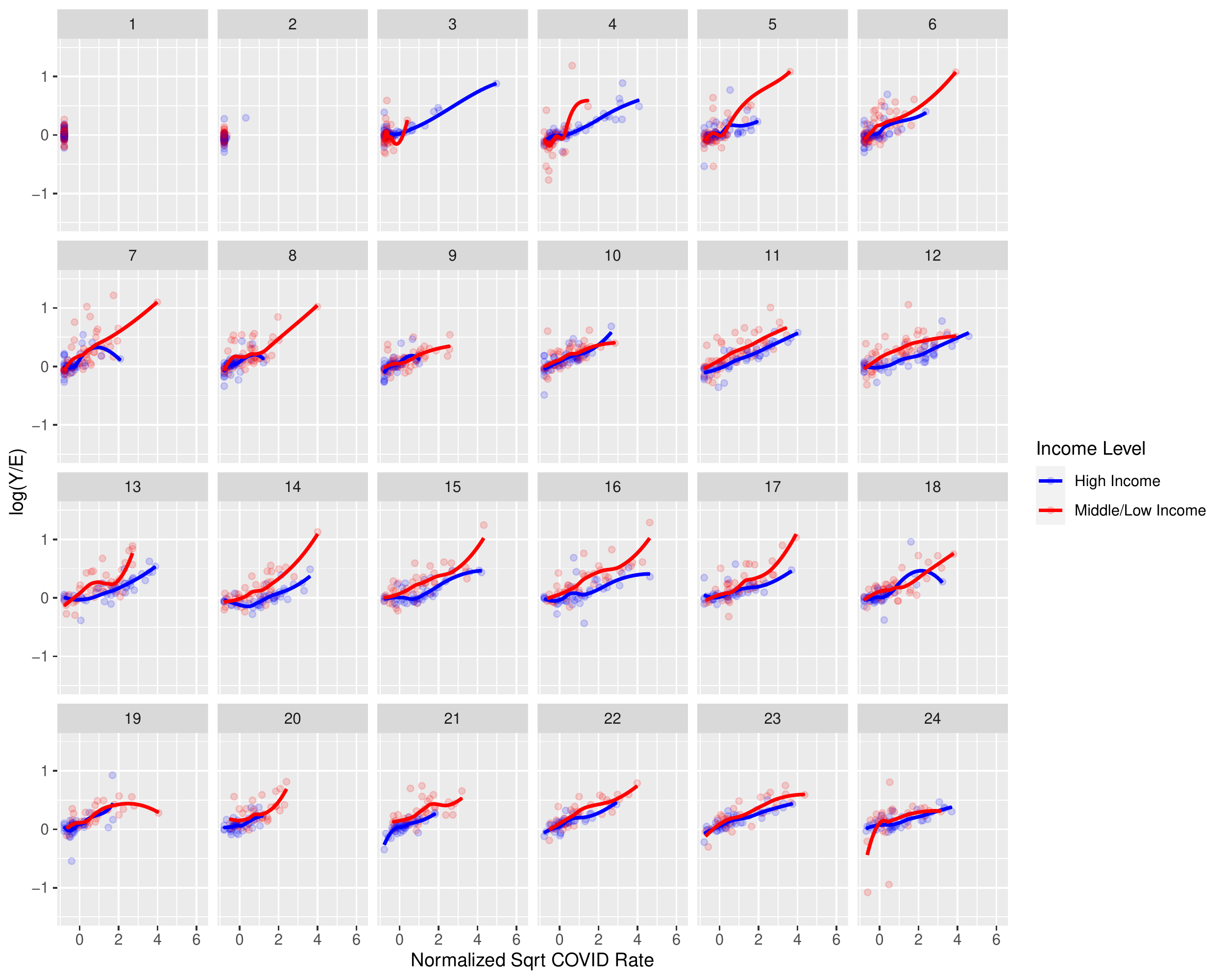}
\caption{Association between $\log(Y/E)$ by month and sqrt COVID-19 death rate, with smoothers.}
\label{fig:covariate_plot_sqrt_covidr_1}
\end{figure}

\begin{figure}[htbp]
\centering
\includegraphics[scale=.4]{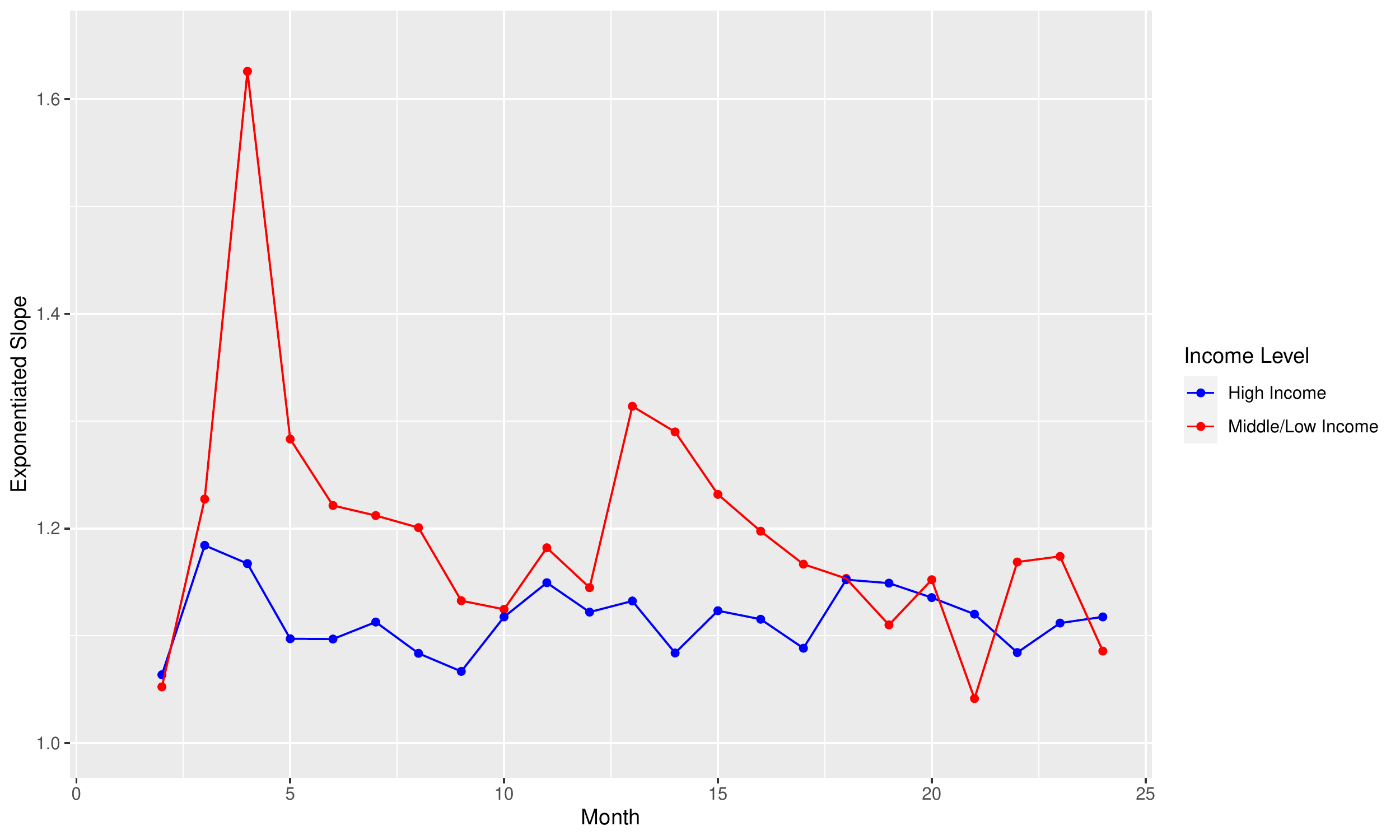}
\caption{Estimates of relative rate parameters  associated with sqrt COVID-19 death rate, by month.}
\label{fig:covariate_plot_sqrt_covidr_3}
\end{figure}

In Figures \ref{fig:covariate_plot_sqrt_covidr_1} and \ref{fig:covariate_plot_sqrt_covidr_3} we see the time-varying associations with the sqrt COVID-19 death rate, for high income and low/middle income countries. The associations are strong and the smoothed versions from the full covariate model can be seen in the top right panel of Figure \ref{fig:TimeVarying2}.

\begin{figure}[htbp]
\centering
\includegraphics[scale=.5]{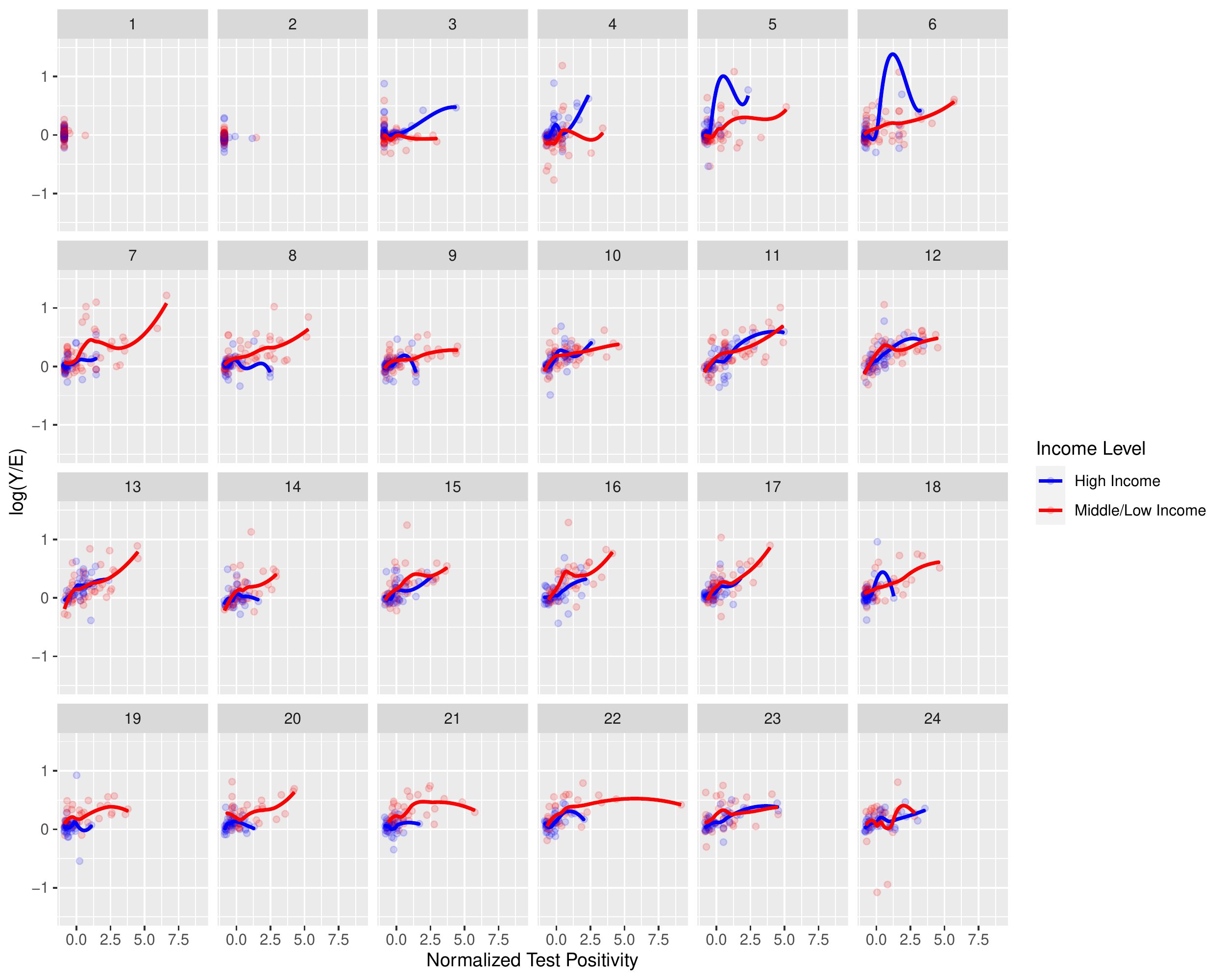}
\caption{Association between $\log(Y/E)$ by month and positive COVID-19 test rate, with smoothers.}
\label{fig:covariate_plot_positive_rate_1}
\end{figure}

\begin{figure}[htbp]
\centering
\includegraphics[scale=.4]{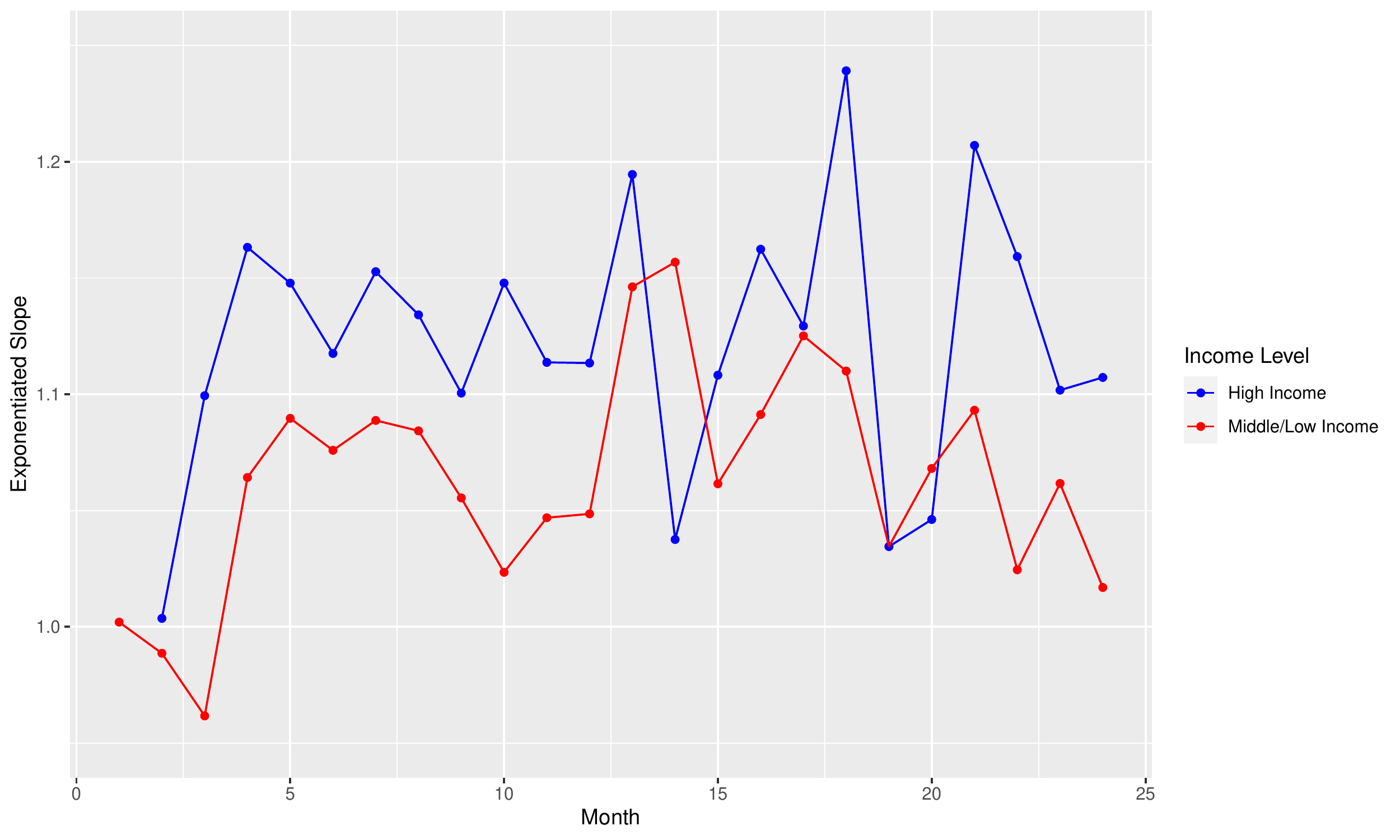}
\caption{Estimates of relative rate parameters  associated with positive COVID-19 test rate, by month.}
\label{fig:covariate_plot_positive_rate_3}
\end{figure}

%and we see a highly variable relationship. 
Figures \ref{fig:covariate_plot_positive_rate_1} and \ref{fig:covariate_plot_positive_rate_3} show the COVID-19 positive test associations, and we see similar and positive though weak associations in both income groups. 

\begin{figure}[htbp]
\centering
\includegraphics[scale=.5]{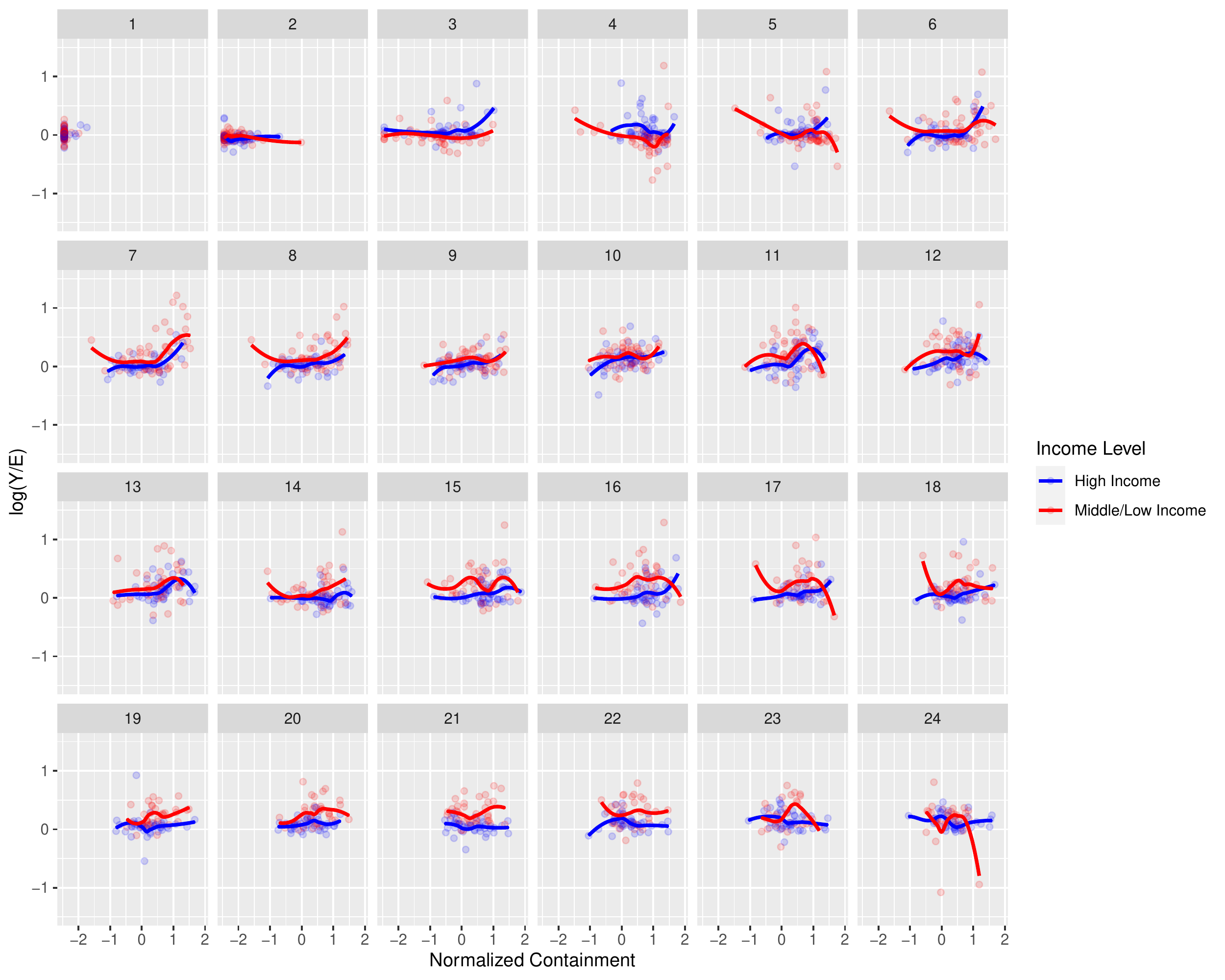}
\caption{Association between $\log(Y/E)$ by month and containment, with smoothers.}
\label{fig:covariate_plot_containment_1}
\end{figure}

\begin{figure}[htbp]
\centering
\includegraphics[scale=.4]{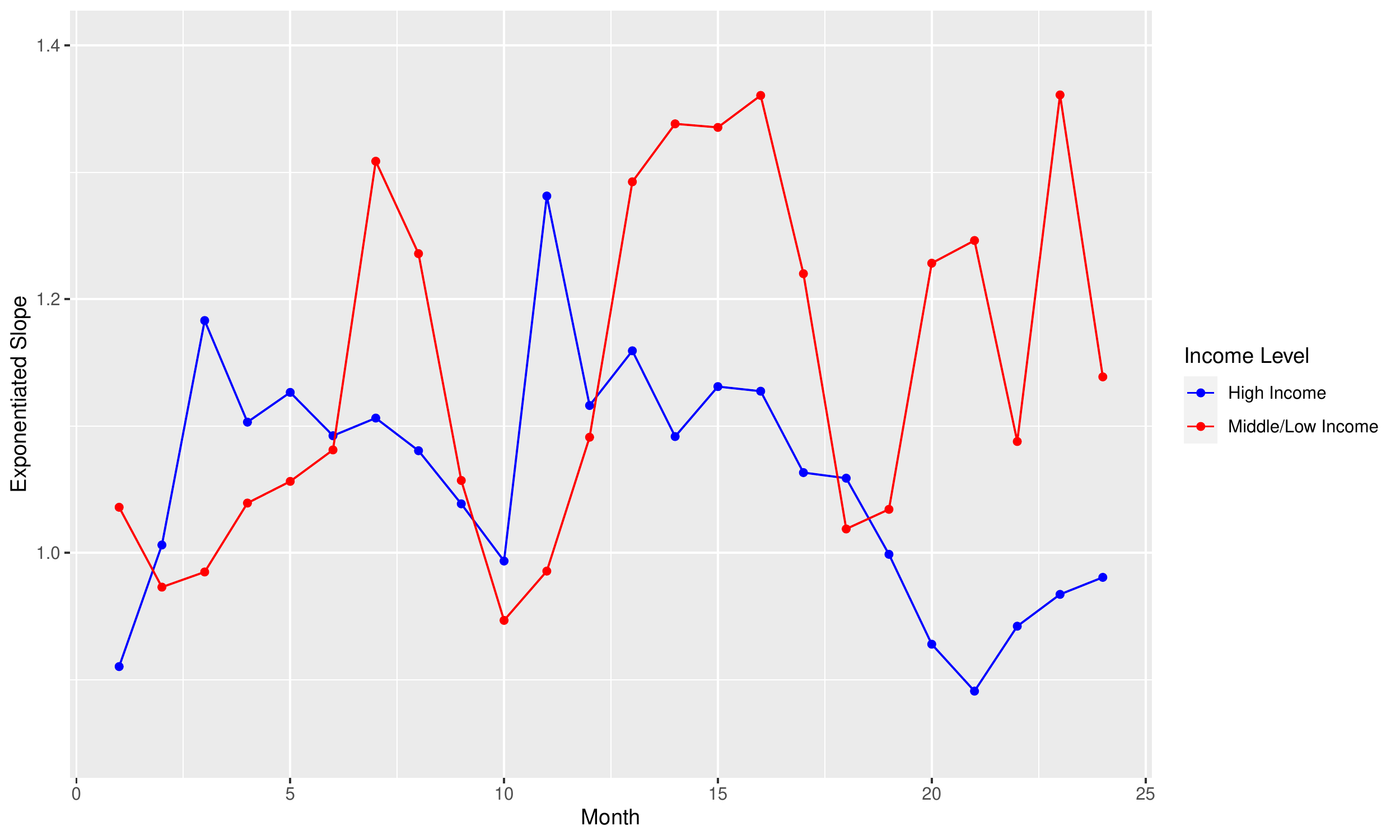}
\caption{Estimates of relative rate parameters  associated with containment, by month.}
\label{fig:covariate_plot_containment_3}
\end{figure}

Figures \ref{fig:covariate_plot_containment_1} and \ref{fig:covariate_plot_containment_3} show the containment associations, and we see again similar and negative relationships in both income groups. 

\begin{figure}[htbp]
\centering
\includegraphics[scale=.4]{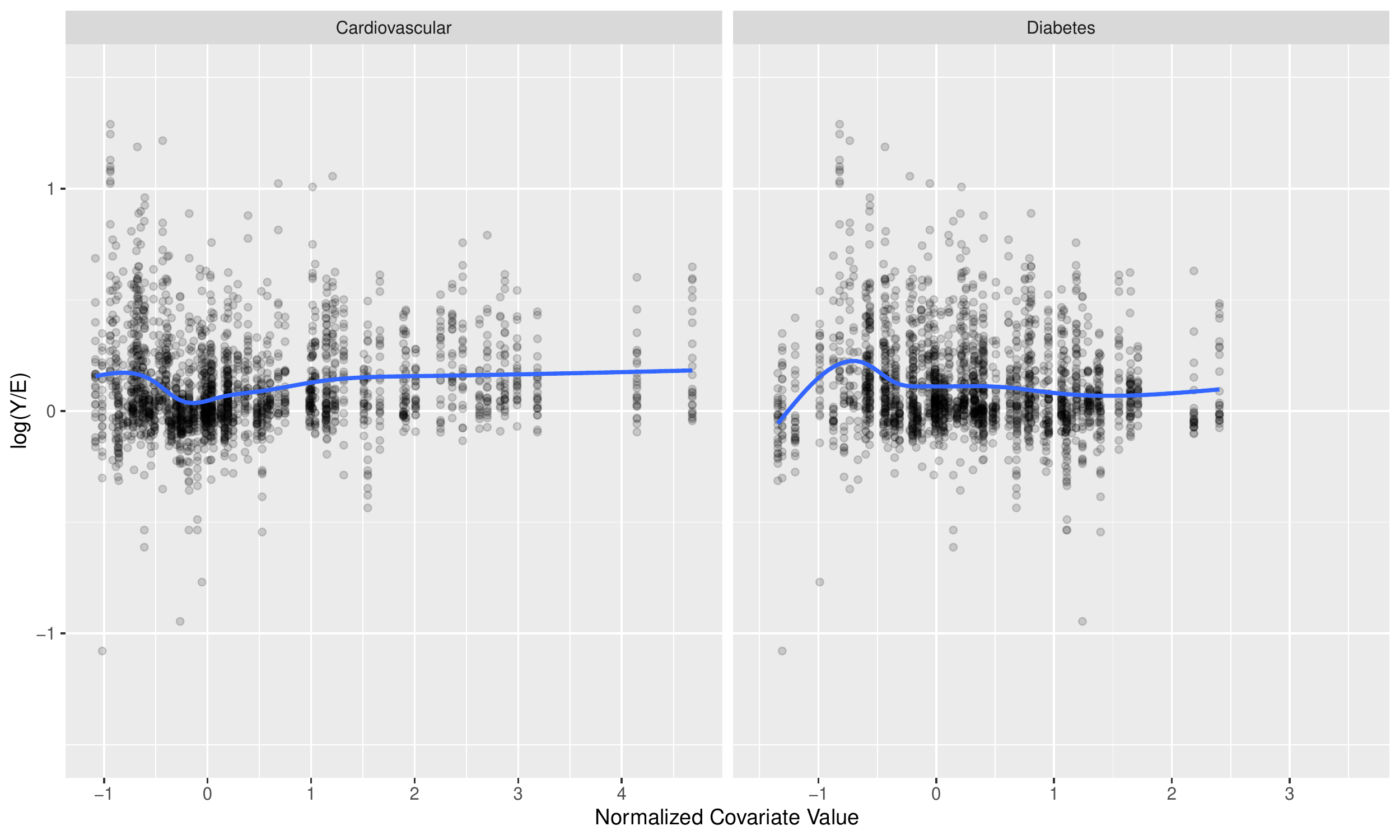}
\caption{Estimates of relative rate parameters associated with the time constant covariates, cardiovascular and diabetes rates from 2019.}
\label{fig:covariate_plot_fixed}
\end{figure}

\begin{figure}[htbp]
\centering
\includegraphics[scale=.6]{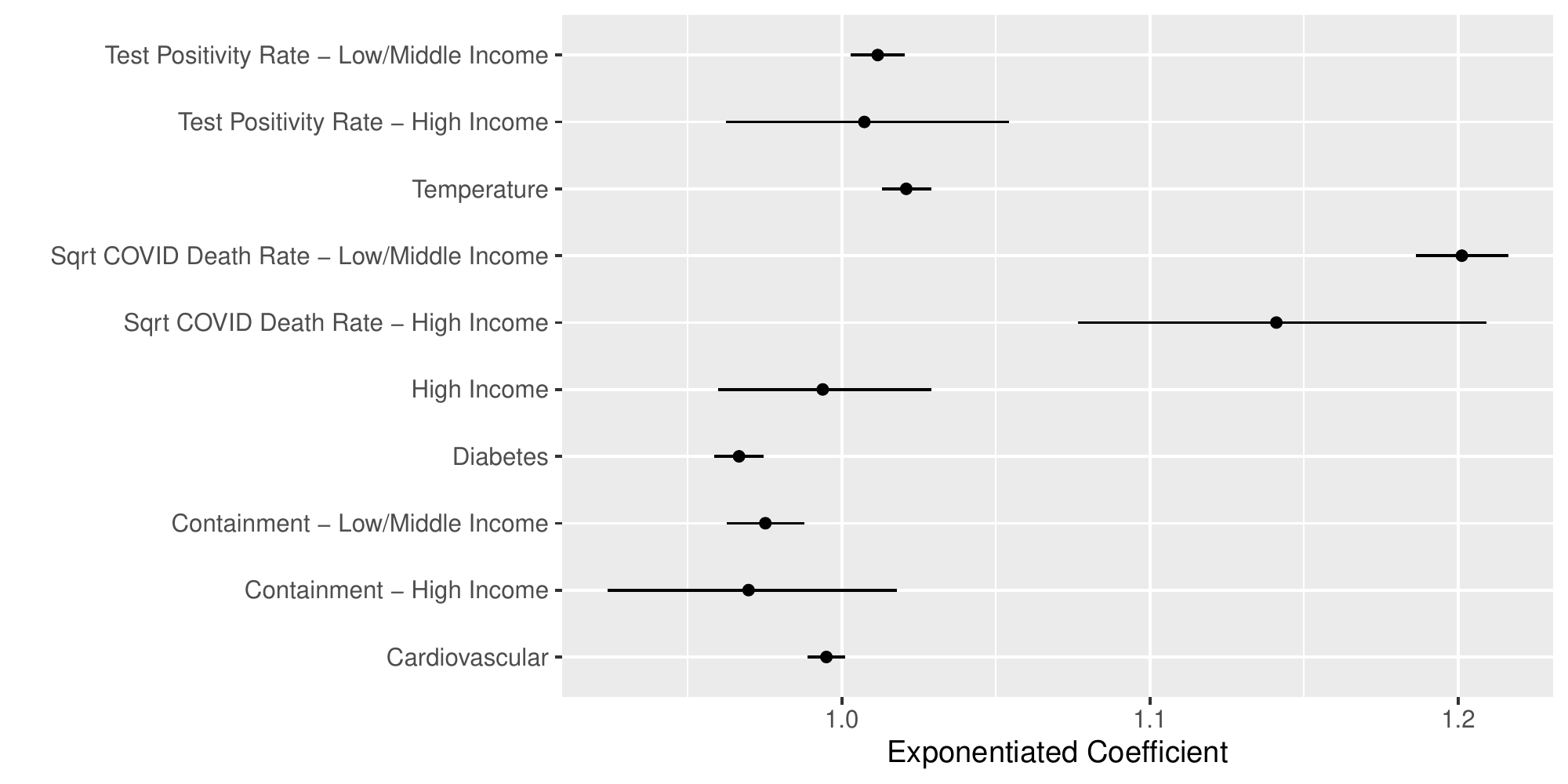}
\caption{Fixed association coefficients, with 95\% interval estimates. We plot the multiplicative change in the relative rate of the excess associated with a 1-unit increase in the relevant covariate.}
\label{fig:Fixed}
\end{figure}

\begin{figure}[htbp]
\centering
\includegraphics[scale=.6]{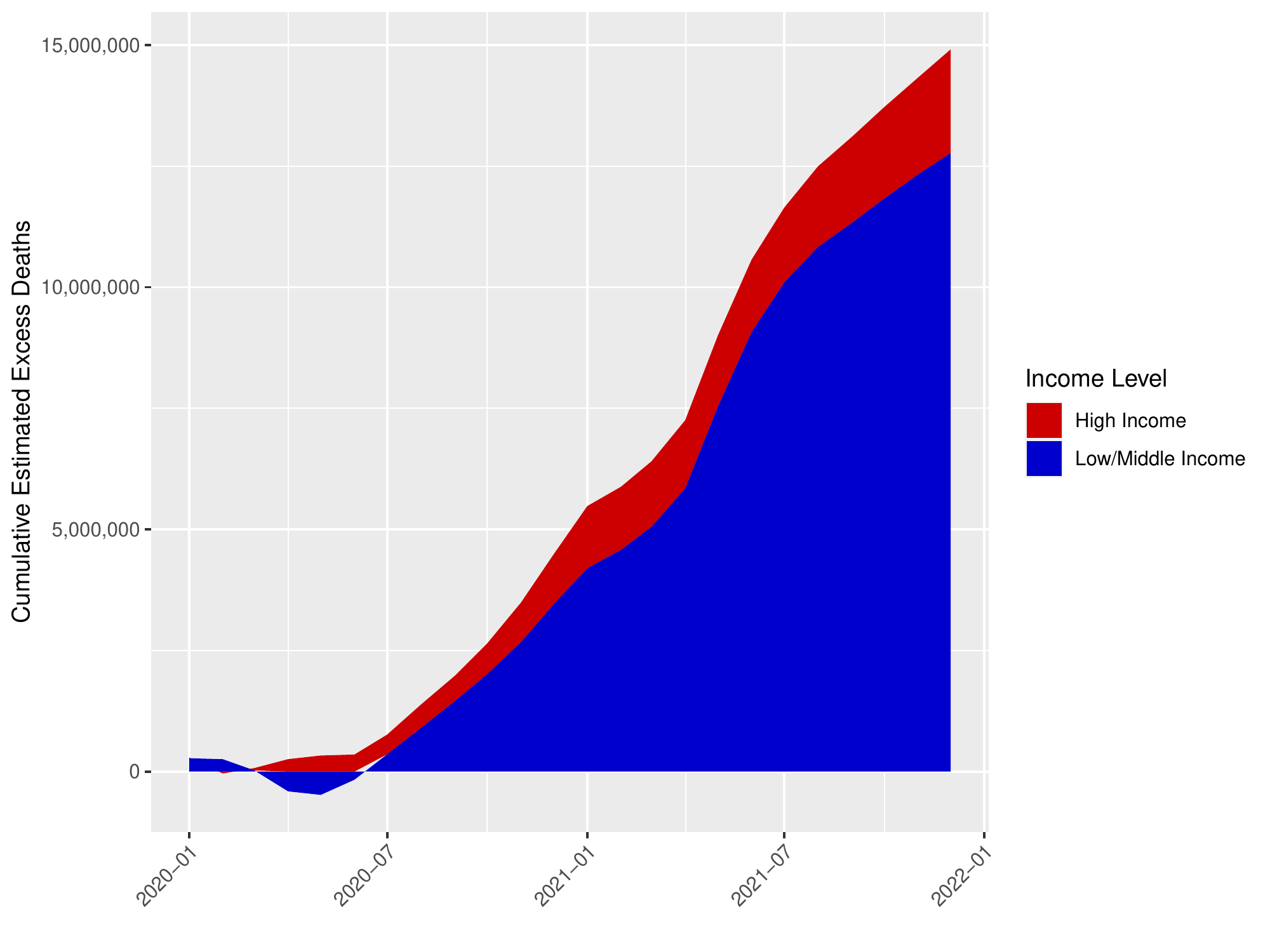}
\caption{Cumulative excess deaths over 2020--2021, by high and low/medium income.}
\label{fig:CumulativeExcessDeaths2}
\end{figure}

Finally, the two constant covariate associations are given in Figure \ref{fig:covariate_plot_fixed}. A high historic diabetes rate is associated with a lower rate. Figure \ref{fig:Fixed} summarizes the fixed effects. In general, for the variables with interactions, the associations are stronger and the intervals narrower, for low/middle income countries, where the pandemic produced more excess deaths (see Figure \ref{fig:CumulativeExcessDeaths2}).

 In Figure \ref{fig:Fixed} we summaries the fixed effect parameters -- we parameterize the model so that the time-varying associations have an overall (average) association and then variation around this with a RW2 (with a sum-to-zero constraint). The strongest association is with the sqrt COVID-19 death rate in low- and middle-income countries, in the expected direction.
The test positivity rate in  low- and middle-income countries is also positivity, as is the temperature association. The sqrt COVID-19 death rate in high-income countries is in the opposite direction to that expected. Containment in low- and middle-income countries has a negative association, as does the historic diabetes rate.

The fitting of the covariate model to the monthly ACM data resulted in the association parameters in Figure \ref{fig:TimeVarying2}. We see that the time-varying aspect seems merited in all cases, apart perhaps from test positivity which is relatively constant, and close to 1. We could have removed this covariate from the model, but we wanted to minimize model changes. The containment and test positivity follow qualitatively similar shape; initially the association is lower so that higher containment and test positivity rates are associated with a relatively reduced mortality rate, relative to that expected. And then, from July 2020, the associations are relatively constant.

We did consider models using lagged covariates, but the fits were similar to the ones shown here. We will continue to explore this issue in future extensions. In general, in future work we would like to investigate flexible yet principled model-building strategies, with an enhanced set of country-level variables.

\begin{figure}[htbp]
\centering
\includegraphics[scale=.5]{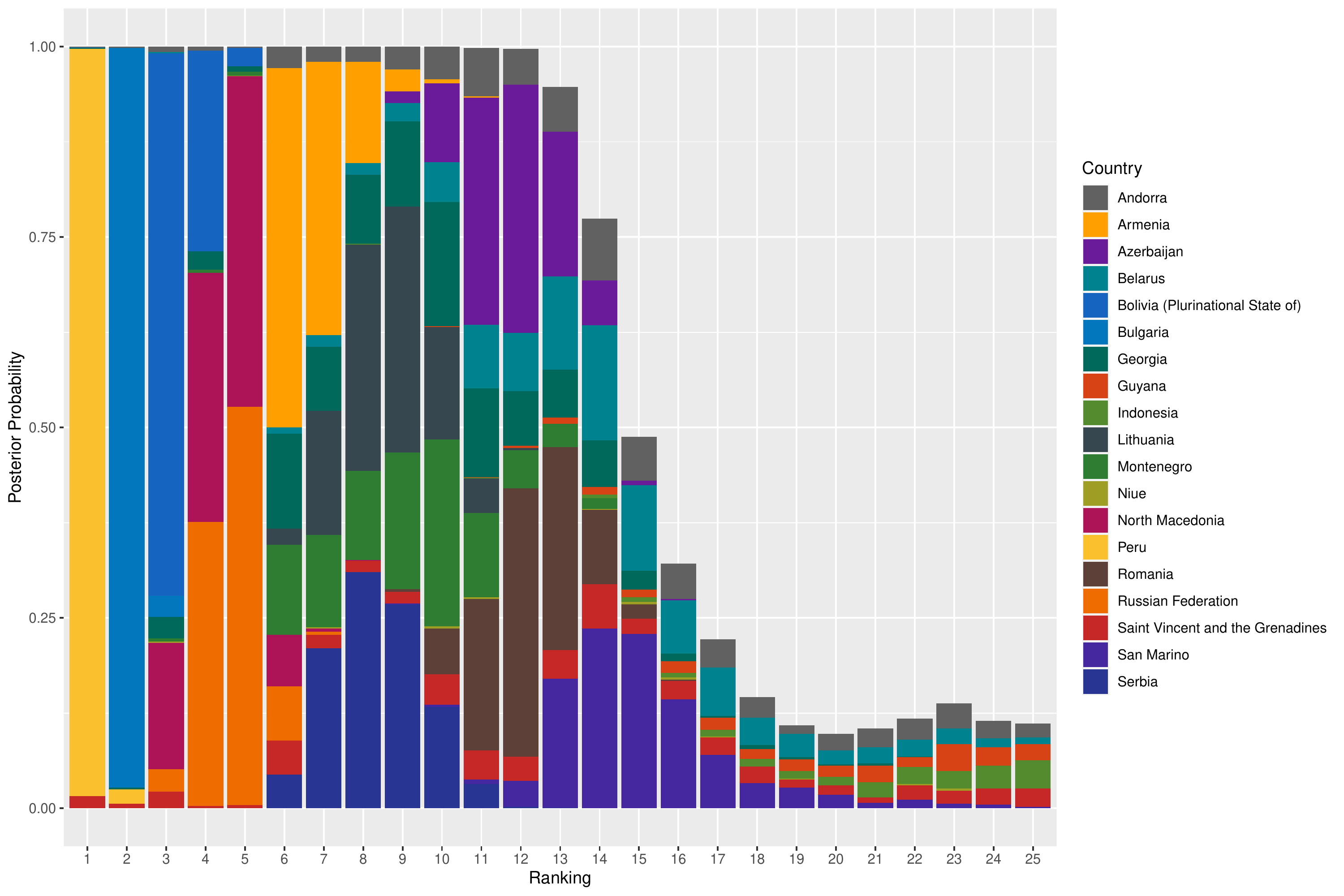}
\caption{Posterior probability of particular countries having high ranks for the greatest excess mortality rate. }
\label{fig:RankingPlot}
\end{figure}

\clearpage
Using samples from the posterior (available from INLA), we can obtain rankings of the countries, in terms of the excess mortality rate. We show these rankings in Figure \ref{fig:RankingPlot}. The posterior probability that Peru has the highest excess rate is very close to 1. Similarly, the second highest rank is for Bulgaria, again with high probability. From the 3rd highest on it is less definitive but Bolivia also clearly has a high excess rate, as do the Russian Federation, North Macedonia, Armenia and Montenegro. The posterior distribution for Saint Vincent and the Grenadines is very wide, and so its probabilities are sprinkled across the rankings. After the first two ranks, there is a lot of ambiguity in the rankings, showing that it is deceptive to simply rank in terms of a single posterior summary such as the median.

\clearpage
In Figure \ref{fig:CumulativeExcessDeaths2}, we show the cumulative excess deaths, by income status. The majority of the excess deaths occur in low/middle income countries. 
%The top plot shows that the mortality in low/middle income countries is relatively.

\begin{figure}[htbp]
\centering
\includegraphics[scale=.6]{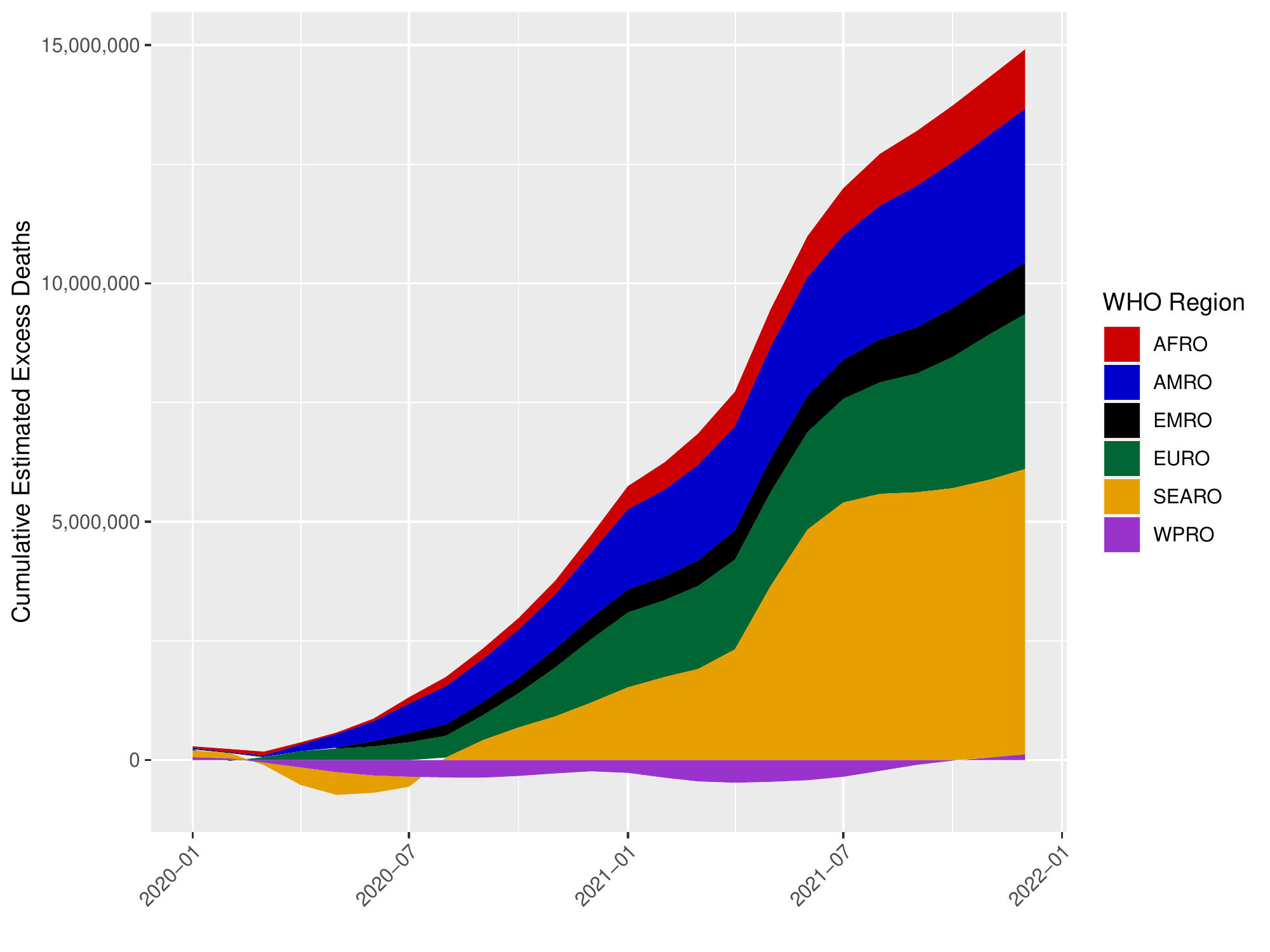}
\caption{Cumulative excess deaths over 2020--2021, by region.}
\label{fig:CumulativeExcessDeaths}
\end{figure}

In Figure \ref{fig:CumulativeExcessDeaths} we plot cumulative excess deaths over 2020--2021, by region. The most startling change is how large the SEARO contribution becomes, because of the addition of India.

\clearpage
\subsection{Argentina}

We now describe each of the analyses carried out for countries with subnational data, starting with Argentina.

\begin{figure}[htbp]
\centering
\includegraphics[scale=.5]{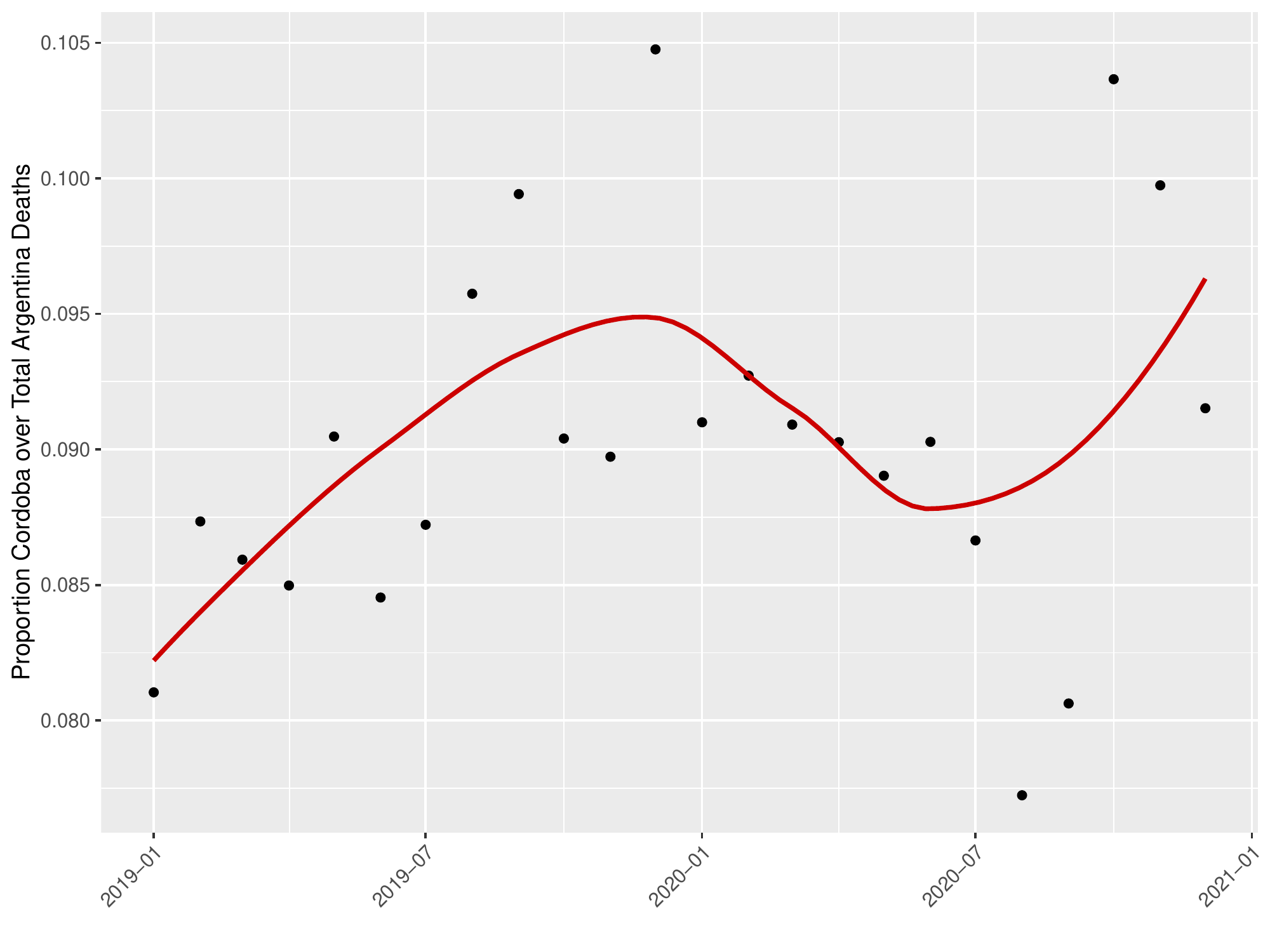}
%\hspace{-.8in}
%\includegraphics[height = 6.3cm]{ClusterLocs_Admin2.pdf}
\caption{Fractions of the total deaths in Cordoba, for 2019 and 2020, with smoother. }
\label{fig:PbyMonth_Argentina}
\end{figure}

We first describe the available data. For 2019 and 2020, we have total national deaths and deaths in the province of Cordoba, both by month. For  2021, there are subnational data for Cordoba.
In Figure \ref{fig:PbyMonth_Argentina} we plot the fraction of Cardoba deaths to national deaths in 2019 and 2020. 
There is some variation, and perhaps a small uptick in the fraction toward the end of 2020,  but overall the fraction is quite constant, and there is not a drastic change in the first year of the pandemic, as compared to the previous year. 

\begin{figure}[htbp]
\centering
\includegraphics[scale=.55]{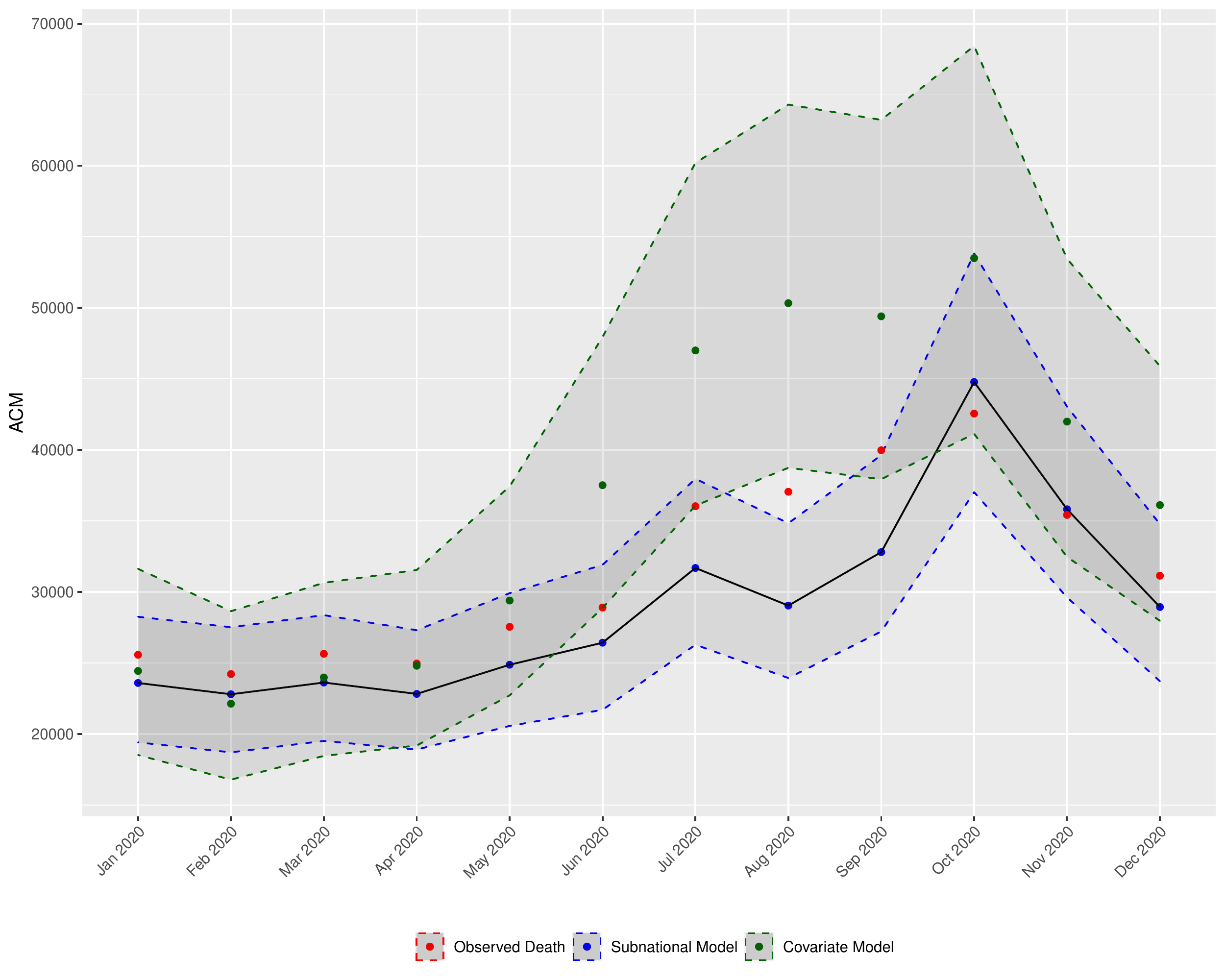}
%\hspace{-.8in}
%\includegraphics[height = 6.3cm]{ClusterLocs_Admin2.pdf}
\caption{Argentina national data and predictions. %Observed deaths for in 2020, and 
Predicted deaths from the subnational model, and from the global negative binomial covariate model for 2021, both with 95\% credible intervals }
\label{fig:Argentina_PredvINLA}
\end{figure}

In Figure \ref{fig:Argentina_PredvINLA} we plot the predictions for 2021 from the subnational binomial model and from the predictive covariate model which is estimated from the countries with national monthly data.  There are some differences between the results, with the covariate model giving wider intervals and higher estimates from April 2020. In the reported results we use the subnational estimates.

\clearpage
\subsection{China}

For China, we have two main data sources. First, the annual national ACM from 2015 to 2021, as reported by the National Bureau of Statistics of China in the 2021 statistical yearbook \citep{china_yearbook2021} and a January 2022 press release \citep{china_nso_pr2022}. Second, we may potentially utilize data on the daily number of deaths registered by the nationally representative Disease Surveillance Points (DSP) system in January--August 2019 and January--August  2020, as reported by \cite{qi_2022}. The DSP registered about 18\% of total deaths in China in 2019 \citep{china_comp_2020} and we  aggregate the DSP data to monthly national counts. 
The data collection sites are constant over time so that the fraction of the population covered is constant also.  To use the subnational model described in Section 5.1 of the main paper we impute {\it monthly} national ACM counts, based on a 12-cell multinomial, conditioned on the {\it annual} national ACM, The monthly probabilities are modeled as a log-linear function of temperature, as we did in the modeling of the expected numbers (described in Section 3 of the main paper). 

To make inference for this model we needed to write our own Markov chain Monte Carlo (MCMC) algorithm (the model is not of the form allowed in INLA, and it requires the generation of discrete counts, which cannot be done in {\tt Stan}).
We denote the apportioned counts in 2019 as the $12 \times 1$ vector $\bY^{2019}$, and treat as known (so that we ignore the fact that we imputed these counts). 
 We let $\bz^{2019}$ and $\bz^{2020}$  represent the monthly DSP counts in years 2019 and 2020, and $p_t$ be the fraction of deaths from the DSP in month $t$. We model $p_t$ as,
logit$(p_t ) = \alpha+\epsilon_t$, with $\epsilon_t \sim \N(0,\sigma_\epsilon^2)$ and let $\ba^{2020}$ be the collection (across months) of fixed covariate predictions based on the global covariate model.
  
\begin{figure}[htbp]
\centering
\includegraphics[scale=.55]{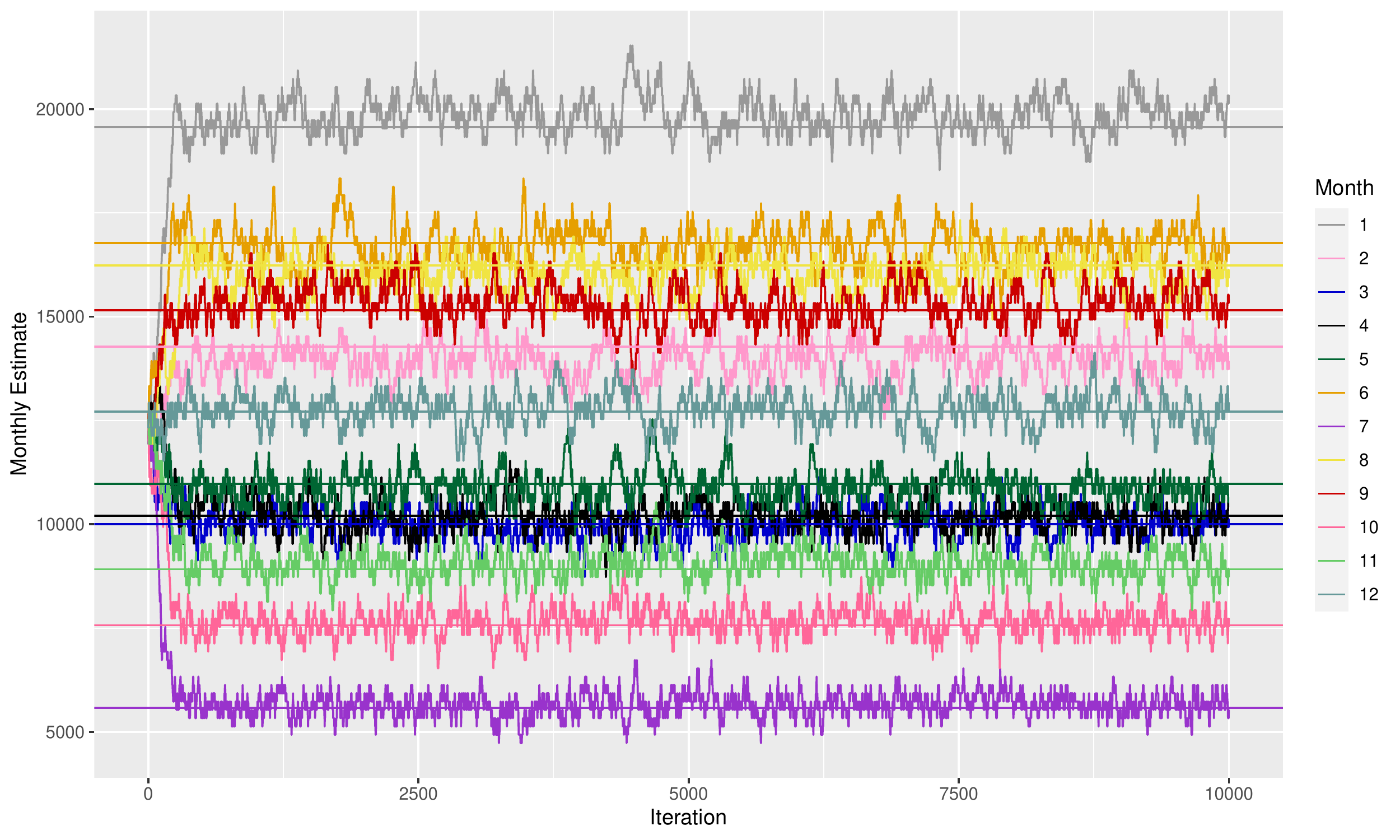}
\caption{Estimated monthly ACM across all 12 time points where the horizontal lines display the true monthly simulated ACM.}\label{fig:MCMCSimulationPlot1}
\end{figure}

\begin{figure}[htbp]
\centering
\includegraphics[scale=.4]{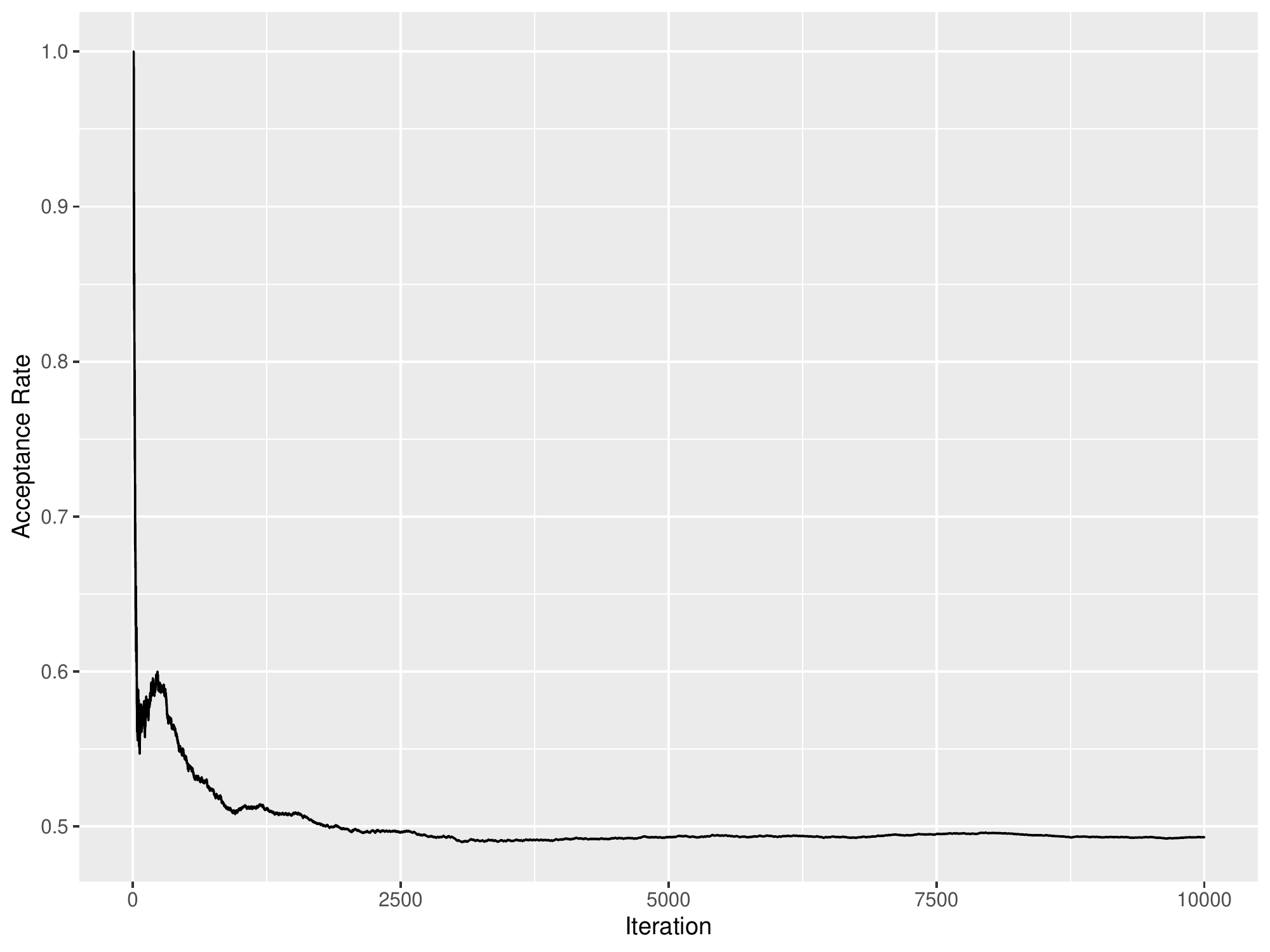}
\caption{Cumulative acceptance rate of proposed monthly ACM vector across iterations within the MCMC chain.}\label{fig:MCMCSimulationPlot2}
\end{figure}

  The unknown monthly ACM counts in 2020 are denoted $\bY^{2020}$. The posterior is,
  \begin{eqnarray*}
  p( \bY^{2020} , \alpha,\sigma_\epsilon^2 | \bz^{2019}, \bz^{2020}, \bY^{2019}, Y_+^{2020}, \ba^{2020} ) &\propto & p( \alpha,\sigma_\epsilon^2 |  \bz^{2019} , \bY^{2019}) \\&\times& p( \bY^{2020}  |\alpha,\sigma_\epsilon^2, \bz^{2019}, \bz^{2020}, Y_+^{2020}  )
  \end{eqnarray*}
  An MCMC scheme alternates between sampling from the conditionals for $p$ and for $\bY^{2020}$. The conditional for $\alpha,\sigma_\epsilon^2$ is 
  \begin{eqnarray*}
    p( \alpha,\sigma_\epsilon^2 |  \bz^{2019} , \bY^{2019} ) &\propto& p(  \bz^{2019}| \alpha,\sigma_\epsilon^2 , \bY^{2019} )\\&\sim& \prod_{t=1}^8 \mbox{Binomial}( Y_t^{2019},p_t) \times \pi(\alpha) \pi( \sigma_\epsilon^2),
    \end{eqnarray*}
    where $p_t=p_t(\alpha,\sigma_\epsilon^2) $. The conditional for $\bY^{2020}$ is more involved:
    \begin{eqnarray*}
    p( \bY^{2020} | \alpha,\sigma_\epsilon^2, \bz^{2019}, \bz^{2020}, Y_+^{2020}, \ba^{2020})  \propto \underbrace{p ( \bz^{2020} | \bY^{2020} , \alpha,\sigma_\epsilon^2 )}_\text{Product of Binomials}  \times \underbrace{p( \bY^{2020} | Y_+^{2020}, \ba^{2020} )}_\text{Multinomial}
    \end{eqnarray*}
For the proposal for the vector $\bY^{2020}$ we need to ensure that the total $Y_+^{2020}$ is respected. The MCMC algorithm we use, is in the spirit of algorithms described in \cite{wakefield:etal:11}, and is given by:
\begin{enumerate}
\item Draw $K$ from a discrete distribution on $1,2,\dots,6$. 
%I would suggest a distribution that gives larger probability towards 1 and smaller probability towards (Number of counts you have)/2. You can use this distribution to calibrate your jumps so that they are not too large (hence more likely to reject your move) and not too small (hence more likely to accept your move but advance less).
\item Draw $K$ counts. Decrease these counts by some fixed discrete number $J$.
\item Draw $K$ of the remaining counts. Increase these counts by $J$.
\item Together with the counts we do not sample, we therefore form a new set of counts with the same total.
    \end{enumerate}
    We first report a simulation study in which we validate the MCMC sampling for the prediction of monthly country ACM from reported annual ACM and monthly subnational ACM.  We conceived the following setup where we simulate,
$$Y_{+,t}=\textnormal{Unif}(5000,20000)$$
with $t=1,\dots,12$ as the true unobserved total monthly ACM and where we have the observed annual total $\sum_{t=1}^{12} Y_{+,t}$ and the observed subnational ACM given by, 
$$z_t=Y_{+,t}\times p_t$$
where $\mbox{logit }p_t = -1.5 + \epsilon_t$ with $\epsilon_t \sim \N(0,0.1)$. We also simulate our covariate model outcomes as $a_c = Y_{+,t} + \delta_{t}$ with $\delta_t \sim \N(0,0.05\times Y_{+,t})$ and where the error has variance fractionally larger than the percent relative error of the covariate model. Following this simulation scheme for the data, through the MCMC sampling approach outlined above and initializing using $Y_{+,t}/12$ for each cell count, we  run 10,000 iterations. We obtain monthly estimates that stabilize around the true values $Y_{+,t}$ as shown in Figure \ref{fig:MCMCSimulationPlot1}, following an acceptance rate that stabilizes just below $0.50$ as shown in Figure \ref{fig:MCMCSimulationPlot2}.

There are reasons to believe that the China DSP system may have become less reliable during the pandemic (Haidong Wang, personal communication), and so we carry out two analyses, one in which we apportion out the annual ACM counts using the DSP data and covariate prior, and the other in which we use the covariate prior only.

\begin{figure}[htbp]
\centering
\includegraphics[scale=.5]{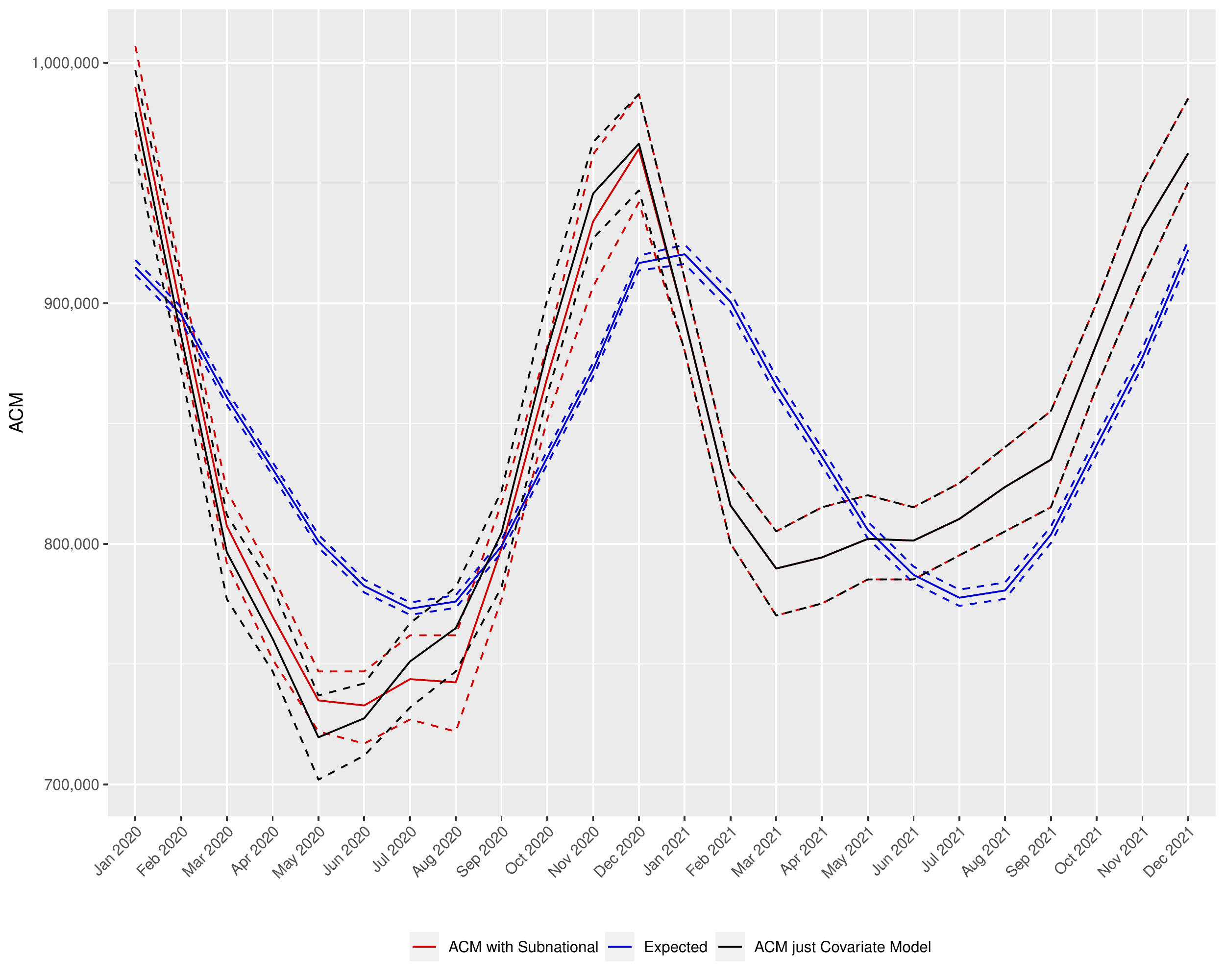}
%\hspace{-.8in}
%\includegraphics[height = 6.3cm]{ClusterLocs_Admin2.pdf}
\caption{Expected numbers for China, along with within-year predictions based on: covariate model only, covariate model and subnational data.}
\label{fig:ChinaModel_Comparisons}
\end{figure}

In Figure \ref{fig:ChinaModel_Comparisons}, we plot the expected numbers, with uncertainty, along with the predictions based on the covariate model only (2020--2021), and the covariate model plus subnational data (2020).
The 2020 estimates are very similar under the two approaches, which is reassuring. For the final reported numbers, we used the covariate model, given the aforementioned question mark over the accuracy of the DSP data.

% \begin{figure}[htbp]
%\centering
%%\includegraphics[scale=.3]{PropCordDeaths1.png}
%%\includegraphics[scale=.3]{PredDeathArgentina.png}
%\includegraphics[scale=.6]{VictoriaFigures/ChinaACMExpecteds.pdf}
%\caption{.}
%\label{fig:ChinaACMExpecteds}
%\end{figure}

\clearpage
\subsection{India}
For India, we use a variety of sources for registered number of deaths at the state and union-territory level. 
Some information was reported directly by the states through official reports and automatic vital registration:\vspace{.2in}

\noindent
Karnataka Birth and Death Registration (URL data 2022-03-05):

\noindent
\url{https://ejanma.karnataka.gov.in/frmTransaction_Details.aspx}
\vspace{.2in}

\noindent
Kerala Civil Registrations (URL date 2022-03-05):

\noindent
\url{https://cr.lsgkerala.gov.in/Pages/Map.php}
\vspace{.2in}

\noindent
Birth and Death Odisha (URL date 2022-03-05):

\noindent
\url{https://www.birthdeath.odisha.gov.in/#/home}
\vspace{.2in}

\noindent
Tamil Nadu Births and Deaths (URL date 2022-03-05):

\noindent
\url{https://www.crstn.org/birth_death_tn/MisRep.jsp}
\vspace{.2in}

%\cite{karnataka_vs,kerala_vs,odisha_vs,tn_vs} 

We also used data obtained by journalists who obtained death registration information through Right To Information requests: 
\cite{s_andhra_nodate,saikia_assam_nodate,noauthor_bihar_nodate,ramani_chhattisgarhs_2021,ramani_excess_2021,ramani_himachal_2021,ramani_excess_2021-1,s_madhya_nodate,ramani_excess_2021-2,ramani_coronavirus_2021,ramani_excess_2021-3,staff_up:_nodate,ramani_excess_2021-4,ramani_coronavirus_2021-1,noauthor_covid_2021}.  
Some of the state level data before and during the pandemic was scaled up to account for incomplete registration 
%in states such as Bihar
where vital registration captures only some of the deaths that occur, as shown in \cite{india_crs_2019}. 
For the historic national totals, the WHO takes the Civil Registration System (CRS) data provided by the Indian government; the CRS is implemented by the Office of the Registrar General of
India (ORGI) housed in the Ministry of Home Affairs,
under the Registration of Births and Deaths Act, 1969. 

\begin{figure}[htbp]
\centering
\includegraphics[scale=.55]{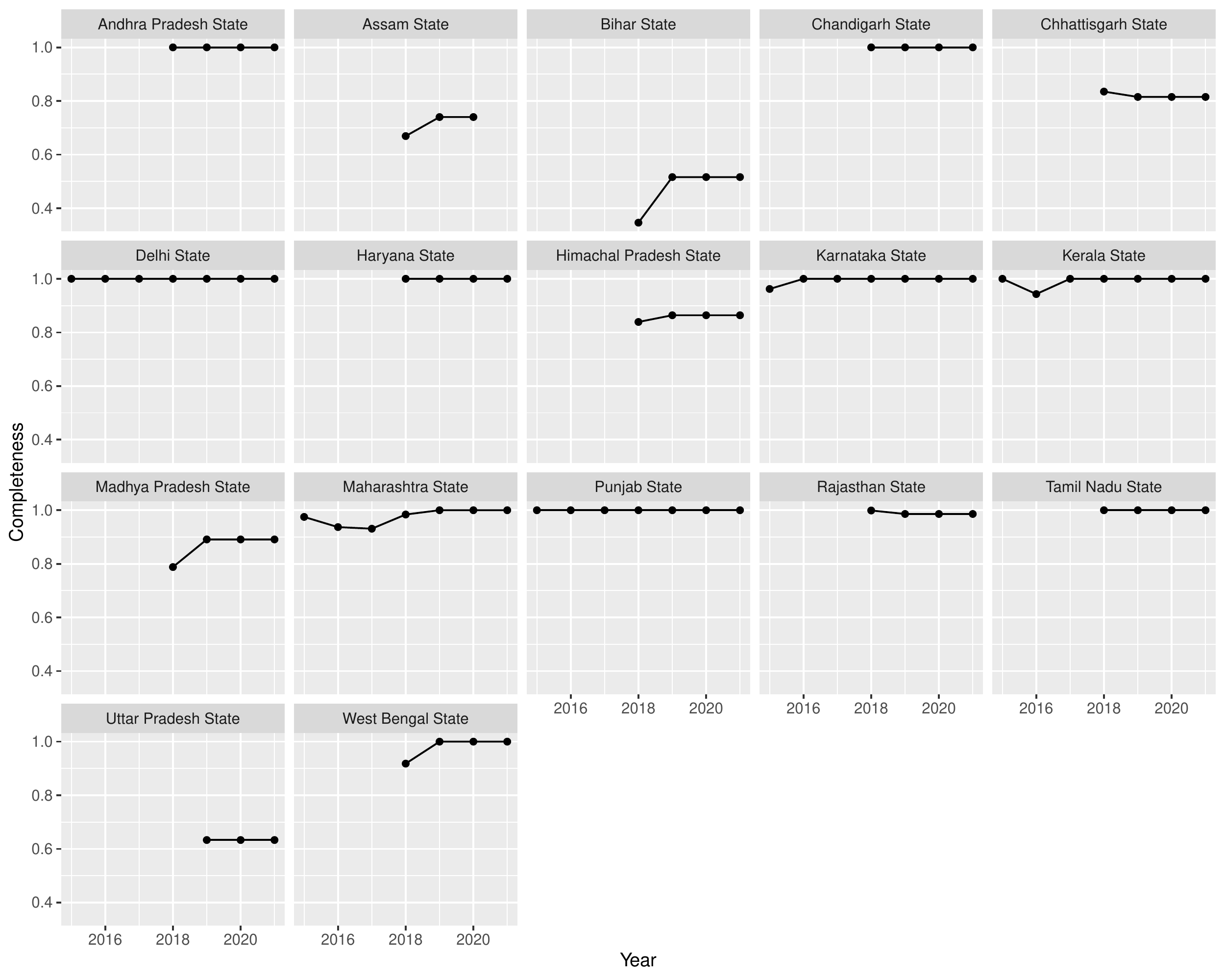}
%\hspace{-.8in}
%\includegraphics[height = 6.3cm]{ClusterLocs_Admin2.pdf}
\caption{Completeness for the 17 states for which we have subnational data during the pandemic.}
\label{fig:india_completeness}
\end{figure}

 It is known that the CRS suffers from under-reporting. \cite{rao2020civil} report that completeness of death registration from 67\% in 2011 to 79\% in 2017. The WHO adjusts the CRS data for under-reporting, using a method that uses  life tables  and data from the Indian Sample Registration System (SRS) which is a random sample of under 1\% of the national population \citep{WHO2020}. In Figure \ref{fig:india_completeness} we plot the completeness by year, along with the completeness we assume for 2020 and 2021 (we use the last completeness estimate). 
 
\begin{figure}[!htbp]
        \center{\includegraphics[scale=0.4]
        {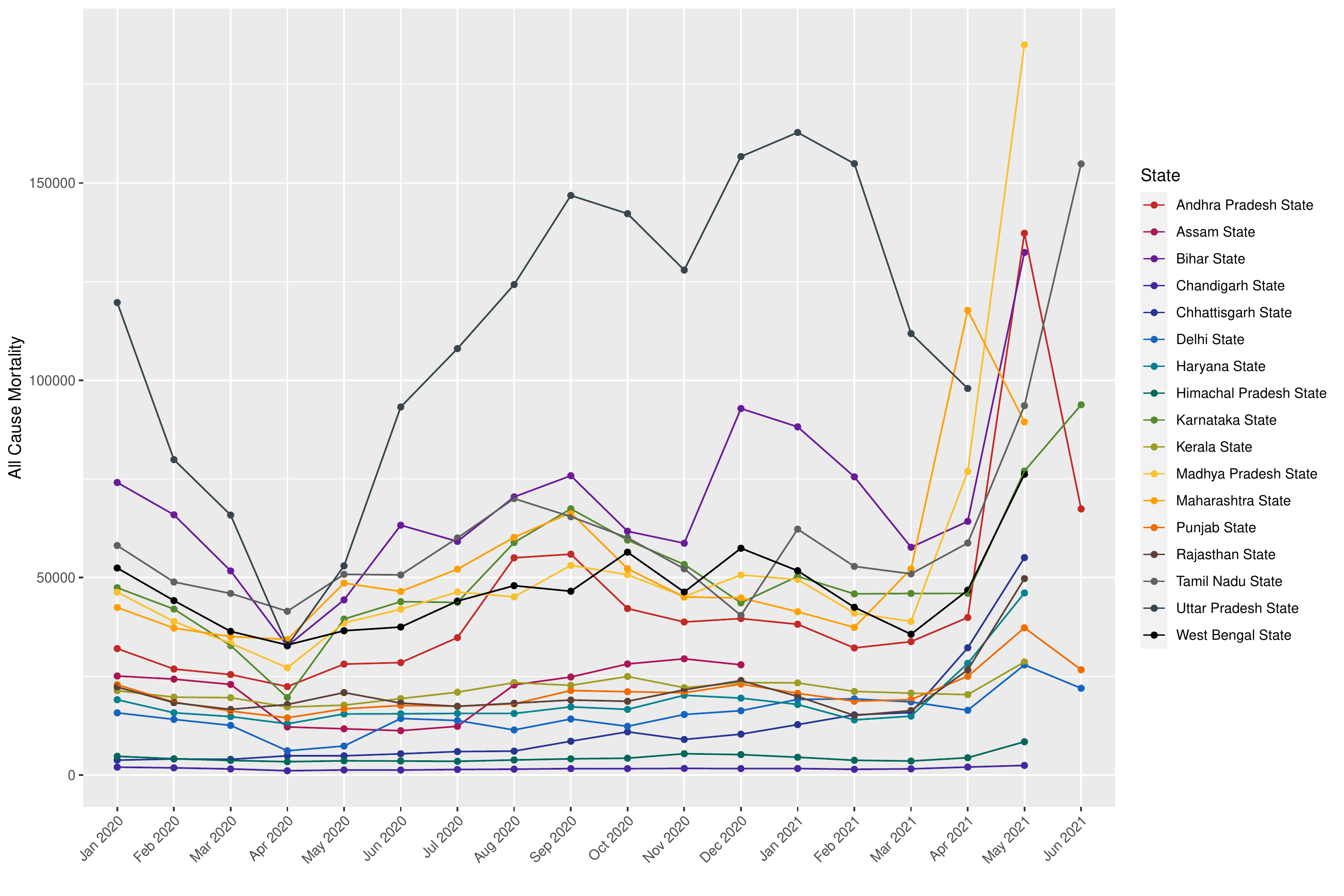}}
        \caption{Plot of reported deaths by Indian States.}\label{fig:IndiaSubnationalDeathPlot} 
\end{figure}

Figure \ref{fig:IndiaSubnationalDeathPlot}  displays the reported deaths by those states that we have data for during the pandemic.

\begin{figure}[htbp]
\centering
\includegraphics[scale=.5]{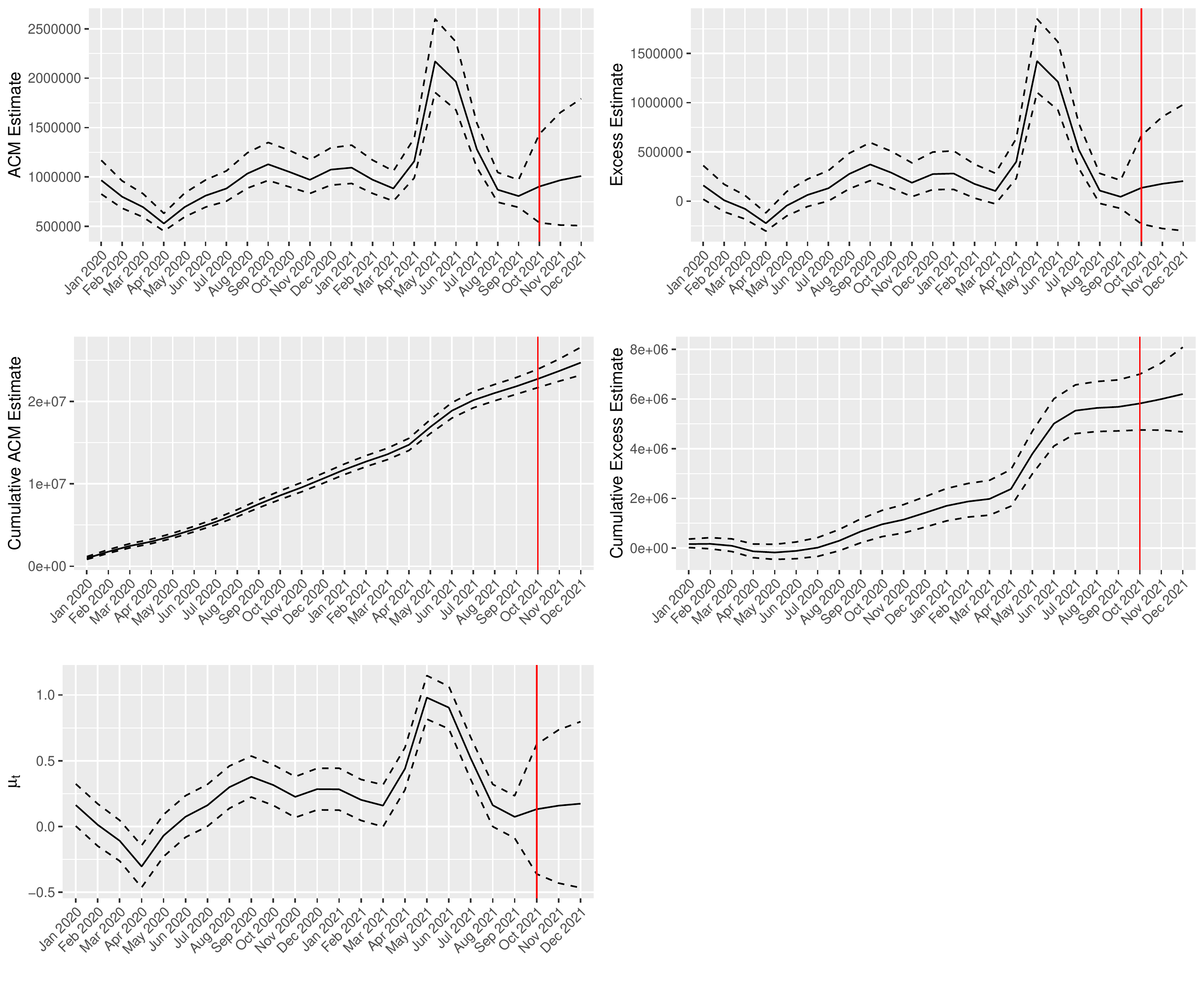}
%\hspace{-.8in}
%\includegraphics[height = 6.3cm]{ClusterLocs_Admin2.pdf}
\caption{AR1 prediction model for India. In the bottom left we plot the AR1 contribution.}
\label{fig:ar1_results}
\end{figure}

 For the final 3 months of 2021 there are data available from a single state only (Tamil Nadu) available, and for these 3 months the counts appear high (perhaps due to late registration), and so we do not use these data and instead use a simple predictive model. 
Specifically, we model $\log(Y_t/E_t)$ (using the estimated $Y_t$ for the first 21 months and weighting by the variance of the estimate) using an autoregressive order 1 (AR1) model  in INLA   (we experimented with various choices, including splines and RW1 and RW2 models), and then predict the final 3 months. 
Figure \ref{fig:ar1_results} shows various  summaries from the AR1 model. We wanted predictions for last 3 months that were relatively neutral, since we have not seen any evidence to suggest large changes in India in this period.

\begin{figure}
\centering
\includegraphics[scale=.5]{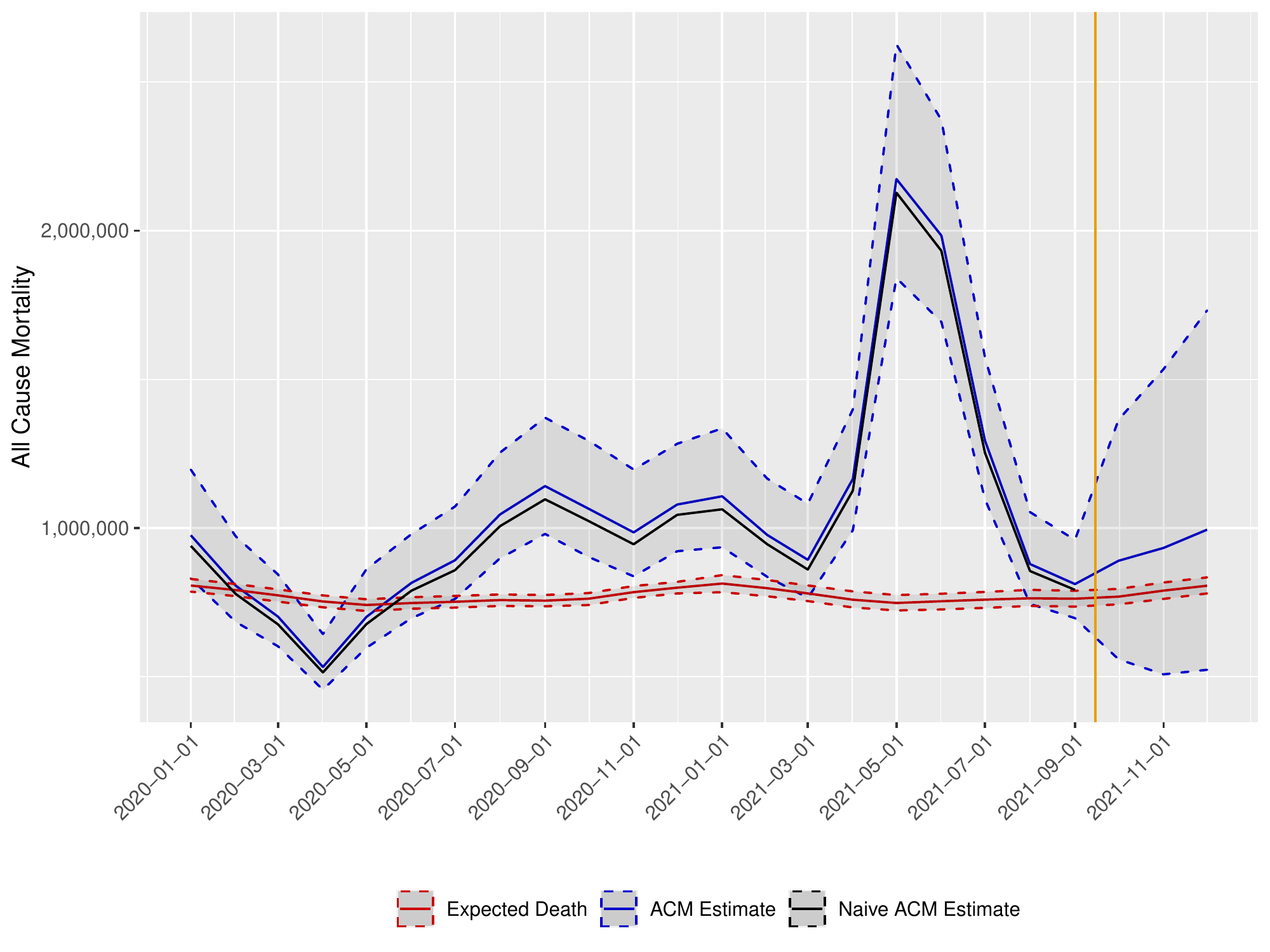}
\caption{Estimates with 95\% uncertainty for India. The final 3 months are based on an AR1 model.}
\label{fig:India_ACMPlot}
\end{figure}

Figure \ref{fig:India_ACMPlot} shows the predicted ACM for India, based on the state level data, along with the expected deaths, both with credible intervals.
	
%The accuracy of our estimates for India depends on a number of factors. First, we need the incompleteness to be corrected for accurately, which we do by basing on historic incompleteness. The estimation of the completeness is described in \cite{india_crs_2019} and uses data from the Sample Registration System (SRS). %Bihar, Madhya Pradesh and Uttar Pradhesh, have low completeness (Over 2015--2019: 0.28--0.52 for Bihar, 0.54--0.89 for Madhya Pradesh, 0.40--0.63 for Uttar Pradhesh) and more unstable fractions (see Figure 5 of the main paper). Assam also has low completeness (0.51--0.74) but the fraction of the total ACM is relatively small and quite constant over time, and so we choose to retain. 
%Second, we need the fractions of deaths within states to be approximately constant over time. It is impossible to assess the latter during the pandemic, since we do not have national data. 

\begin{figure}[htbp]
\centering
\includegraphics[scale=.4]{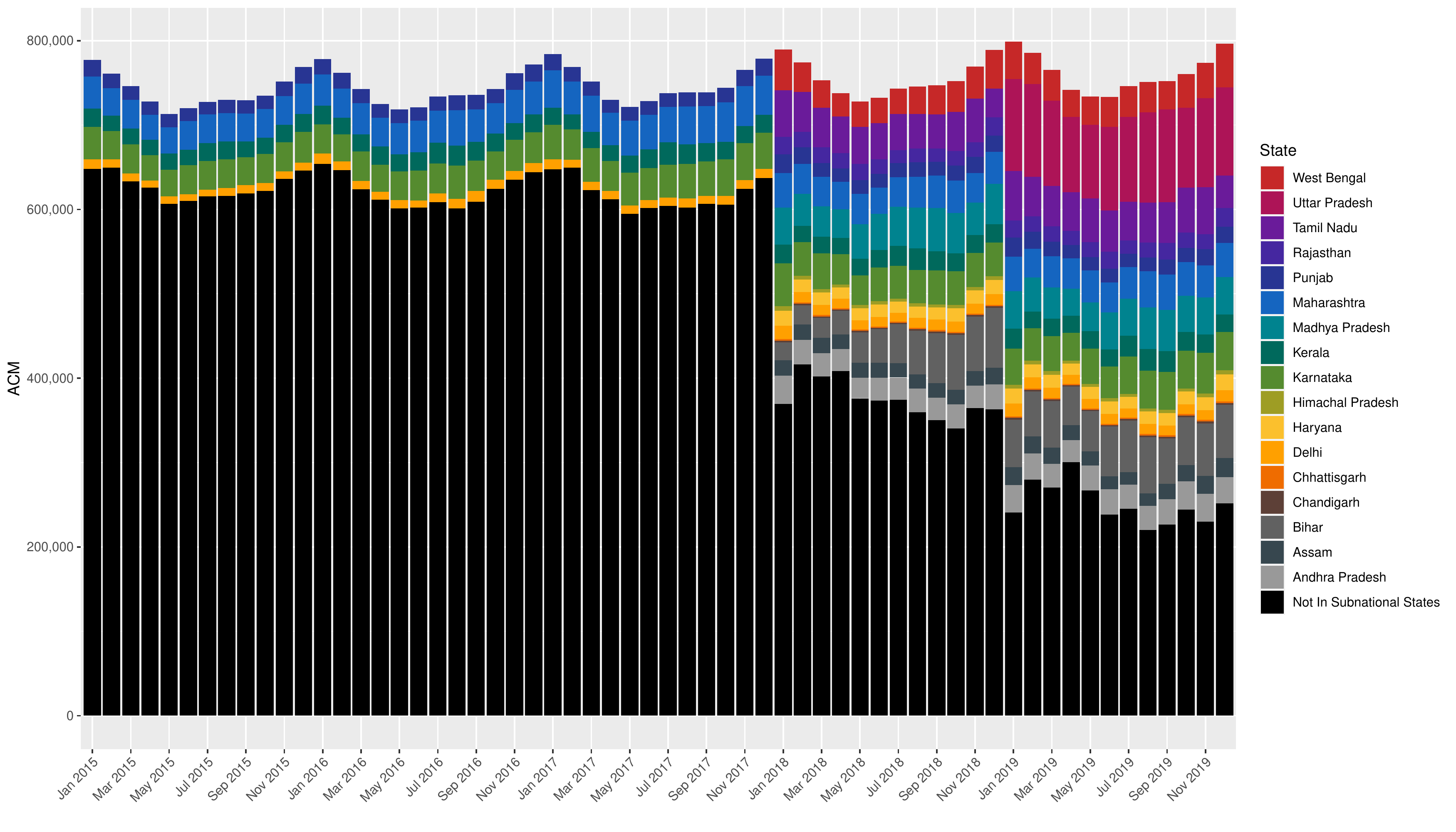}
%\hspace{-.8in}
%\includegraphics[height = 6.3cm]{ClusterLocs_Admin2.pdf}
\caption{All-cause mortality by month in pre-pandemic period, 2015--2019. Black rectangles are totals while colored rectangles are the states for which we have data in the pandemic.}
\label{fig:IndiaPlot3}
\end{figure}

\begin{figure}[htbp]
\centering
\includegraphics[scale=.4]{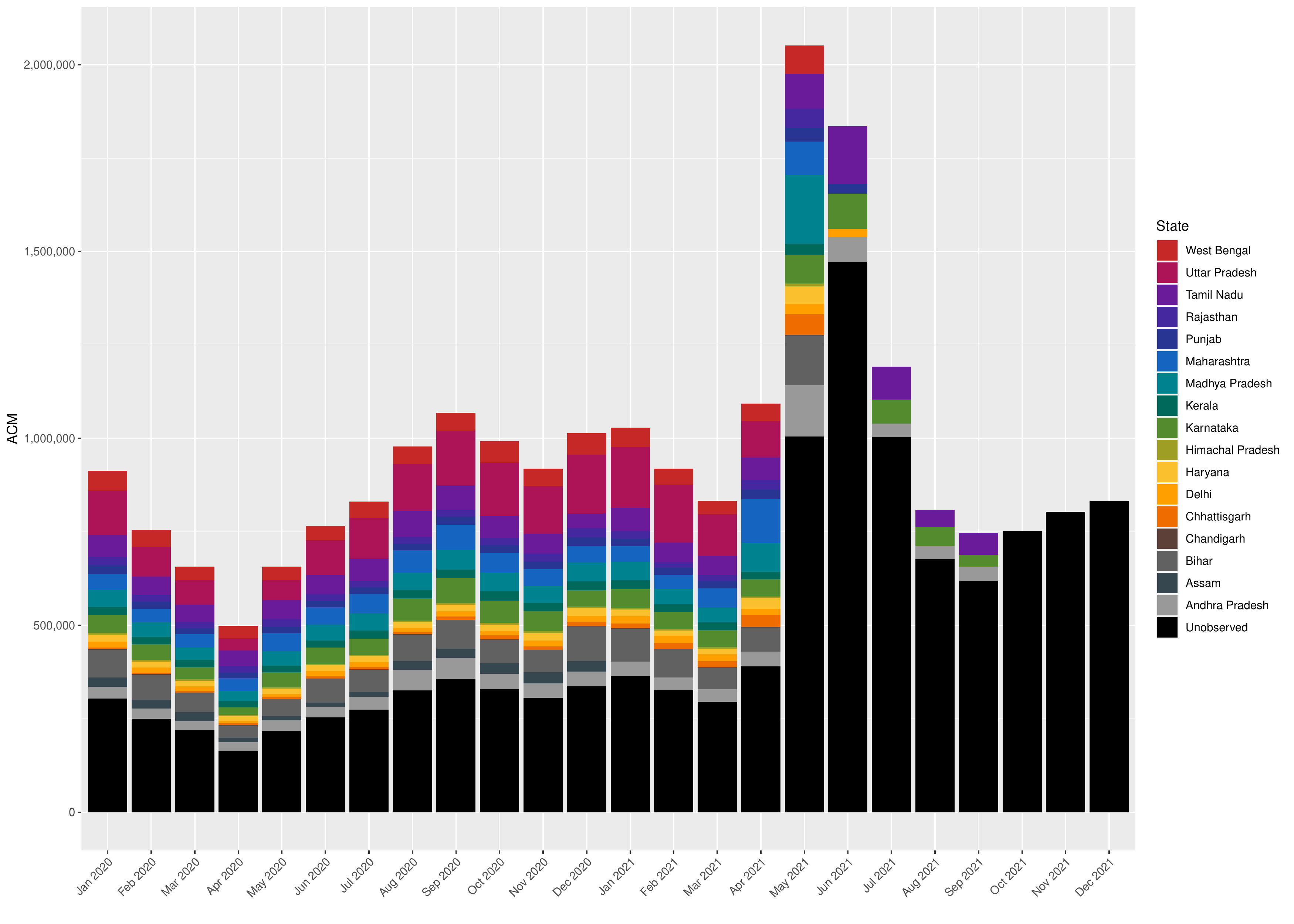}
%\hspace{-.8in}
%\includegraphics[height = 6.3cm]{ClusterLocs_Admin2.pdf}
\caption{All-cause mortality by month in the pandemic. Black rectangles are estimated while colored rectangles are observed.}
\label{fig:IndiaPlot2}
\end{figure}

\begin{figure}[htbp]
\centering
\includegraphics[scale=.6]{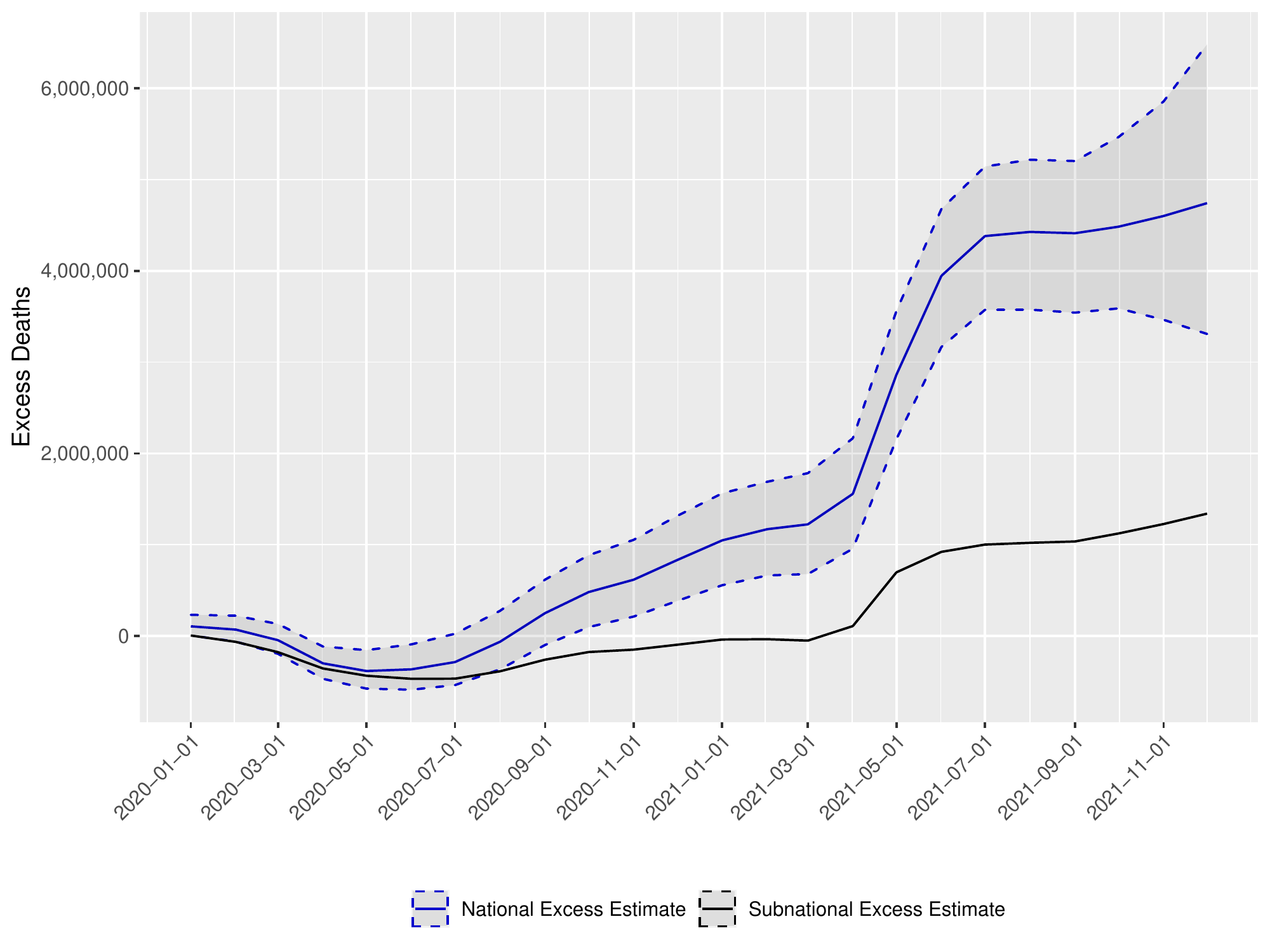}
%\hspace{-.8in}
%\includegraphics[height = 6.3cm]{ClusterLocs_Admin2.pdf}
\caption{Cumulative excess over January 2020 to December 2021. The black curve is the contribution from the observed state level data.}
\label{fig:India_ExcessPlot4}
\end{figure}

Figure \ref{fig:IndiaPlot3} shows the ACM counts by states, and the remainders, which sum to the total ACM, by month. Figure \ref{fig:IndiaPlot2} shows the pandemic period, with the crucial change being that now the black rectangles are estimated, based on the fraction of deaths in each state, as estimated from the data in Figure \ref{fig:IndiaPlot3}. Figure \ref{fig:India_ExcessPlot4} shows the cumulative excess over January 2020--December 2021, along with the estimated contribution from the state-level data only.

Recall that the subnational model is built on the assumption that the proportion of deaths in a state is approximately constant over time.
Having data from many states does protect from requiring a constant fraction assumption for all states, since the level of bias depends on the cumulative effects of departures from constancy across all states.

\begin{figure}[htbp]
\centering
\includegraphics[scale=.4]{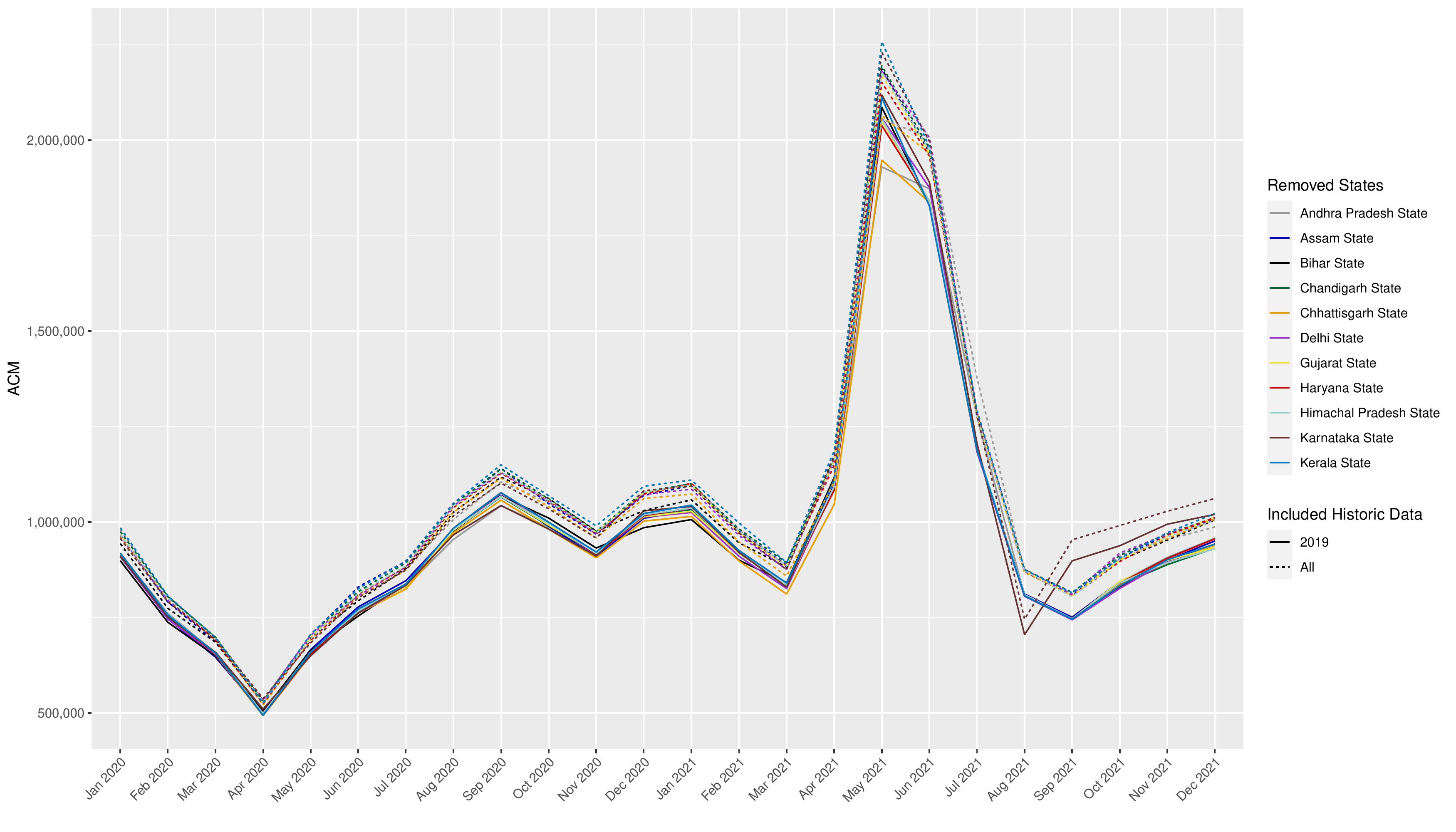}
%\hspace{-.8in}
%\includegraphics[height = 6.3cm]{ClusterLocs_Admin2.pdf}
\caption{Time series of ACM with different states excluded, and with different years of data used for the expected numbers.}
\label{fig:India_Sensitivity2}
\end{figure}

\begin{figure}[htbp]
\centering
\includegraphics[scale=.4]{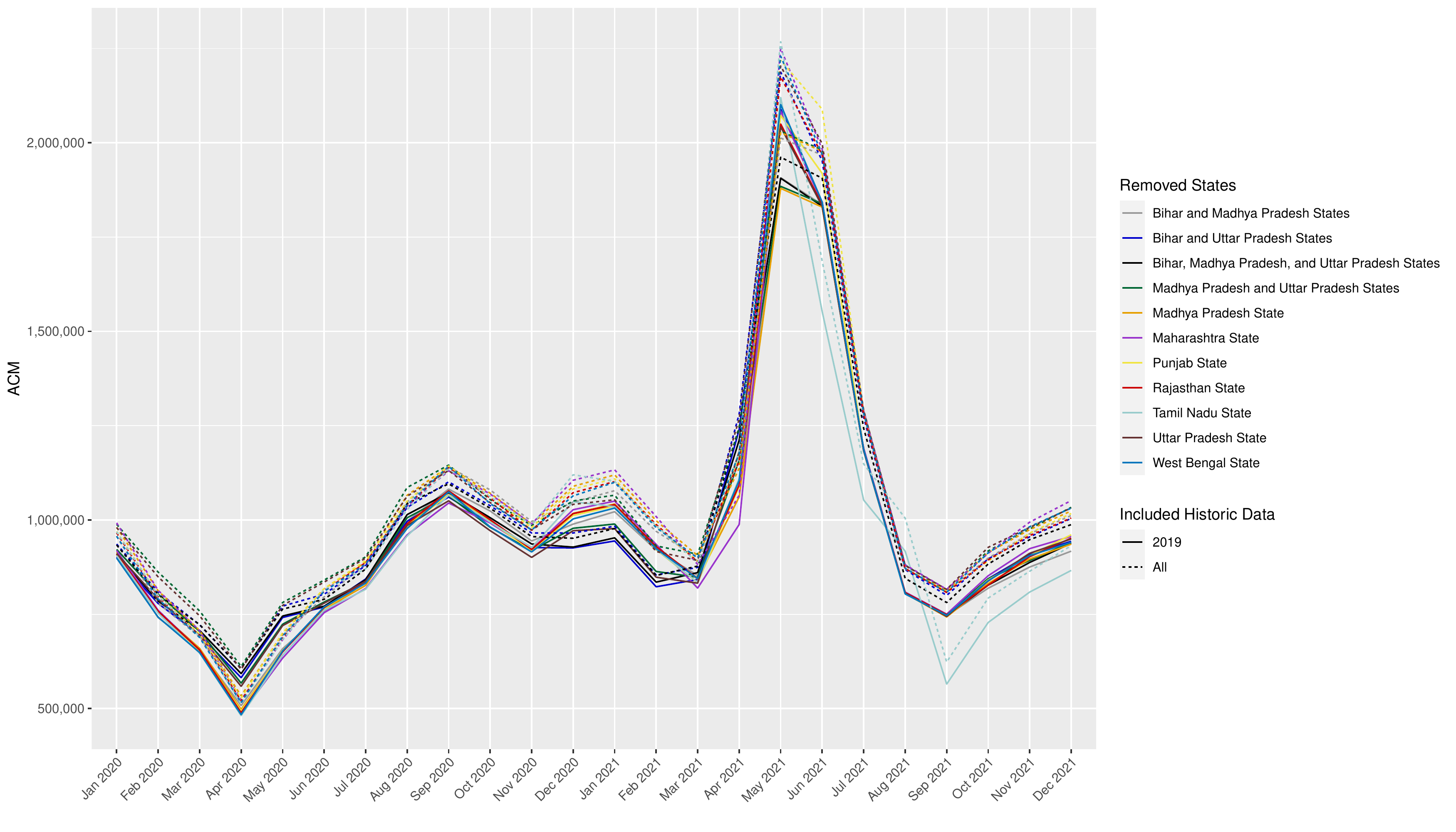}
%\hspace{-.8in}
%\includegraphics[height = 6.3cm]{ClusterLocs_Admin2.pdf}
\caption{Time series of ACM with different states excluded, and with different years of data used for the expected numbers.}
\label{fig:India_Sensitivity3}
\end{figure}

\begin{figure}[htbp]
\centering
\includegraphics[scale=.4]{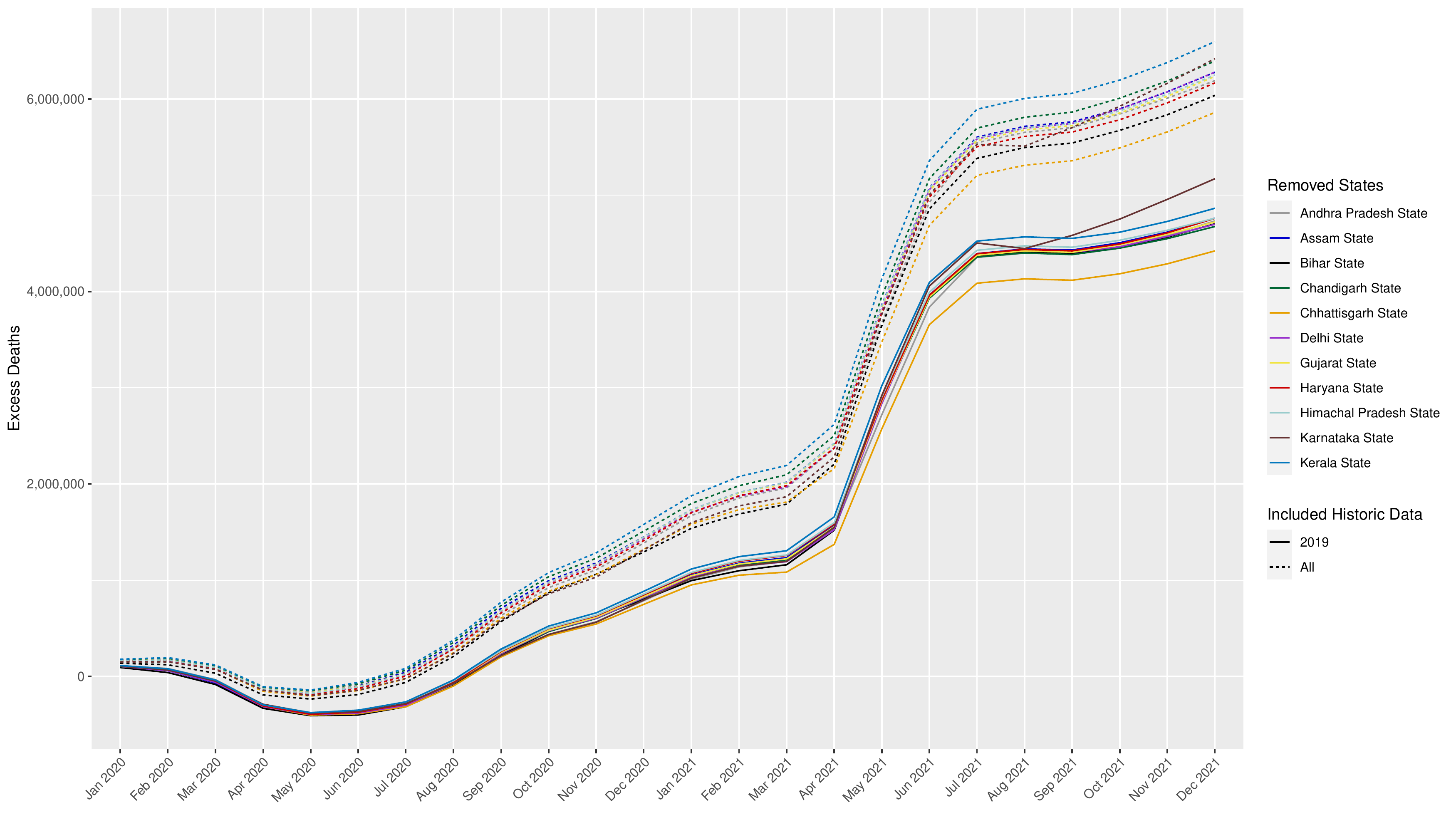}
%\hspace{-.8in}
%\includegraphics[height = 6.3cm]{ClusterLocs_Admin2.pdf}
\caption{Cumulative excess with different states excluded, and with different years of data used for the expected numbers. Based on all the data, we estimate that India's excess is 4.7 million deaths, with a 95\% credible interval of (3.31, 6.48) million. 
}
\label{fig:IndiaCumulative_Sensitivity2}
\end{figure}

\begin{figure}[htbp]
\centering
\includegraphics[scale=.4]{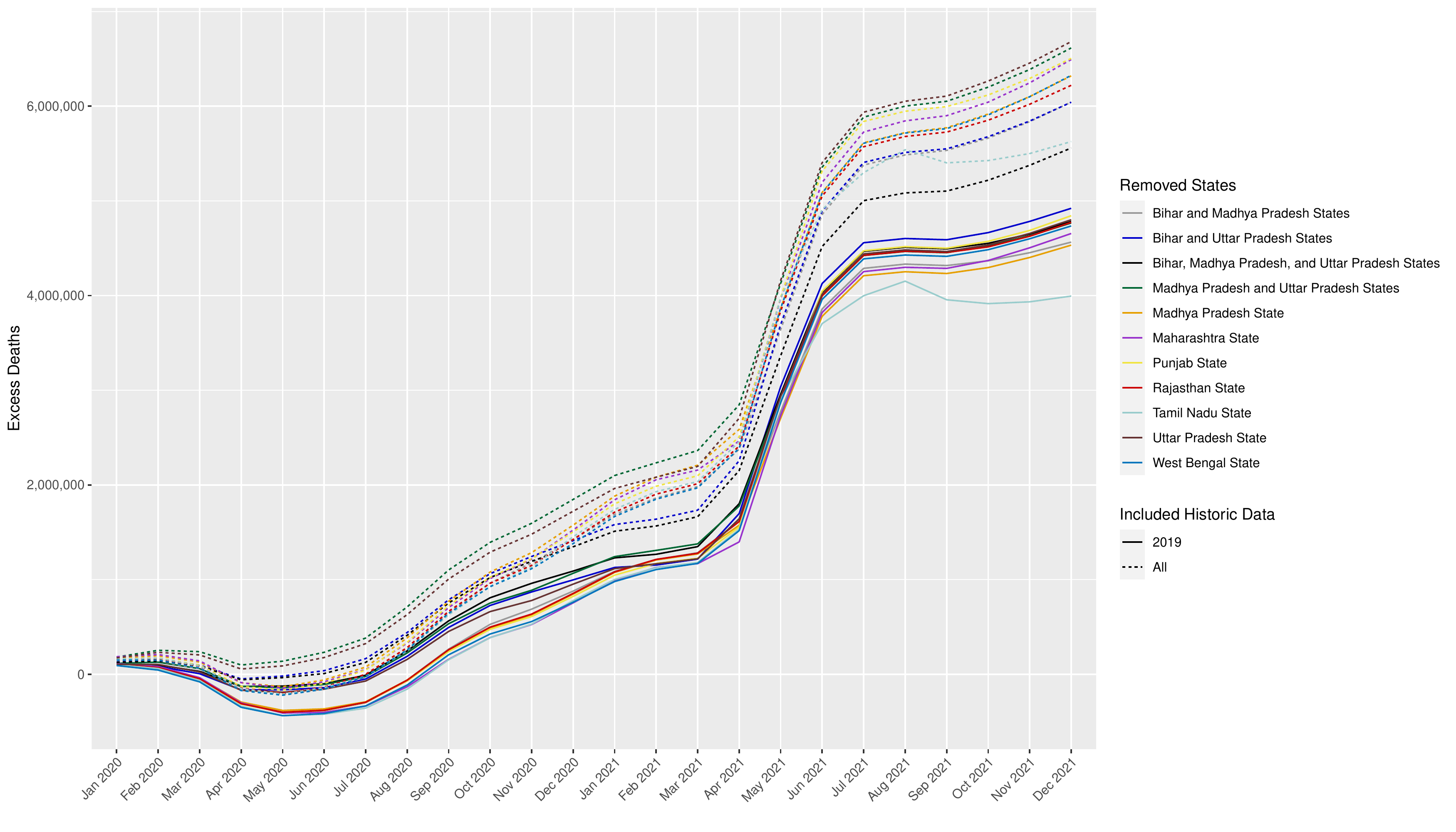}
%\hspace{-.8in}
%\includegraphics[height = 6.3cm]{ClusterLocs_Admin2.pdf}
\caption{Cumulative excess with different states excluded, and with different years of data used for the expected numbers. Based on all the data, we estimate that India's excess is 4.7 million deaths, with a 95\% credible interval of (3.31, 6.48) million. }
\label{fig:IndiaCumulative_Sensitivity3}
\end{figure}

To address the sensitivity we carried out analyses based on different subsets of the data  and estimating the fractions of deaths based on data from 2015--2019, or from 2019 only. Figures \ref{fig:India_Sensitivity2} and  \ref{fig:India_Sensitivity3} show the time series of estimated national ACM for different subsets of states, and based on different years of subnational data. 
Figures \ref{fig:IndiaCumulative_Sensitivity2} and \ref{fig:IndiaCumulative_Sensitivity3} show the cumulative national ACM versions of these plots. For monthly and cumulative plots we split into two sets to make the plots simpler to read. The cumulative totals are substantively similar, offering evidence that the excess result is not being driven by data from any one state.

%For India, we remove one state at a time and examine the sensitivity of the results, in terms of both the monthly time series of excess, and the cumulative by month.



%There was a large jump in mortality in May and June of 2021, with our estimates being based on data from 16 and 5 states in these two months. The 5 states with data in June 2021 all had full completeness, but 5 of the 16 states in May 2021 had completeness \cite{india_crs_2019} which was less than 100\% (Bihar, Chhattisgarh, Himachal Pradesh, Madhya Pradesh, Rajasthan), and estimated from historic data, which is a difficult task.
%In Table \ref{tab:datamortsummary} we show the multinomial subnational model estimates of total excess mortality from January 2020 through June 2021 using: (1) all the data, and then (2) excluding the five states whose completeness was less than 1, and then also excluding data from Andhra Pradesh, since in the historic data it also had completeness less than 1 in 2015 and 2016, and it is also likely to be influential, since we see in Figure \ref{fig:IndiaSubnationalDeathPlot}  that it has large death counts in the May--June 2021 months. The increased uncertainty in the final two months is due to the greater uncertainty at higher mean levels in the sampling model, and the reduction in available information (less states) as seen in Figure \ref{fig:IndiaSubnationalDeathPlot}.

\begin{figure}[htbp]
\centering
\includegraphics[scale=.3]{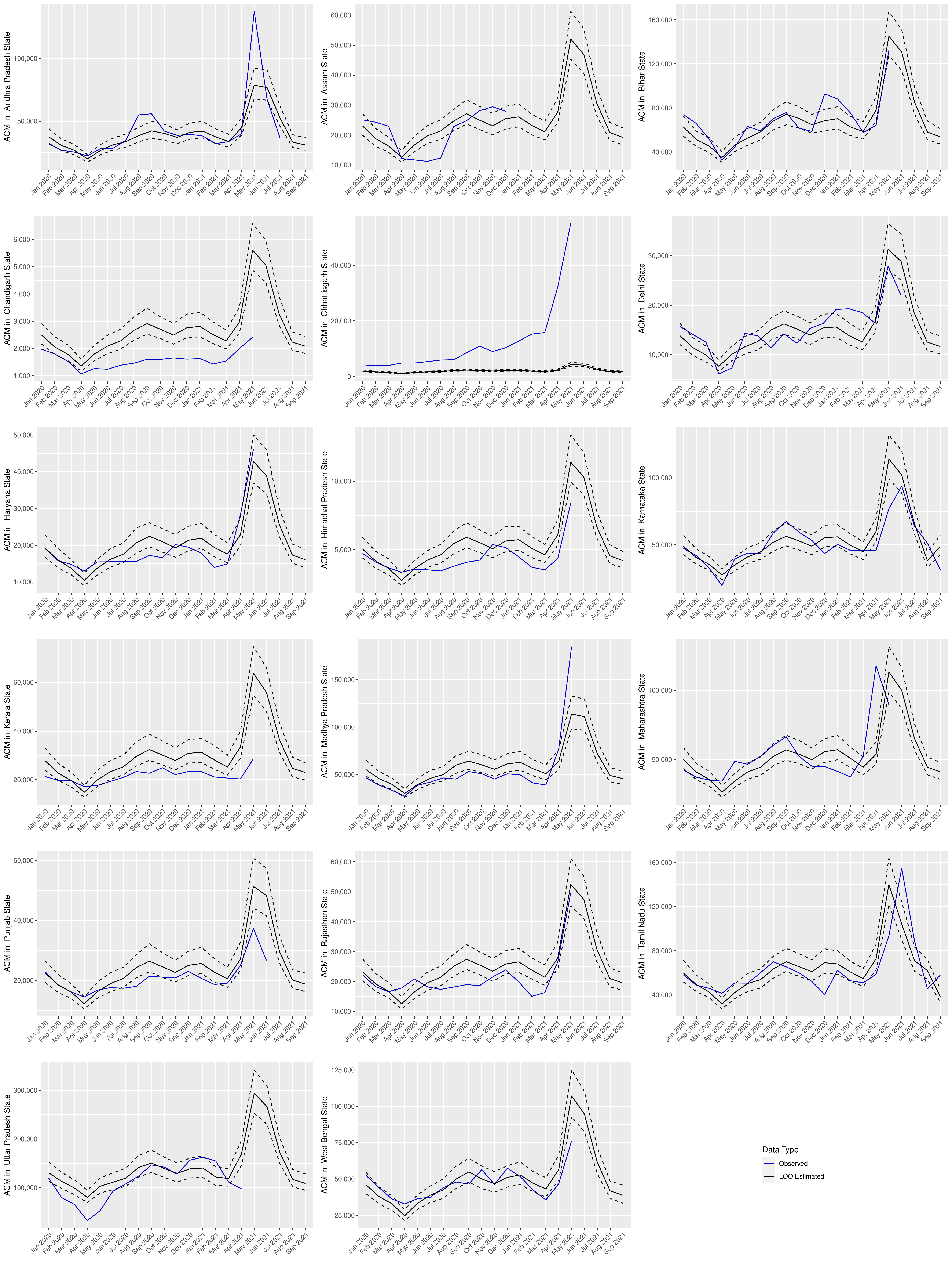}
\caption{Observed ACM by month and predicted, based on a leave-one-out procedure, in which the state ACM is predicted based on the data from all other states.}\label{fig:IndiaStateLOO1}
\end{figure}

\begin{figure}[htbp]
\centering
\includegraphics[scale=.3]{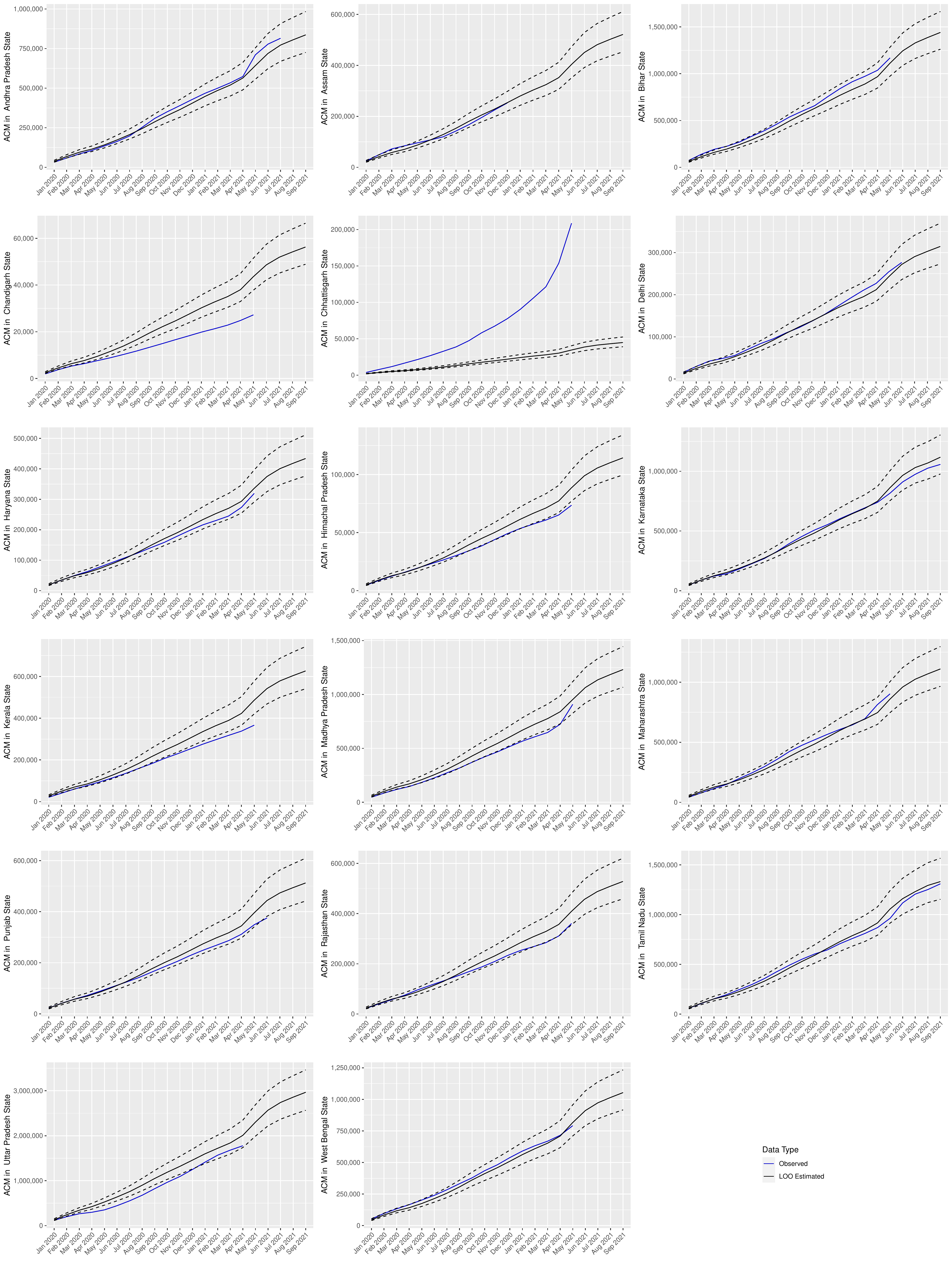}
\caption{Cumulative ACM by month and predicted, based on a leave-one-out procedure, in which the state ACM is predicted based on the data from all other states.}\label{fig:IndiaStateCum}
\end{figure}

To further assess the Indian subnational model, we carry out the following leave-one-out strategy. We systematically leave one state out at a time, and then estimate the monthly national distribution of ACM, by month. We then apply the estimate of the fraction of the left out state to the total, and compare with the observed ACM. Figures \ref{fig:IndiaStateLOO1} and \ref{fig:IndiaStateCum} show the resultant time series and cumulative totals. The predictions are better for some states than others but they are not systematically higher or lower, which gives some reassurance that our model is robust. Chandigarh is low and Chhattisgarh is high, but these states are responsible for a small proportion of the total. 

%\begin{figure}[htbp]
%\centering
%\includegraphics[scale=.3]{VictoriaFigures/IndiaStateLOO2.pdf}
%\caption{.}\label{fig:IndiaStateLOO2}
%\end{figure}

%As another sensitivity exercise we remove one state at a time, and then estimate the number of deaths we would see in this state using the estimated fraction of the total deaths in that state and the predictive distribution of the national ACM.

\clearpage

In  Table \ref{tab:India}  we give our estimates (based on all data for the first 21 months and with expected numbers calculates from 2019) and also give the estimates from IHME, The Economist, \cite{jha2022covid} and three estimates from \cite{anand2021three}.  The  \cite{jha2022covid}  estimate is based on a nationally representative telephone survey, a government survey that covers 0.14 million adults and the Government of India's data from facility-based deaths and CRS deaths in 10 states. \cite{anand2021three} use three methods: Indian States' CRS (method 1), international age-specific infection fatality rates applied to Indian demography (method 2) and seroprevalence and a household survey (method 3).
 % \cite{zimmermann2021sars} also provides an in-depth study of the pandemic in India, and summarizes estimates of the estimated excess number of deaths.
  
%The official COVID-19 death count was downloaded for the relevant period from \url{https://covid19.who.int/} and is 489K, which is around a tenth of the estimated excess mortality.

There is reasonable agreement between the different estimates, which is reassuring, given the different data sources used. Along with the sensitivity and leave-one-out analyses, this provides further evidence that the model is reasonable and the subnational data (taken as a whole) are representative, so that, when combined, they provide a reliable excess mortality estimate.

\begin{table}[htp]
\begin{center}
\begin{tabular}{l|l|l}
\textbf{Approach} & \textbf{Estimate $(\times 10^6)$} &Period\\ \hline
%%Multinomial Model: Full Data & 5001130 (4081270 , 5990251)  \\  \hline
Naive  & 5.04 (4.48, 5.59)&Jan 20--Dec 21\\ 
WHO  &  4.74 (3.31, 6.48)&Jan 20--Dec 21  \\ 
%%Multinomial Model: Excluding Bihar, Chhattisgarh, Himachal Pradesh, Madhya Pradesh, Rajasthan States & 4220937 (1234006 , 5527979) \\ \hline
%%Multinomial Model: Excluding  States & 4.2 (1.2, 5.5) \\ 
%%Multinomial Model: Excluding Bihar, Chhattisgarh, Himachal Pradesh, Madhya Pradesh, Rajasthan, Andhra Pradesh States &  3829601 (1383550 , 4979239) 
%Multinomial Model: Excluding 6 States &   ()&Jan 21--Dec 22 \\ \hline
%%Reported COVID-19& 0.40\\ \hline
The Economist &4.86  (1.70, 8.47)&Jan 20--Dec 21 \\
IHME &4.07  (3.71, 4.36)&Jan 20--Dec 21 \\
 \hline
Naive  & 4.29 (4.00, 4.59)&June 20--June 21\\ 
WHO & 4.33 (2.85, 6.13)&June 20--June 21  \\  
\cite{jha2022covid}& 3.23 (3.06, 3.39)& June 20--June 21\\ \hline
Naive & 3.96 (3.62, 4.29) &April 20--June 21 \\
WHO & 3.99  (2.40, 5.95) &April 20--June 21 \\
\cite{anand2021three} Method 1 & 3.4 & April 20--June 21 \\
\cite{anand2021three} Method 2 & 4.0 & April 20--June 21 \\
\cite{anand2021three} Method 3 & 4.9 & April 20--June 21 \\ \hline
%\cite{leffler_2021} & 2.69 (1.98, 3.57) & Jan 20--Aug 2021
\end{tabular}
\end{center}
\caption{ The parentheses give 95\% uncertainty intervals. The \cite{jha2022covid} estimate is for excess COVID-19 deaths. The naive estimates are based on the ACM estimates $Y_{t,1}/\widehat p_t$ where $Y_{t,1}$ is the observed ACM from the available states, and $\widehat p_t$ is the estimated fraction  of deaths available in  month $t$.%\cite{anand2021three}  methods are:...
%The 5 states that are excluded are:  Bihar, Chhattisgarh, Himachal Pradesh, Madhya Pradesh, Rajasthan. The 6 states version excludes, in addition, Andhra Pradesh.
}
\label{tab:India}
\end{table}

\clearpage
\subsection{Indonesia}

\begin{figure}[htbp]
\centering
\includegraphics[scale=.5]{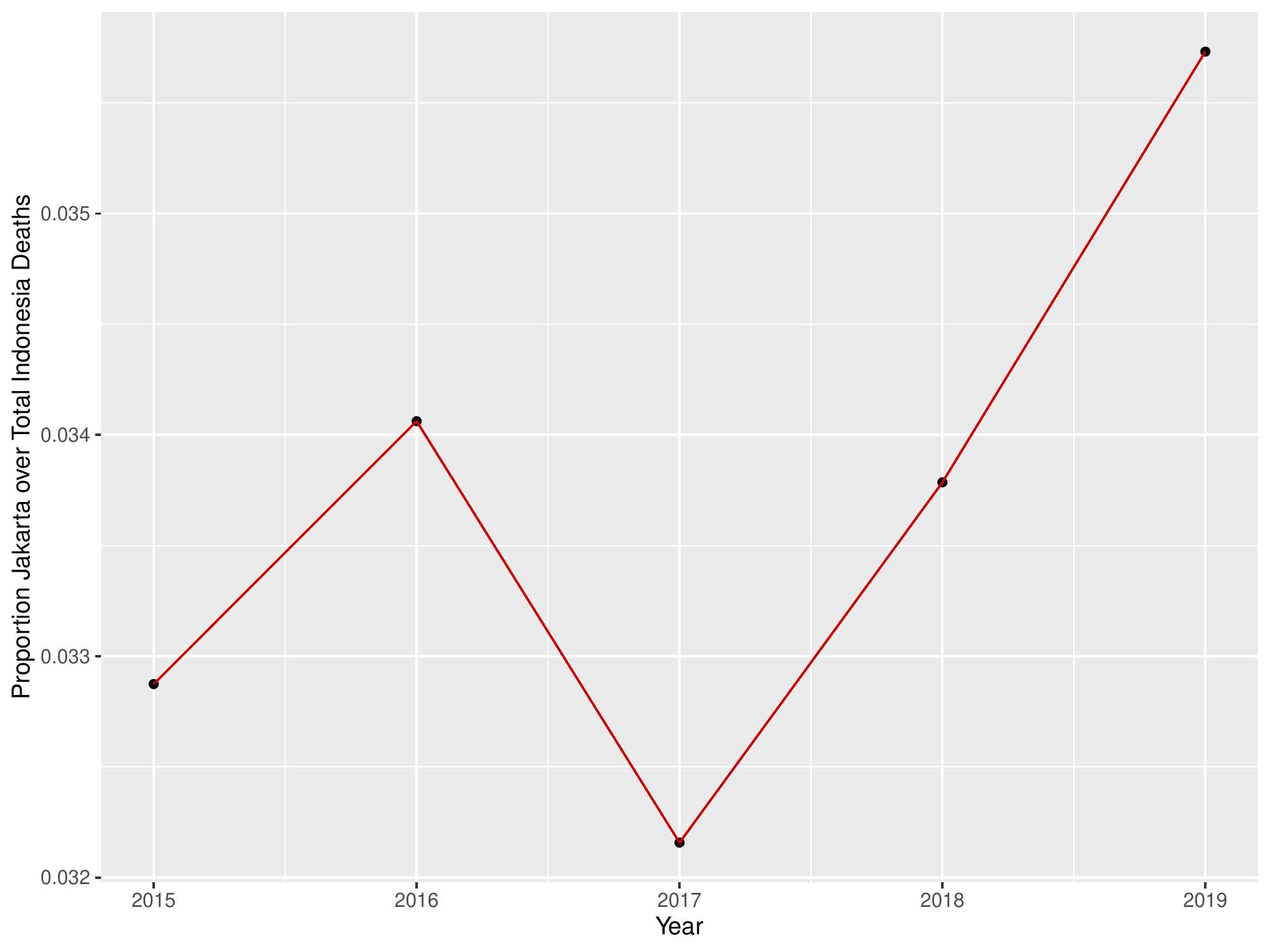}
\caption{Proportion of ACM counts in Jakarta in 2015--2019.}\label{fig:PbyMonth_Indonesia}
\end{figure}

\begin{figure}[htbp]
\centering
\includegraphics[scale=.45]{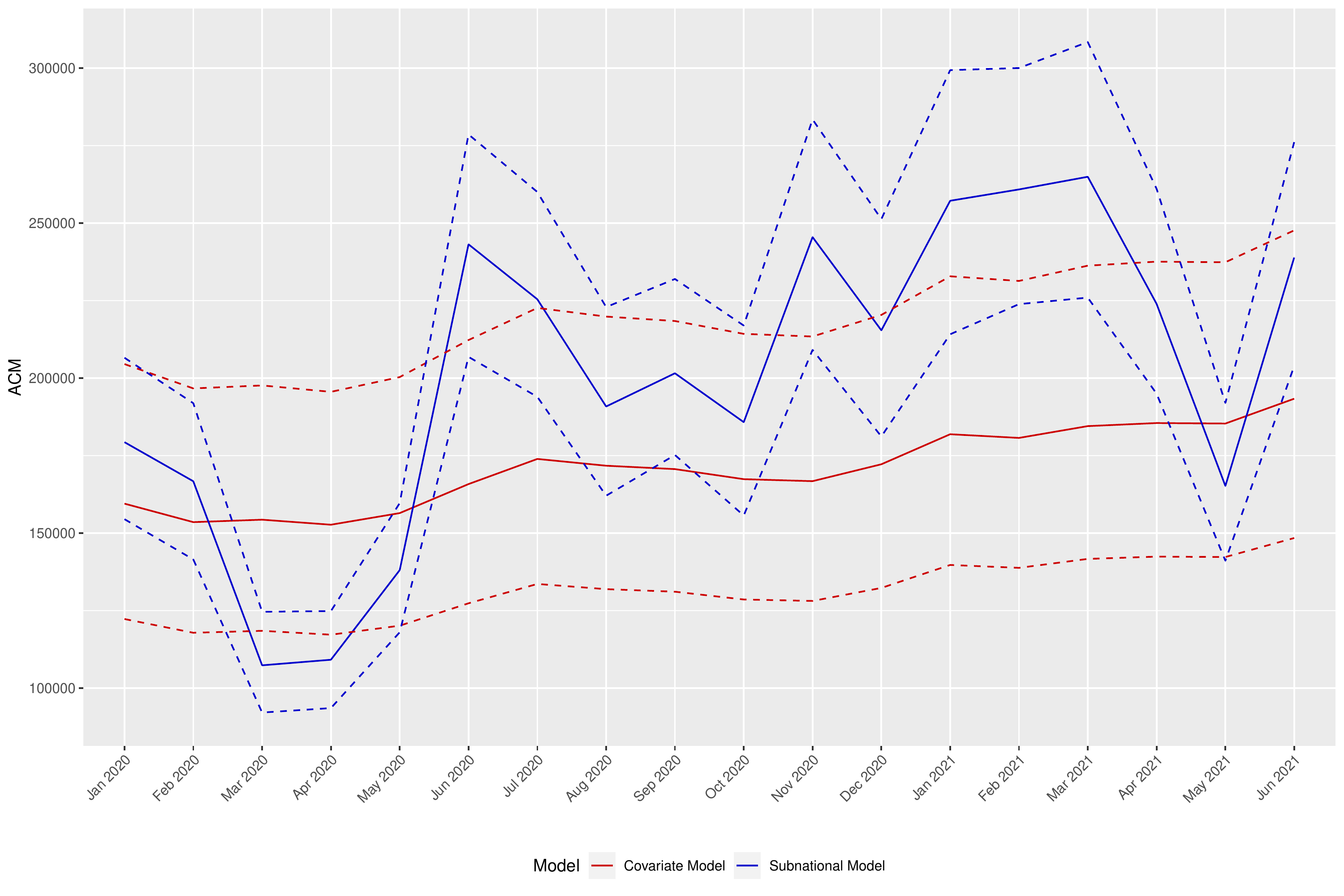}
\includegraphics[scale=.45]{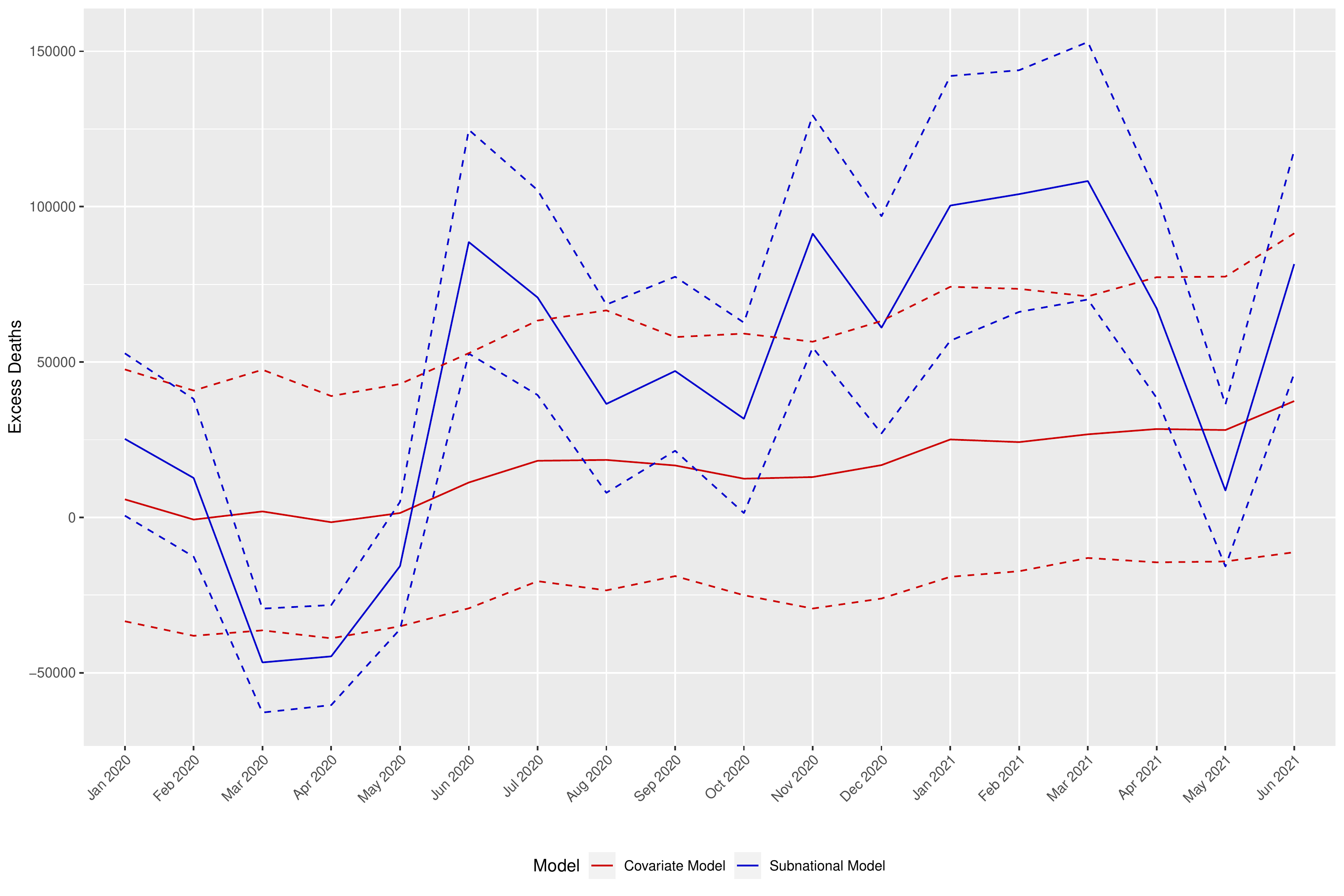}
\caption{Estimated ACM (top) and excess ACM (bottom) for Indonesia from covariate model and subnational data from Jakarta.}\label{fig:Indonesia}
\end{figure}

We have the annual number of national deaths in Indonesia as published by the Global Burden of Disease (GBD) Study from 2015--2019 \cite{gbd_2019}. The subnational data consist of the monthly number of deaths from 2015--2021 from Jakarta, Indonesia. The proportions of deaths in Jakarta in 2015--2019 are shown in Figure \ref{fig:PbyMonth_Indonesia}. Note that this is only just above 3\%.   An alternative to using the Jakarta data, is to use the global covariate model.
In Figure \ref{fig:Indonesia} we show the comparison between these two analyses (over the period January 2020--June 2021, in the last 6 months of 2021 there are no subnational data, and so the covariate model is used). For the covariate model the point and 95\% interval estimate are 495K (256K, 726K) while for the subnational they are 1,024K (752K 1,287K). There are large differences in the two analyses, and in the main paper the subnational estimates were the ones we used when reporting regional and global estimates, as WHO liked to use observed  data from a country, when available. We believe it is important to see both estimates, however, because of the small proportion of the population observed in an urban setting.

%\begin{figure}[htbp]
%\centering
%\includegraphics[scale=.5]{VictoriaFigures/Indonesia_PredvINLA.pdf}
%\caption{Left: Proportions. Right: Estimates of excess mortality, with 95\% uncertainty, for Indonesia in 2020-2021.}\label{fig:Indonesia_PredvINLA}
%\end{figure}

 \clearpage
\subsection{Turkey}
We have monthly annual deaths in Turkey as published by the Turkish National Statistics Office from 2015--2019. The subnational data correspond to monthly ACM counts from 2018--2021 in 24 provinces and cities from all over Turkey, 
as published by Guclu Yaman. More details, for 21 of the locations are at \url{https://gucluyaman.com/tr/excess-mortality-in-turkey/}.

\begin{figure}[htbp]
\centering
\includegraphics[scale=.6]{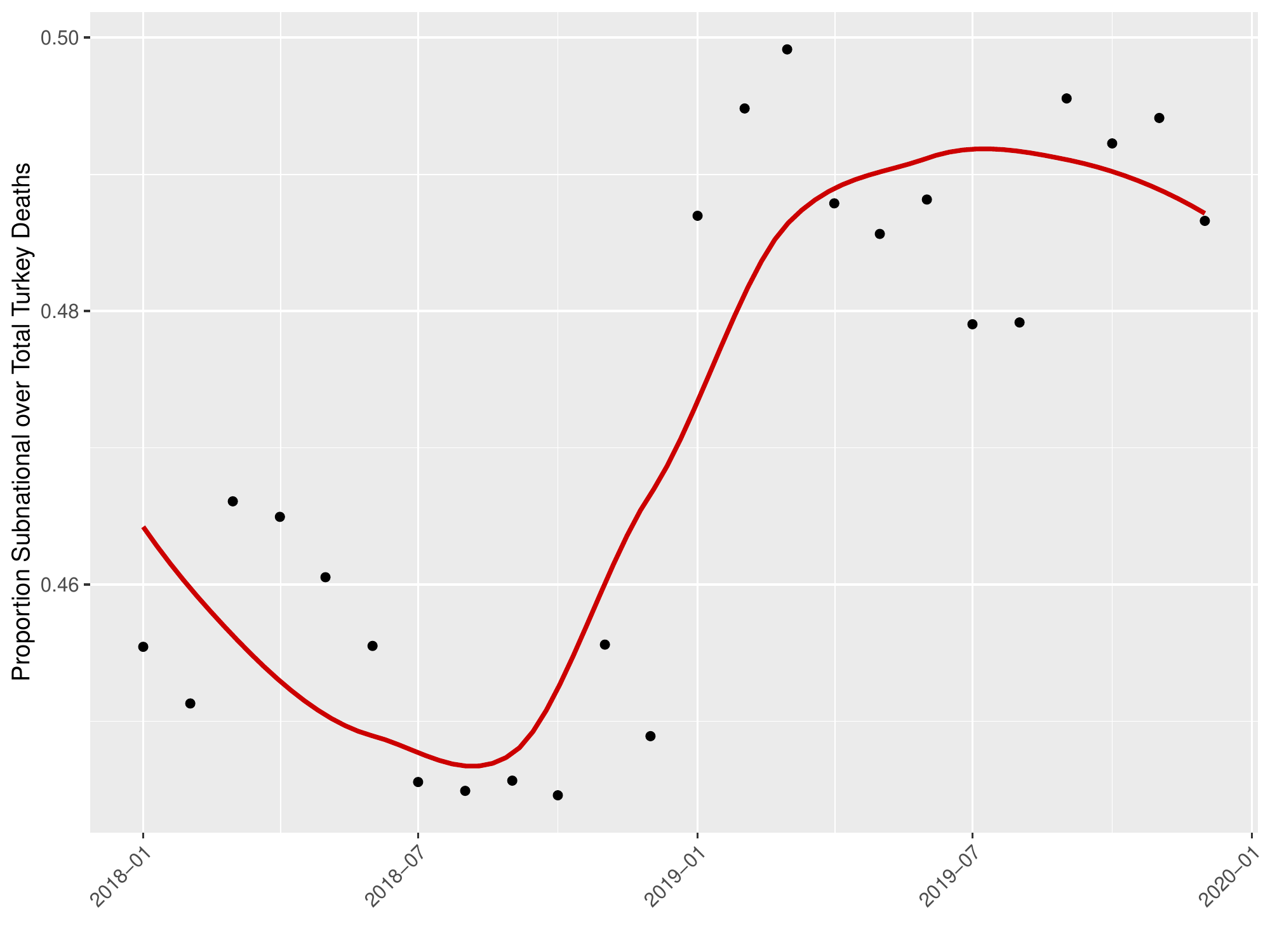}
%\hspace{-.8in}
%\includegraphics[height = 6.3cm]{ClusterLocs_Admin2.pdf}
\caption{Proportion of ACM counts in subregions of Turkey in 2015--2019.}
\label{fig:PbyMonth_Turkey}
\end{figure}

\begin{figure}[htbp]
\centering
\includegraphics[scale=.7]{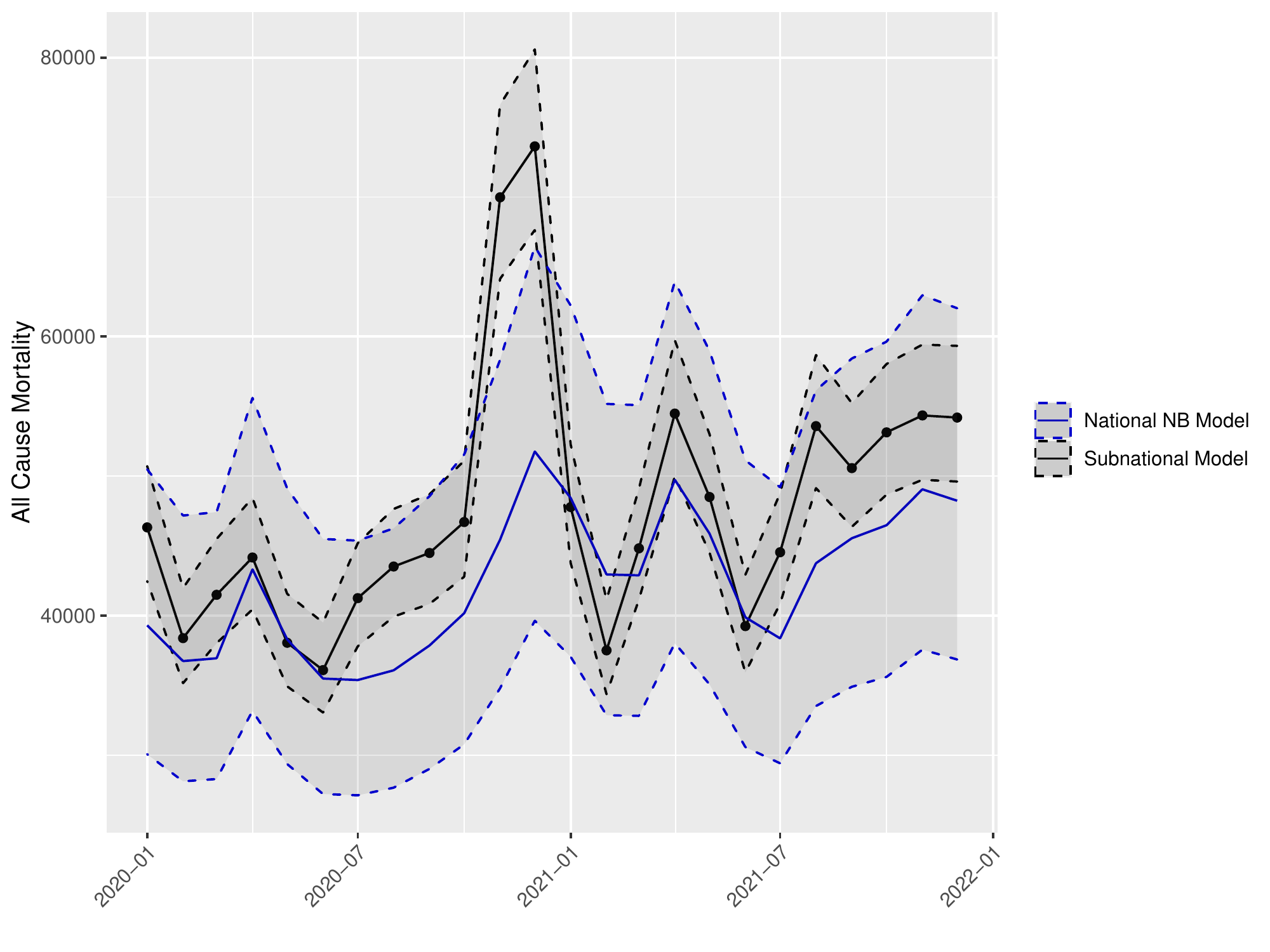}
%\hspace{-.8in}
%\includegraphics[height = 6.3cm]{ClusterLocs_Admin2.pdf}
\caption{Predicted ACM, with 95\% uncertainty, for Turkey from covariate model and subnational model.}
\label{fig:Turkey_PredvINLA}
\end{figure}

Figure \ref{fig:PbyMonth_Turkey} shows the fractions of deaths in the 23 subnational areas over 2018--2019.
Using the binomial subnational model, and the global covariate model, we obtain the estimates shown in Figure \ref{fig:Turkey_PredvINLA} -- it is reassuring that they are so similar. In the main paper, we used the subnational estimates.
% A more detailed list of sources is here (but less up to date, showing only 21 sub-national units): https://gucluyaman.com/tr/excess-mortality-in-turkey/

\clearpage
\subsection{Additional Analyses for Germany and Sweden}

\begin{figure}[htbp]
\centering
\includegraphics[scale=.4]{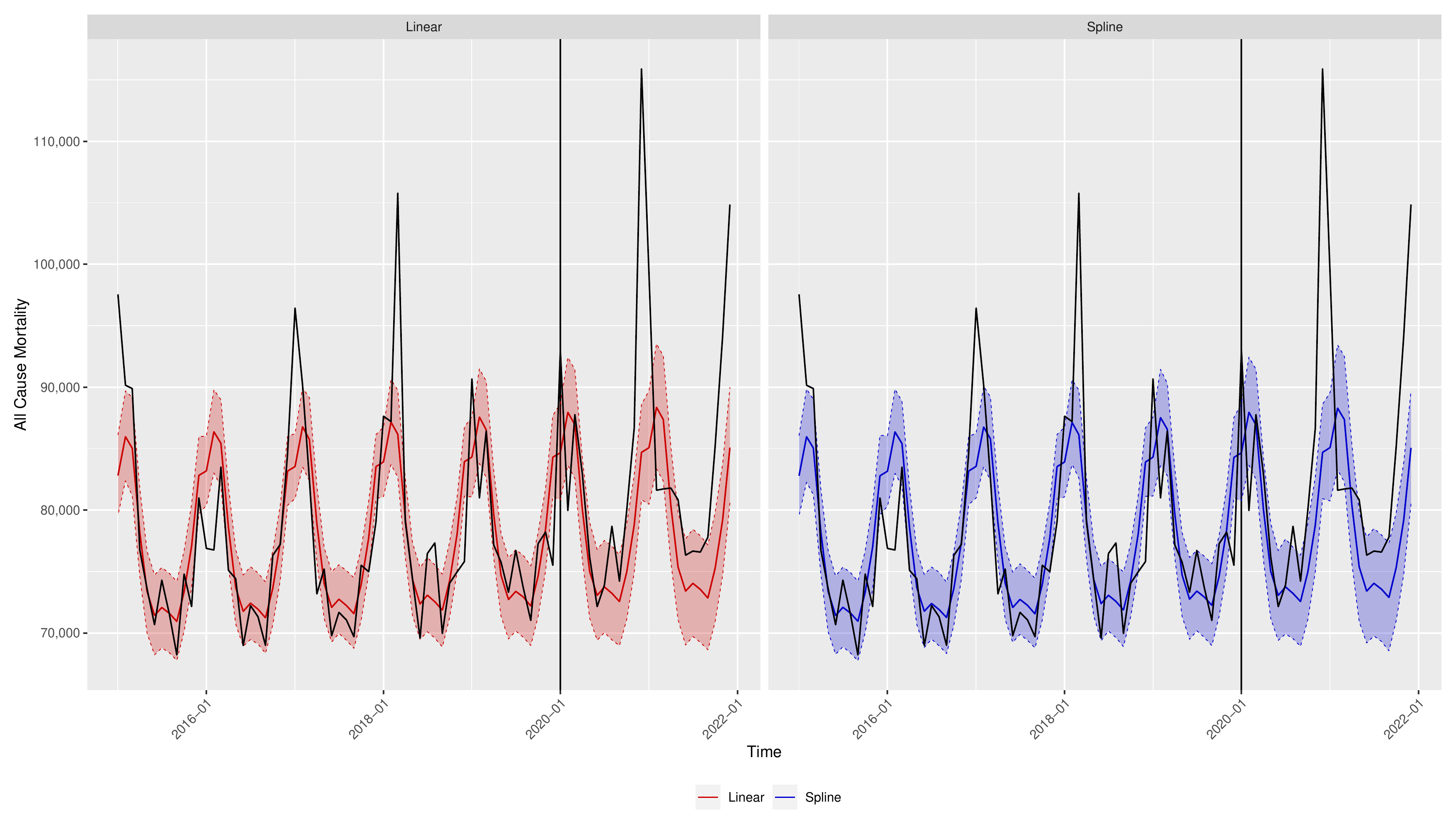}
%\hspace{-.8in}
%\includegraphics[height = 6.3cm]{ClusterLocs_Admin2.pdf}
\caption{Expected ACM counts and observed ACM (the black lines) for Germany, using unadjusted ACM data.}
\label{fig:germany_linear}
\end{figure}

After the official release of the WHO results, there was attention on Germany and Sweden which has lead us to examine our models and data sources for those countries. The original excess estimate for Germany was 195K (161K, 229K). This estimate was obtained using a negative binomial model with a thin-plate spline for the annual trend (on the linear predictor scale). A scaling factor to account for completeness was also used, which lead the ACM counts in 2016--2018 to be scaled up. Unfortunately for the adjusted data the spline fit was unduly influenced by a lower count in 2019 which caused the spline prediction in the pandemic to be too low, and the excess correspondingly to be too high. We removed the completeness adjustment
%, as we see in the right-hand panel of Figure \ref{fig:germany_linear}. 
and replaced the spline term for the annual trend with a  linear term, and this produced the much more reasonable series in the left-hand panel of Figure \ref{fig:germany_linear}.
 %(the intervals are also narrower with the linear model, as expected, given the simpler form).  
This produces a revised excess estimate of 122K with a 95\% interval of (101K, 143K). With the adjusted data, the spline also produced a reasonable fit (right-hand panel of Figure \ref{fig:germany_linear}) and very similar estimates.
% The magnitude of the difference between the two estimates shows the sensitivity to the expected numbers model used, and a priority for future work is a systematic evaluation of models for producing expected numbers.

\begin{figure}[htbp]
\centering
\includegraphics[scale=.4]{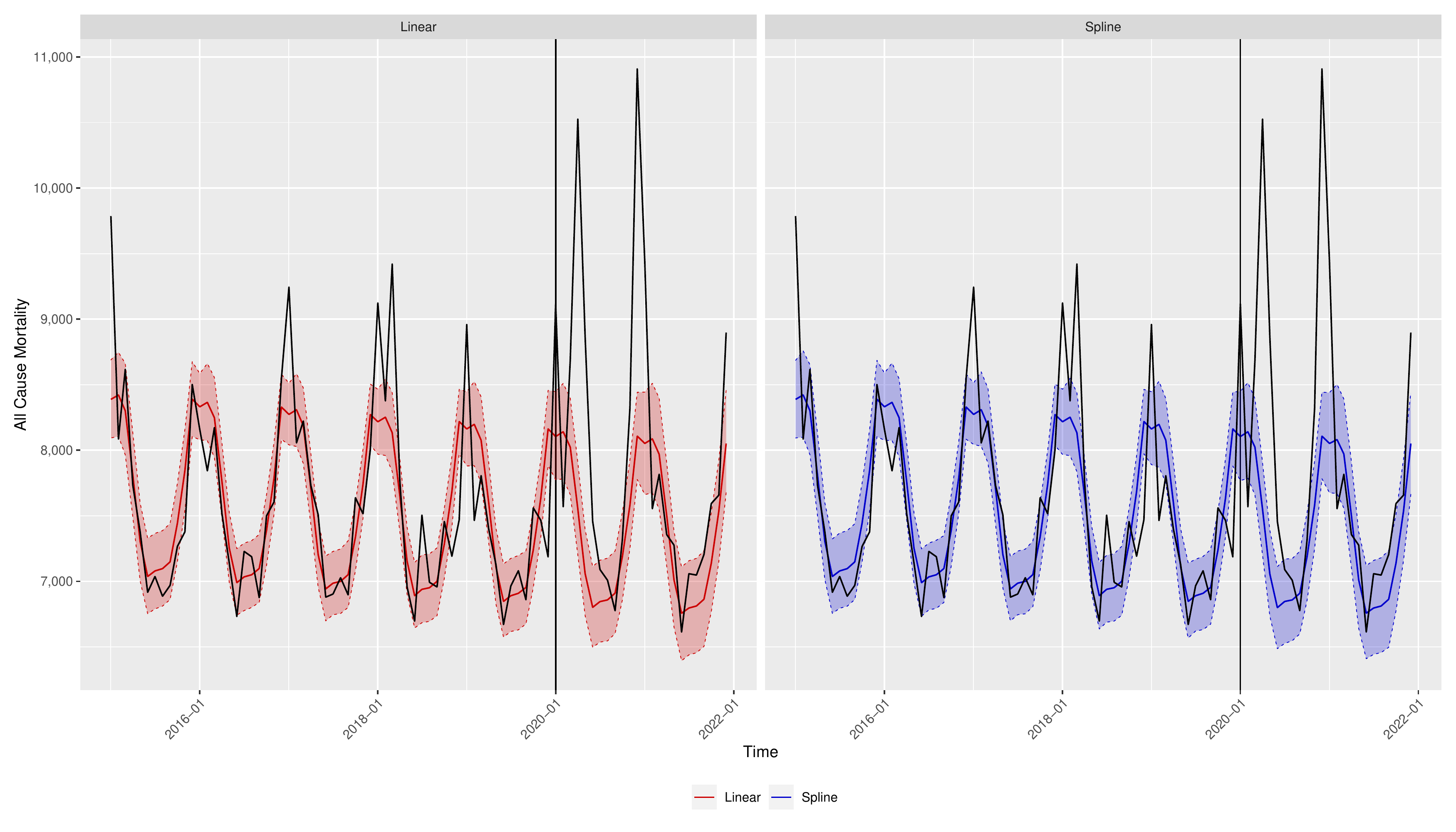}
%\hspace{-.8in}
%\includegraphics[height = 6.3cm]{ClusterLocs_Admin2.pdf}
\caption{Expected ACM counts and observed ACM (the black lines) for Sweden, using unadjusted ACM data.}
\label{fig:sweden_linear}
\end{figure}

%n general, the WHO carry out checks for completeness of death registration.
For Sweden, the WHO made a completeness adjustment for the 2019 mortality count (which was lower than the previous year), which resulted in an increase in the count, and this same adjustment was also applied to the 2020 and 2021 counts. The original excess estimate for Sweden, using the adjusted data, was 11.3K (9.9K, 12.7K). With hindsight the adjustment was unnecessary and so we present here an analysis with the unadjusted data, and replacing the spline term for the annual trend with a linear term.
The revised estimates with the unadjusted data are 13.4K (11.7K, 15.2K), so an increase over the previous analysis. With the unadjusted data the spline model gave similar excess estimates. The fits are shown in Figure \ref{fig:sweden_linear}. For the next update of estimates, we will revisit the completeness process that was used to produce the counts used in the various analyses, and also examine different models for calculating the expected numbers.

\clearpage
\section{Supplementary Materials: Model Assessment}\label{sec:modelcomparison}

The sampling model we assume is,
\begin{eqnarray*}
Y_{c,t} | \theta_{c,t} \sim \mbox{NegBinl}( \widehat{E}_{c,t}\theta_{c,t},\widehat{ \tau}_{c,t})
\end{eqnarray*}
with known overdispersion parameter $\widehat{ \tau}_{c,t}$  and mean  $\E[ Y_{c,t} | \theta_{c,t}] = \widehat E_{c,t}\theta_{c,t}$ and $\v( Y_{c,t} | \theta_{c,t}  ) = \widehat{E}_{c,t}\theta_{c,t}\left(1+\widehat{E}_{c,t}\theta_{c,t}/ \widehat{ \tau}_{c,t}\right)$ with
\begin{eqnarray}\label{eq:log-linearsup}
\log \theta_{c,t} &=& \alpha + 
\sum_{b=1}^B  \beta_{bt}  X_{bct} + \sum_{g=1}^G \gamma_g Z_{gc} + \epsilon_{c,t},
%\log \theta_{c,t} &=& \alpha^\text{\tiny{H}}_{t} 1( c \in \mbox{ high }) + \alpha^\text{\tiny{N}}_{t} 1( c \in \mbox{ not high }) \\
%&+& \beta^\text{\tiny{H}}_{t} 1( c \in \mbox{ high }) X_{c,t} + \beta^\text{\tiny{N}}_{t} 1( c \in \mbox{ not high }) X_{c,t}
\end{eqnarray}
and $\epsilon_{c,t} \sim_{iid} \N(0,\sigma_{\epsilon}^2)$.

We wish to assess whether the covariate model provides a good fit to the data. To this end we perform a number of model checks.

\subsection{Fitted Values}
\begin{figure}[htp]
\centering
\includegraphics[scale=.2]{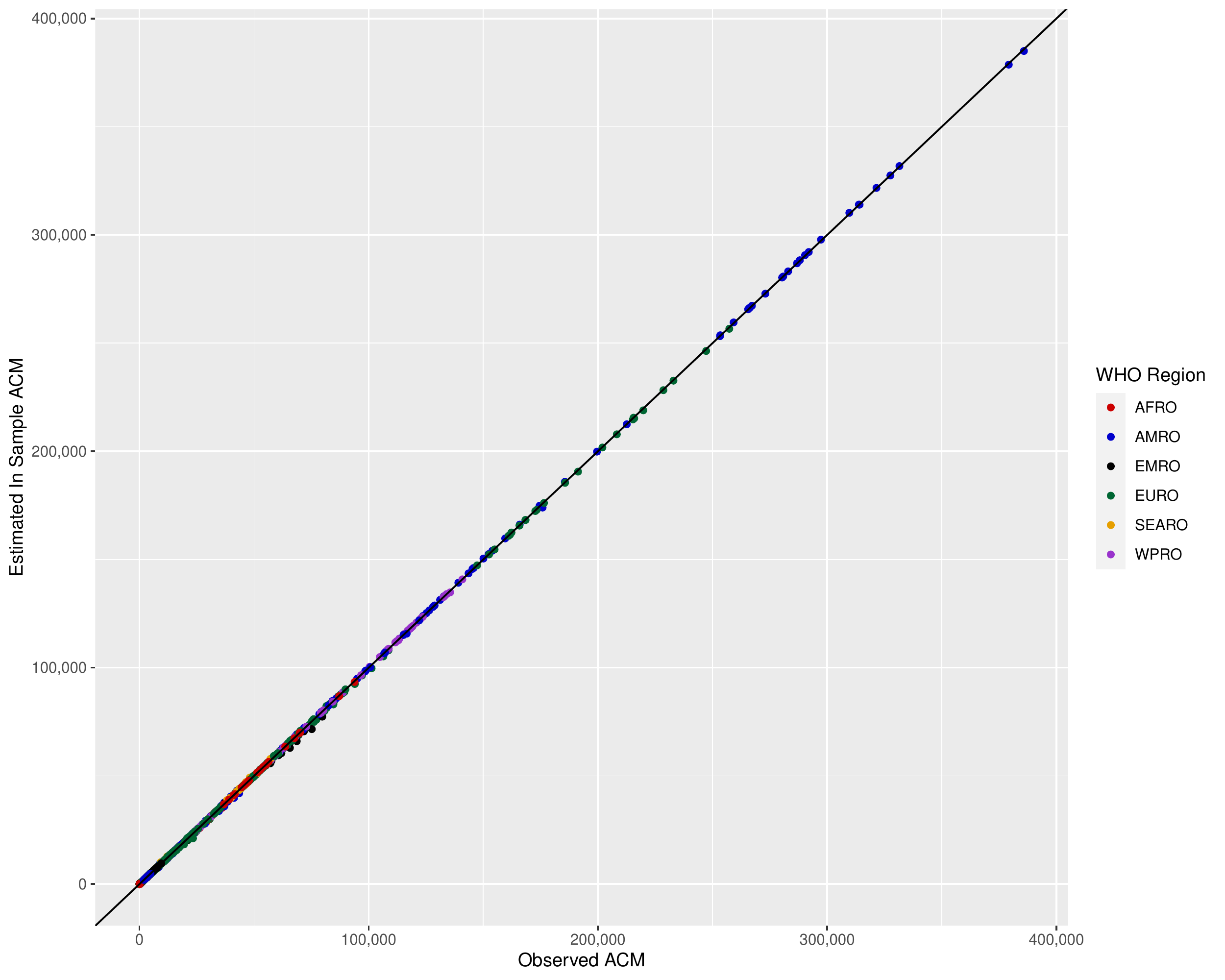}
\caption{In-sample observed versus predicted, color-coded by region. }
\label{fig:InSampleObservedVsPredicted}
\end{figure}

\begin{figure}[htp]
\centering
\includegraphics[scale=.2]{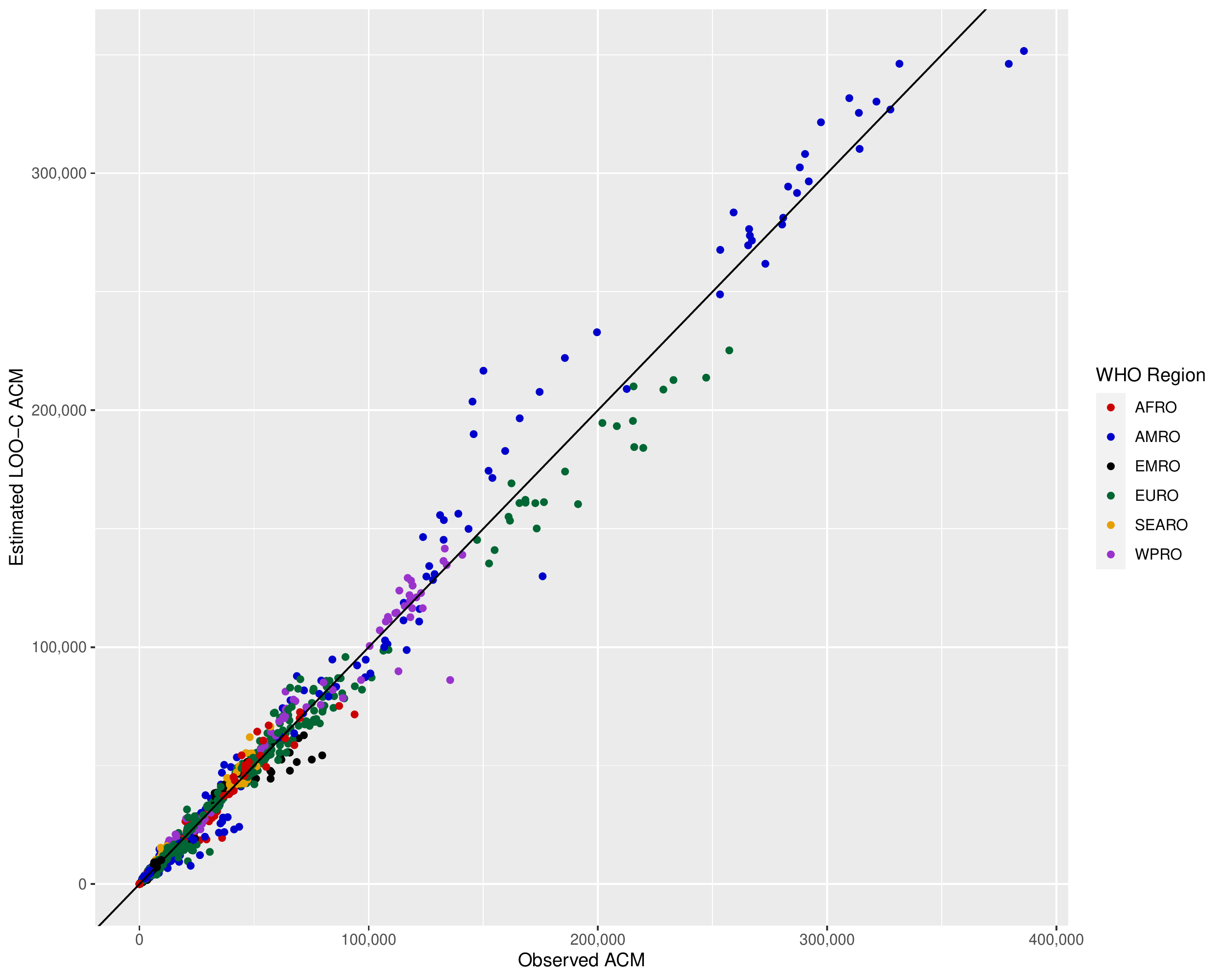}
\includegraphics[scale=.2]{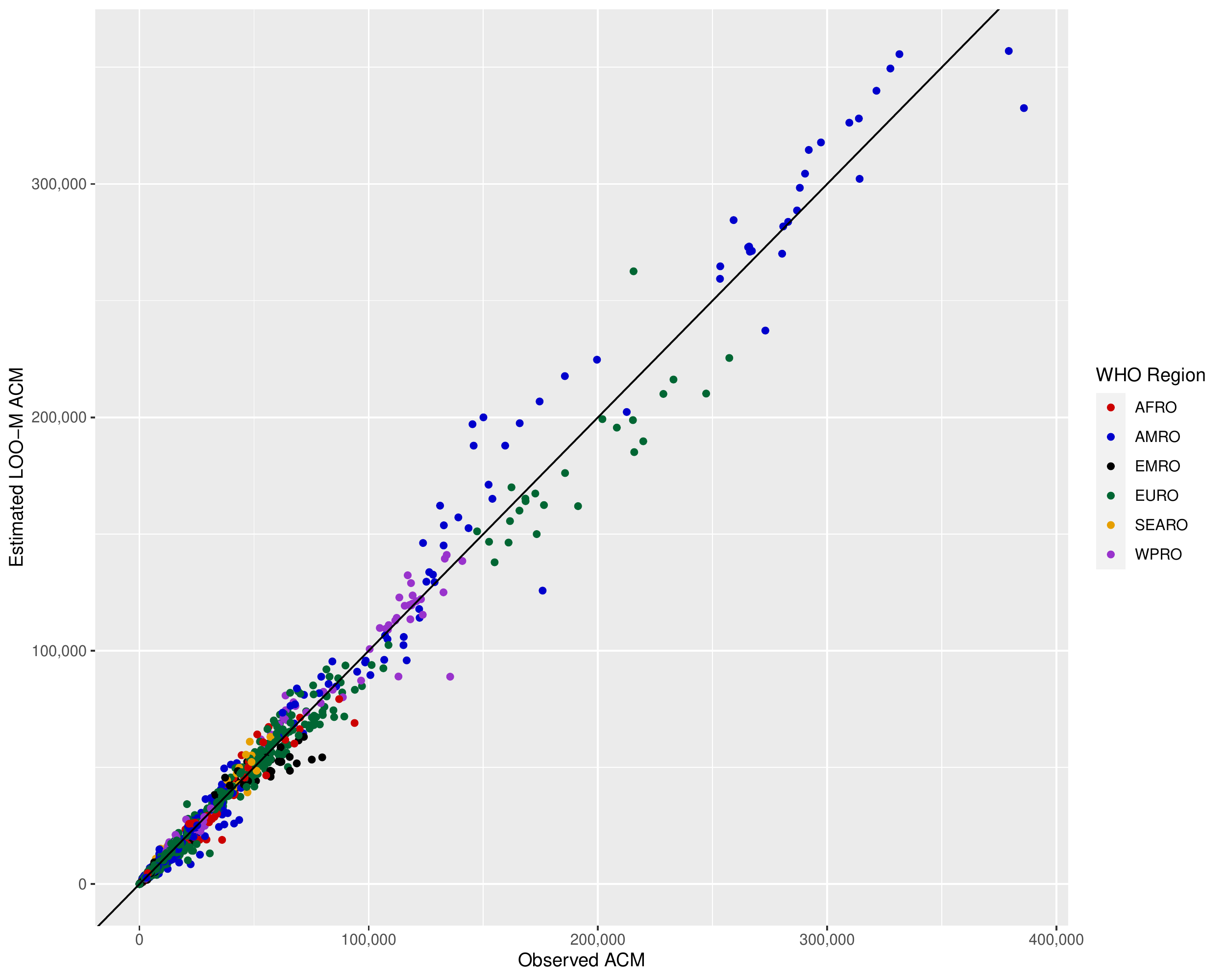}
\caption{Out-of-sample observed versus predicted: Left: country removed. Right: month removed. Color-coded by region. }
\label{fig:CVObservedVsPredicted}
\end{figure}

We plot fitted values $\widehat y_{c,t}$ versus observed values $y_{c,t}$. The fitted values are given by
$\widehat Y_{c,t}=  \widehat E_{c,t}\widehat \theta_{c,t}$ where $\widehat \theta_{c,t}$ is the posterior median. We color code the points by region.
These plots are created both for in-sample in Figure \ref{fig:InSampleObservedVsPredicted}, and out-of-sample (via cross-validation in which data from either a complete country or a complete month are removed) in Figure \ref{fig:CVObservedVsPredicted}). As we would expect, the in-sample points almost all lie on the line of equality (since we include the random effects $\epsilon_{c,t}$ for each country-time point combination). The out-of-sample versions are less good, as expected, but nothing jumps out as being particularly aberrant.

\subsection{Standardized Residuals}
%\clearpage
We plot standardized residuals versus time (to see if there are is systematic model misspecification over time) and versus the log of the fitted values (to assess whether the mean-variance relationship is adequate), in both cases color coded by region. Standardized residuals are:
$$r_{c,t} = \frac{y_{c,t} - \widehat y_{c,t}}
{\sqrt{ \widehat E_{c,t}\widehat \theta_{c,t}\left(1+ \widehat E_{c,t}\widehat \theta_{c,t}/\widehat{ \tau}_{c,t}\right)}}.
$$

\begin{figure}[htbp]
\centering
\includegraphics[scale=.25]{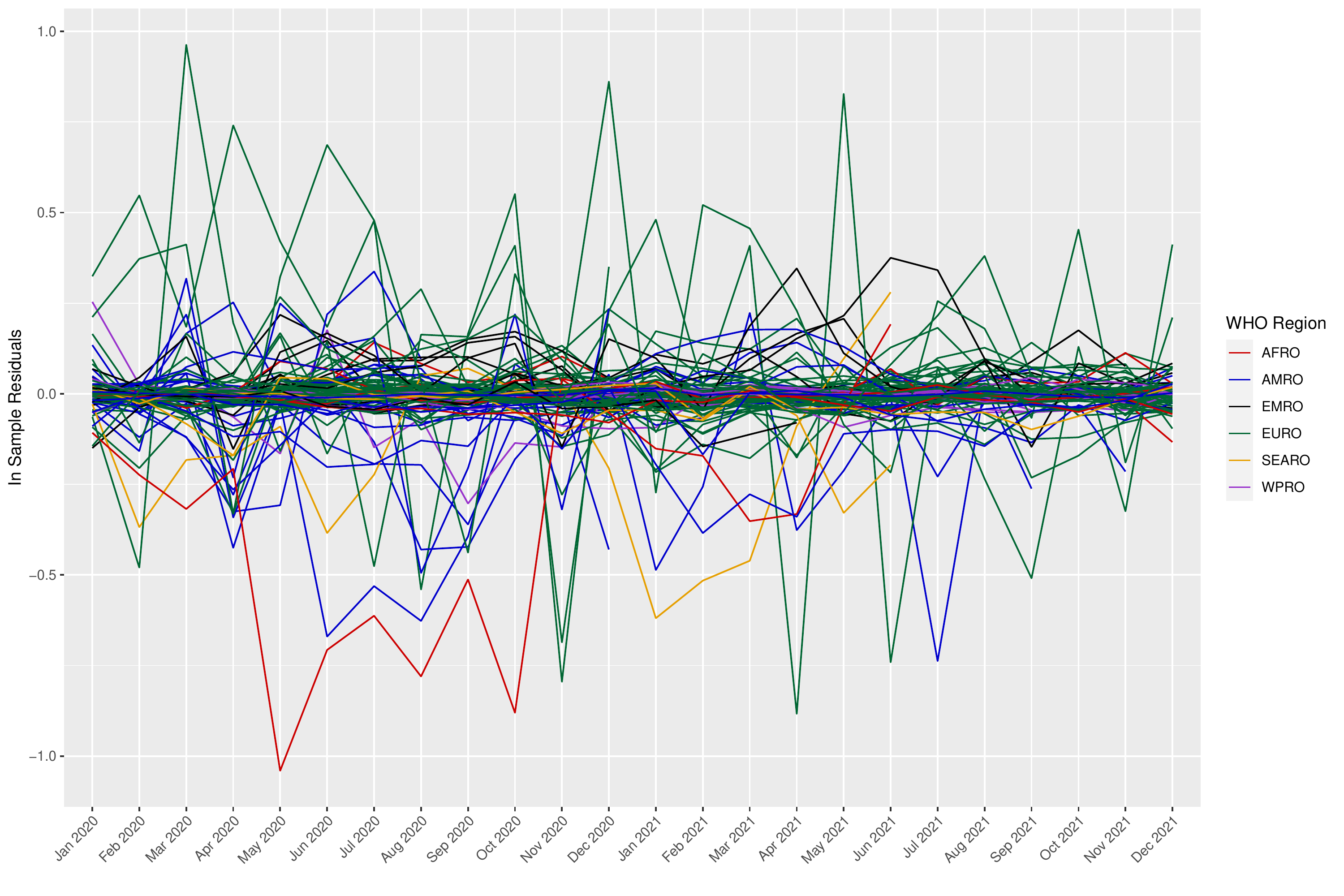}
\caption{In-Sample standardized residuals over time, color-coded by region. }
\label{fig:InSampleResidualsVsTime}
\end{figure}

\begin{figure}[htbp]
\centering
\includegraphics[scale=.25]{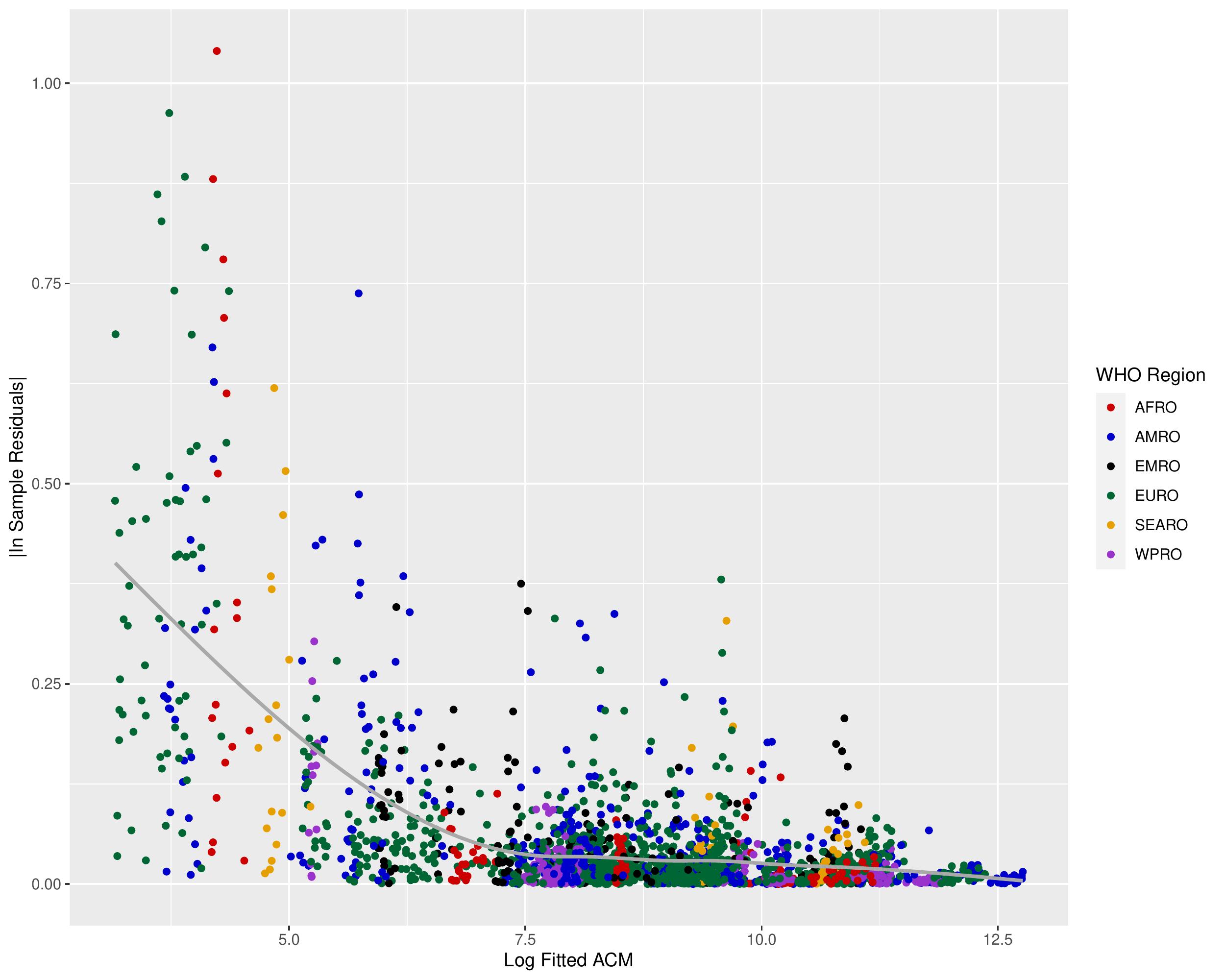}
\includegraphics[scale=.25]{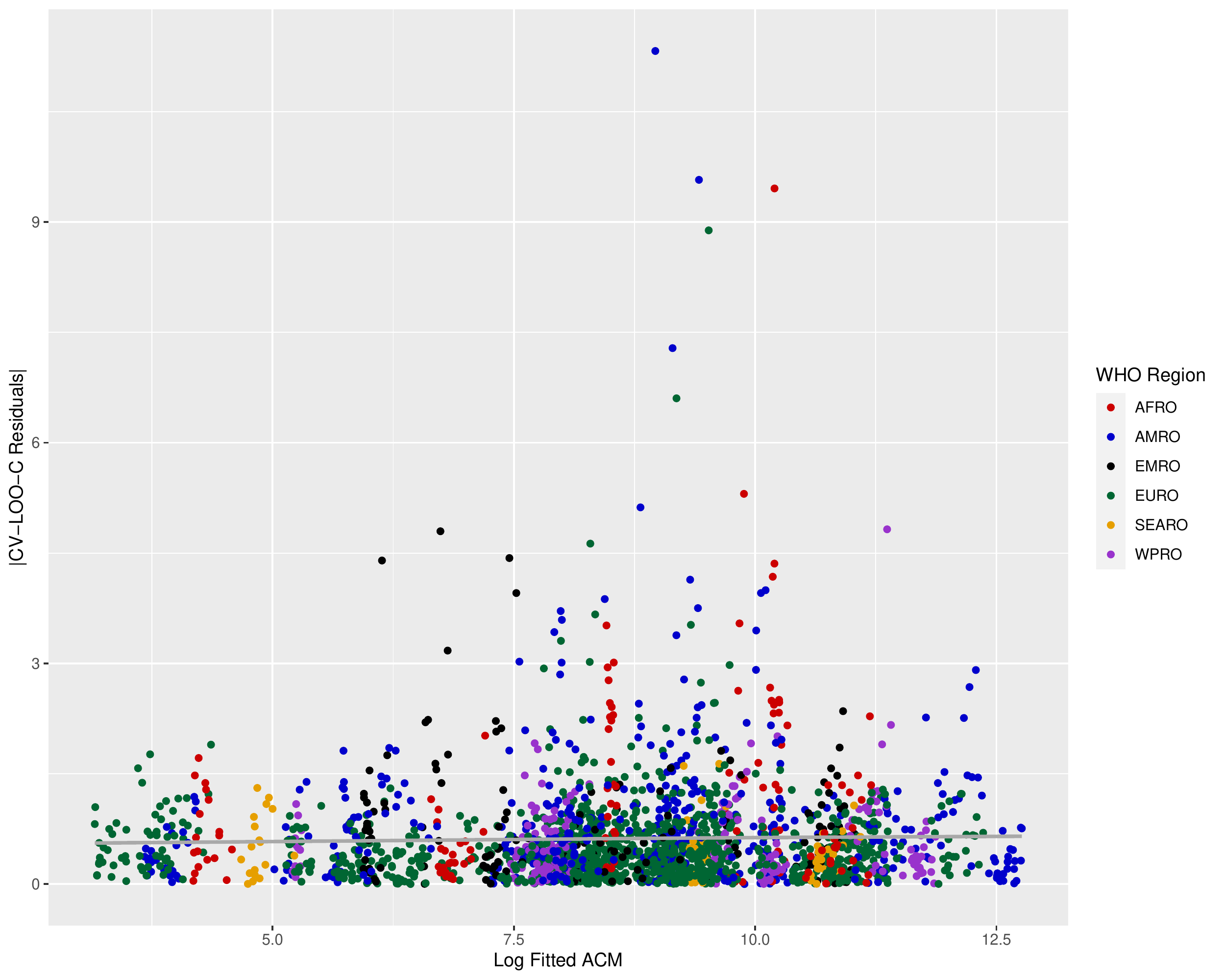}
%\caption{}
%\label{fig:AbsValueCVResidVsLogFitted}
%\end{figure}
%
% \begin{figure}[htbp]
%\centering
\caption{Absolute values of residuals versus log fitted values. Left: In-sample versions. Right: Leave-one-country out versions. Smoothers are drawn on each.}
\label{fig:AbsValueInSampleResidVsLogFitted}
\end{figure}

Figure \ref{fig:InSampleResidualsVsTime} shows in-sample standardized residuals over time -- it is difficult to make any definitive statements from this plot. There are some relatively high residuals for low fitted values, but the total number of such points is small, and their contribution to the overall excess picture is small (the countries with values below 5 in log fitted ACM are (in order from left to right in the plot) San Marino, Monaco, Saint Kitts and Nevis, Andorra, Antigua and Barbuda, Seychelles and the Maldives). Figure \ref{fig:AbsValueInSampleResidVsLogFitted} shows absolute standardized residuals versus log fitted values, with a smoother added, for both in-sample and leave-one-out residuals.  In both  plots, there is no systematic pattern for larger values, which is important, as this would indicate a deficiency in the mean-variance relationship.

\clearpage
\subsection{Bias and RMSE}
We assess the errors in our model, also using CV, over the countries with ACM data. Let $r_{c,t} = Y_{c,t}/N_{c,t}$ be the observed ACM rate and $\widehat{r}_{c,t}=\widehat{Y}_{c,t}  /N_{c,t}$ where $\widehat Y_{c,t} = \mbox{PostMedian}({Y}_{c,t} | \by_{-ct})$ is the estimated rate. We report the relative bias of the ACM rate,
\begin{equation}\label{eq:error1}
 \frac{1}{\sum_c |M_c|} \sum_c \sum_{t \in M_c} \frac{\widehat r_{c,t}  - r_{c,t}}{r_{c,t}},
\end{equation}
 and the absolute version of this quantity,
\begin{equation}\label{eq:error2}
\frac{1}{\sum_c |M_c|} \sum_c \sum_{t \in M_c} \frac{|\widehat r_{c,t}  - r_{c,t}|}{r_{c,t}}.
\end{equation}
These measures can be calculated with the estimated rates being based on data with either a complete country's worth of data or a complete months worth of data being left out.

%\begin{eqnarray}
%\mbox{R-RATE-C} &=&100 \times \frac{1}{M}  \sum_c \sum_{t \in M_c}\left(
%\frac{\widehat \delta_{c,t}/N_{c,t} - \delta_{c,t}/N_{c,t} }{\delta_{c,t}/N_{c,t} }
%\right)\\
%&=&100 \times  \frac{1}{M}  \sum_c \sum_{t \in M_c}\left(
%\frac{\widehat \delta_{c,t} - \delta_{c,t} }{\delta_{c,t}}
%\right) \\&=& 100 \times  \frac{1}{M}  \sum_c \sum_{t \in M_c}\left(
%\frac{\widehat Y_{c,t} - y_{c,t} }{\delta_{c,t}}
%\right)
%\end{eqnarray}
%where $\widehat \delta_{c,t}$ is obtained from a model in which the data from country $c$ is removed, and $M=\sum_c |M_c|$\ with $M_c$ the set of time periods available for each country.

We also calculate the root mean squared error (RMSE) of the fit:
$$
\sqrt{\frac{1}{\sum_c |M_c|} \sum_c \sum_{t \in M_c} (\widehat r_{c,t}  - r_{c,t})^2}
$$
again using the two cross-validation schemes (by month and by country).
We also estimate the coverage of the predictive intervals from these CV exercises, at the 50\%, 80\% and 95\% levels.

Table \ref{tab:datamortsummary} shows the summaries. The relative biases are small (just under 2\%), while the absolute relative biases are around 10\%. The RMSE measures are around 1.25 $\times 10^3$. Under both CV schemes, the coverages are a little higher than the nominal for the 50\% and 80\% levels and a little lower than the nominal for the 95\% level.

\begin{table}[htp]
\begin{center}
\begin{tabular}{l|l|c}
CV Level & Measure& Performance\\ \hline
Country& Relative Bias & 1.98 \\
Country& Absolute Relative Bias & 10.08 \\
Country &RMSE ($\times 1000$) & 1.25 \\
Country &Coverage: 50\% Interval & 59.3\\
Country &Coverage: 80\% Interval & 82.7\\
Country &Coverage: 95\% Interval & 91.6\\
\hline
Month &Relative Bias & 1.84 \\
Month &Absolute Relative Bias  & 10.18 \\
Month &RMSE ($\times 1000$)& 1.24 \\
Month &Coverage: 50\% Interval & 57.8\\
Month &Coverage: 80\% Interval & 83.7\\
Month &Coverage: 95\% Interval & 92.9\\
\end{tabular}
\end{center}
\caption{Leave one country and month out model assessment measures. Relative bias and absolute relative bias are expressed as percentages.}\label{tab:datamortsummary}
\end{table}

%\begin{table}[htp]
%
%\begin{center}
%
%\begin{tabular}{l|c|c|c|c|}
%
%Metric&X-Validation&  Covariate Model \\ \hline
%
%Abs Err of ACM Rate &Monthly&10.08 \\
%Rel Err of ACM Rate &Monthly& \\
%Abs Err of ACM Rate &Yearly& \\
%Rel Err of ACM Rate &Yearly& \\
%\end{tabular}
%\end{center}
%\caption{Leave one month and country model assessment measures.}\label{tab:datamortsummary}
%
%\end{table}
%
%\begin{tabular}{l|c|c|c|c|c|c|}
%
%& Baseline & Model 3 & Model 4 & Final Model \\ \hline
%
%A-ERR-C & 654.32 & 406.40 & 356.54 & 375.4\\
%
%R-ERR-C & 0.1272 & 0.0768 & 0.0714 & 0.0731\\
%A-ERR-M & 658.67 & 413.74 & 417.10 & 396.31\\
%R-ERR-M & 0.1268 & 0.08435 & 0.0822 & 0.0758
%\end{tabular}

%IHME use two model assessment measures for the excess $\delta_{c,t}=Y_{c,t}-E_{c,t}$.

%Mean relative error of predicted excess mortality rate, as used in   \cite{wang:2022:covid}:

% \begin{figure}[htbp]
%\centering
%\includegraphics[scale=.3]{VictoriaFigures/CVObservedVsPredicted_Month.pdf}
%\caption{Out-of-sample (month removed) observed versus predicted, color-coded by region. }
%\label{fig:PbyMonth_Argentina}
%\end{figure}

\clearpage
\section{Supplementary Materials: Data Types By Country}

On the following pages we list the countries for which we produced excess mortality estimates, along with the regions within which they lie, and the type of data they have available.
%\clearpage
\begin{table}[h]
\begin{tabular}{ |p{6cm}|p{3cm}|p{3cm}|  }
\hline
\multicolumn{3}{|c|}{Country List} \\
\hline
Country & WHO Region & Data Type \\
\hline
Afghanistan & EMRO & No Data \\
Albania & EURO & Full National \\
Algeria & AFRO & Partial National \\
Andorra & EURO & Partial National \\
Angola & AFRO & No Data \\
Antigua and Barbuda & AMRO & Partial National \\
Argentina & AMRO & Partial National / Subnational Data \\
Armenia & EURO & Full National \\
Australia & WPRO & Full National \\
Austria & EURO & Full National \\
Azerbaijan & EURO & Full National \\
Bahamas & AMRO & No Data \\
Bahrain & EMRO & No Data \\
Bangladesh & SEARO & No Data \\
Barbados & AMRO & Partial National \\
Belarus & EURO & Partial National \\
Belgium & EURO & Full National \\
Belize & AMRO & Partial National \\
Benin & AFRO & No Data \\
Bhutan & SEARO & No Data \\
Bolivia (Plurinational State of) & AMRO & Full National \\
Bosnia and Herzegovina & EURO & Full National \\
Botswana & AFRO & No Data \\
Brazil & AMRO & Full National \\
Brunei Darusssalam & WPRO & Partial National \\
Bulgaria & EURO & Full National \\
Burkina Faso & AFRO & No Data \\
Burundi & AFRO & No Data \\
Cabo Verde & AFRO & No Data \\
Cambodia & WPRO & No Data \\
Cameroon & WPRO & No Data \\
Canada & AMRO & Partial National \\
Central African Republic & AFRO & No Data \\
Chad & AFRO & No Data \\
Chile & AMRO & Full National \\
China & WPRO & Annual Data \\
Colombia & AMRO & Full National \\
Comoros & AFRO & No Data \\
Congo & AFRO & No Data \\
Cook Islands & WPRO & No Data \\
Costa Rica & AMRO & Partial National \\
C\^ote d'Ivoire & AFRO & No Data \\
Croatia & EURO & Full National \\
Cuba & AMRO & Partial National \\
Cyprus & EURO & Full National \\
Czechia & EURO & Full National \\
Democratic People's Republic of Korea & SEARO & No Data \\
Democratic Republic of the Congo & AFRO & No Data \\
Denmark & EURO & Full National \\
\hline
\end{tabular}
\end{table}
\begin{table}
\begin{tabular}{ |p{6cm}|p{3cm}|p{3cm}|  }
\hline
\multicolumn{3}{|c|}{Country List} \\
\hline
Country & WHO Region & Data Type \\
\hline
Djibouti & EMRO & No Data \\
Dominica & AMRO & No Data \\
Dominican Republic & AMRO &  Partial National \\
Ecuador & AMRO & Full National \\
Egypt & EMRO & Partial National \\
El Salvador & AMRO & Partial National \\
Equatorial Guinea & AFRO & No Data \\
Eritrea & AFRO & No Data \\
Estonia & EURO & Full National \\
Eswatini & AFRO & No Data \\
Ethiopia & AFRO & No Data \\
Fiji & WPRO & No Data \\
Finland & EURO & Full National \\
France & EURO & Full National \\
Gabon & AFRO & No Data \\
Gambia & AFRO & No Data \\
Georgia & EURO & Partial National \\
Germany & EURO & Full National \\
Ghana & AFRO & No Data \\
Greece & EURO & Full National \\
Grenada & AMRO & Annual Data \\
Guatemala & AMRO & Full National \\
Guinea & AFRO & No Data \\
Guinea-Bisssau & AFRO & No Data \\
Guyana & AMRO & No Data \\
Haiti & AMRO & No Data \\
Honduras & AMRO & No Data \\
Hungary & EURO & Full National \\
Iceland & EURO & Full National \\
India & SEARO & Subnational Data \\
Indonesia & SEARO & Subnational Data \\
Iran (Islamic Republic of) & EMRO & Full National \\
Iraq & EMRO & Partial National \\
Ireland & EURO & Full National \\
Israel & EURO & Full National \\
Italy & EURO & Full National \\
Jamaica & AMRO & Partial National \\
Japan & WPRO & Full National \\
Jordan & EMRO & Partial National \\
Kazakhstan & EURO & Full National \\
Kenya & AFRO & Full National \\
Kiribati & WPRO & No Data \\
Kuwait & EMRO & Partial National \\
Kyrgyzstan & EURO & Full National \\
Lao People's Democratic Republic & WPRO & No Data \\
Latvia & EURO & Full National \\
Lebanon & EMRO & Full National \\
Lesotho & AFRO & No Data \\
Liberia & AFRO & No Data \\
Libya & EMRO & No Data \\
Lithuania & EURO & Full National \\
Luxembourg & EURO & Full National \\
Madagascar & AFRO & No Data \\
\hline
\end{tabular}
\end{table}

\begin{table}
\begin{tabular}{ |p{6cm}|p{3cm}|p{3cm}|  }
\hline
\multicolumn{3}{|c|}{Country List} \\
\hline
Country & WHO Region & Data Type \\
\hline
Malawi & AFRO & No Data \\
Malaysia & WPRO & Partial National \\
Maldives & SEARO & Partial National \\
Mali & AFRO & No Data \\
Malta & EURRO & Full National \\
Marshall Islands & WPRO & No Data \\
Mauritania & AFRO & No Data \\
Mauritius & AFRO & Full National \\
Mexico & AMRO & Full National \\
Micronesia (Federated States of) & WPRO & No Data \\
Monaco & EURO & Full National \\
Mongolia & WPRO & Full National \\
Montenegro & EURO & Partial National \\
Morocco & EMRO & No Data \\
Mozambique & AFRO & No Data \\
Myanmar & SEARO & No Data \\
Namibia & AFRO & No Data \\
Nauru & WPRO & No Data \\
Nepal & SEARO & No Data \\
Netherlands & EURO & Full National \\
New Zealand & WPRO & Full National \\
Nicaragua & AMRO & Partial National \\
Niger & AFRO & No Data \\
Nigeria & AFRO & No Data \\
Niue & WPRO & No Data \\
North Macedonia & EURO & Full National \\
Norway & EURO & Full National \\
Oman & EMRO & Full National \\
Pakistan & EMRO & No Data \\
Palau & WPRO & No Data \\
Panama & AMRO & Partial National \\
Papua New Guinea & WPRO & No Data \\
Paraguay & AMRO & Full National \\
Peru & AMRO & Full National \\
Phillipines & WPRO & Partial National \\
Poland & EURO & Full National \\
Portugal & EURO & Full National \\
Qatar & EMRO & Full National \\
Republic of Korea & WPRO & Full National \\
Republic of Moldova & EURO & Full National \\
Romania & EURO & Full National \\
Russian Federation & EURO & Full National \\
Rwanda & AFRO & No Data \\
Saint Kitts and Nevis & AMRO & Annual Data \\
Saint Lucia & AMRO & No Data \\
Saint Vincent and the Grenadines & AMRO & Annual Data \\
Samoa & WPRO & No Data \\
San Marino & EURO & Full National \\
Sao Tome and Principe & AFRO & No Data \\
Saudi Arabia & EMRO & No Data \\
Senegal & AFRO & No Data \\
\hline
\end{tabular}
\end{table}

\begin{table}
\begin{tabular}{ |p{6cm}|p{3cm}|p{3cm}|  }
\hline
\multicolumn{3}{|c|}{Country List} \\
\hline
Country & WHO Region & Data Type \\
\hline
Serbia & EURO & Full National \\
Seychelles & AFRO & Partial National \\
Sierra Leone & AFRO & No Data \\
Singapore & WPRO & Full National \\
Slovakia & EURO & Full National \\
Slovenia & EURO & Full National \\
Solomon Islands & WPRO & No Data \\
Somalia & EMRO & No Data \\
South Africa & AFRO & Full National \\
South Sudan & AFRO & No Data \\
Spain & EURO & Full National \\
Sri Lanka & SEARO & Annual Data \\
Sudan & EMRO & No Data \\
Suriname & AMRO & Partial National \\
Sweden & EURO & Full National \\
Switzerland & EURO & Full National \\
Syrian Arab Republic & EMRO & No Data \\
Tajikistan & EURO & Partial National \\
Thailand & SEARO & Full National \\
The United Kingdom & EURO & Full National \\
Timor-Leste & SEARO & No Data \\
Togo & AFRO & Full National \\
Tonga & WPRO & No Data \\
Trinidad and Tobago & AMRO & No Data \\
Tunisia & EMRO & Partial National \\
Turkey & EURO & Subnational Data \\
Turkmenistan & EURO & No Data \\
Tuvalu & WPRO & No Data \\
Uganda & AFRO & No Data \\
Ukraine & EURO & Full National \\
United Arab Emirates & EMRO & No Data \\
United Republic of Tanzania & AFRO & No Data \\
United States of America & AMRO & Full National \\
Uruguay & EURO & Full National \\
Uzbekistan & EURO & Full National \\
Vanuatu & WPRO & No Data \\
Venezuela (Bolivarian Republic of) & AMRO & No Data \\
Viet Nam & WPRO & Annual Data \\
Yemen & EMRO & No Data \\
Zambia & AFRO & No Data \\
Zimbabwe & AFRO & No Data \\
\hline
\end{tabular}
\caption{Countries, regions and data scenarios}
\end{table}

\clearpage

\bibliographystyle{natbib} 
\bibliography{/Users/jonno/Dropbox/BibFiles/spatepi}